\documentclass[12pt]{article}

\usepackage{fullpage}
\usepackage{amsmath}
\usepackage{amssymb}
\usepackage{graphicx}
\usepackage{subfigure}
\usepackage{color,soul} 

\newcommand{\cf}{\mathrm{CF}}
\newcommand{\cfnetworksmajor}[2]{(1-\gamma)(1-2/3\,e^{-#1})+\gamma(1-2/3\,e^{-#2})}
\newcommand{\cfnetworksminor}[2]{(1-\gamma)(1/3)e^{-#1}+\gamma(1/3)e^{-#2}}
\newcommand{\cftreemajor}[1]{1-2/3\,e^{-#1}} 
\newcommand{\cftreeminor}[1]{1/3\,e^{-#1}} 

\newcommand{\beginsupplement}{%
        \setcounter{table}{0}
        \renewcommand{\thetable}{S\arabic{table}}%
        \setcounter{figure}{0}
        \renewcommand{\thefigure}{S\arabic{figure}}%
     }

\newcommand*{\titleGP}{\begingroup 
\centering 
\vspace*{\baselineskip} 

\rule{\textwidth}{1.6pt}\vspace*{-\baselineskip}\vspace*{2pt} 
\rule{\textwidth}{0.4pt}\\[.5\baselineskip] 

{\LARGE Inferring phylogenetic networks with maximum \\ [0.5\baselineskip]
  pseudolikelihood under incomplete lineage sorting} \\ [0.4\baselineskip] 

\rule{\textwidth}{0.4pt}\vspace*{-\baselineskip}\vspace{3.2pt} 
\rule{\textwidth}{1.6pt}\\[\baselineskip] 

{\scshape 
Claudia Sol\'{i}s-Lemus${}^1$ and C\'{e}cile An\'{e}${}^{1,2}$}

{\itshape ${}^1$Department of Statistics, ${}^2$ Department of Botany\par}

{\itshape University of Wisconsin - Madison\par}
\endgroup}

\begin{document}
\titleGP

\begin{abstract}
Phylogenetic networks are necessary to represent the tree of life
      expanded by edges to represent events such as horizontal gene
      transfers, hybridizations or gene flow. Not all species follow
      the paradigm of vertical inheritance of their genetic
      material. While a great deal of research has flourished into the
      inference of phylogenetic trees, statistical methods to infer
      phylogenetic networks are still limited and under
      development. The main disadvantage of existing methods is a lack
      of scalability. Here, we present a statistical method to infer
      phylogenetic networks from multi-locus genetic data in a
      pseudolikelihood framework. Our model accounts for
      incomplete lineage sorting through the coalescent model, and
      for horizontal inheritance of genes through reticulation nodes in
      the network.
      Computation of the pseudolikelihood is fast and simple, and it
      avoids the burdensome calculation of the full likelihood which
      can be intractable with many species. Moreover, estimation at
      the quartet-level has the added computational benefit that it is
      easily parallelizable. Simulation studies comparing our method
      to a full likelihood approach show that our pseudolikelihood
      approach is much faster without compromising accuracy. We
      applied our method to reconstruct
      the evolutionary
      relationships among swordtails and platyfishes
      (\textit{Xiphophorus}: Poeciliidae),
      which is characterized by widespread hybridizations.
\end{abstract}

Evolutionary relationships are typically visualized in a tree,
which implicitly assumes vertical
transfer of genetic material from ancestors to descendants.  However,
not all species follow this paradigm. If 
genes can be
horizontally transferred between some organisms, a tree is
not a good representation of their history.  Such reticulate
events include hybridization, horizontal gene transfer or
migration with gene flow, and 
require methods to infer phylogenetic networks. 
While a great deal of research has flourished for the inference of
phylogenetic trees from different types of data, methods to infer
phylogenetic networks are still limited and under development.

There are mainly two kinds of phylogenetic networks: implicit and
explicit.  Implicit networks --also called split networks-- describe
the discrepancy in gene trees, or other sources of data, and methods
are well developed to reconstruct these networks
\cite{Huson2010,Spillner2013,Than2008, Yang2014}.  These methods tend
to be fast.  However, implicit networks lack biological interpretation
as the internal nodes do not represent ancestral species.
Explicit networks, on the other hand,
represent explicit reticulation events and each node
represents an ancestral species.
Combinatorial methods to infer
explicit networks (which we call phylogenetic networks here)
are fast but ignore gene tree error and incomplete lineage sorting (ILS)
as a possible source of gene tree discordance (e.g. \cite{gambette2012}).
Model-based methods are most accurate but can be computationally
challenging. They calculate the
likelihood of an observed gene tree given a species network taking
into account both reticulation and ILS
\cite{Yu2012a, Meng2009, Strimmer2000}.
Their scope was expanded in \cite{Yu2014}
to search for the most likely
phylogenetic network based on multi-locus data
(see also \cite{Nguyen2015} for a different likelihood framework, where
sites instead of genes are treated as independent and ILS is ignored).
The likelihood-based method in \cite{Yu2014}, implemented in PhyloNet,
provides a solid theoretical framework to estimate the
maximum likelihood phylogenetic network from a set of gene
trees. It has several advantages:
it incorporates uncertainty on the gene trees
estimated from sequence data, accounts for a background level of
gene tree discordance due to ILS,
and controls the complexity of the network with a cross validation step.
However, its likelihood computation is heavy and becomes
intractable when increasing the number of taxa
or the number of hybridizations, making this method practical
for small scenarios of up to about 10 species and 4 hybridizations
in the network.

Here, we provide a fast statistical method to estimate phylogenetic
networks from multi-locus data.
We first present the theory for the pseudolikelihood of a
network. We do so by deriving the proportion of the genome that
has each 4-taxon tree (quartet concordance factors)
as expected under the coalescent model extended by
hybridization events, and we prove the generic identifiability of the
model.
We then use the observed quartet
concordance factors as inferred from the multi-locus data
to estimate the species network. Our method SNaQ (Species Networks
applying Quartets) is implemented in our open-source software package
PhyloNetworks in Julia 
and publicly available at \verb+https://github.com/crsl4+.

Like PhyloNet, our method can incorporate uncertainty in estimated
gene trees and gene tree discordance due to ILS. Our pseudolikelihood
has computational advantages. It is simpler and more scalable to many
species, compared to the full likelihood.  It also scales to a large
number of loci because estimation of gene trees can be highly
parallelized, then summarized by only 3 tree frequencies on each 4-taxon subsets
used as input in the pseudolikelihood.
In simulations, our method showed good performance and scaled to scenarios
for which PhyloNet could not run. 
We also used SNaQ to infer the evolutionary
relationships between \textit{Xiphophorus} fishes,
from 1,183 loci across 24 taxa. Our results were
congruent with \cite{Cui2013}
and refined the placement of some hybridizations found in that study.
The analyses here presented show that SNaQ can enable
scientists to incorporate organisms to the ``tree of life'' in parts
that are more net-like than tree-like, and thus, complete a broader
picture of evolution.

\section*{Models}
\subsection*{Phylogenetic networks}
\label{methods}
Intuitively, a phylogenetic network is a
phylogenetic tree with added hybrid edges, causing some nodes
to have two parents (but see \cite{francisSteel2015}).
Phylogenetic networks can describe various biological processes
causing gene flow from one population to another
such as hybridization, introgression, or horizontal gene transfer.
Hybridization occurs when individuals from 2 genetically distinct
populations interbreed, resulting in a new separate population.
Introgression, or introgressive hybridization, is the integration of
alleles from one population into another existing population,
through hybridization and backcrossing.
Genes are horizontally transferred when acquired by a population
through a process other than reproduction, from a possibly distantly
related population. 
Although these three processes are biologically different,
we do not make the distinction when modeling them with a network.
In other words, our model takes into
account all three biological scenarios, but those
scenarios are not distinguishable in the estimated phylogenetic
network unless more biological information is provided.

Just like phylogenetic trees, networks can be rooted or
unrooted.  A \textit{rooted phylogenetic network} on
taxon set $X$ is a connected directed acyclic graph with vertices
$V=\{r\} \cup V_L \cup V_H \cup V_T$, edges $E=E_H \cup E_T$ and a
bijective leaf-labeling function $f:V_L \rightarrow X$ with the
following characteristics. The root $r$ has indegree $0$ and
outdegree $2$. Any leaf $v \in V_L$ has indegree $1$ and outdegree
$0$. Any tree node $v \in V_T$ has indegree $1$ and outdegree
$2$. Any hybrid node $v \in V_H$ has indegree $2$ and outdegree
$1$. A tree edge $e \in E_T$ is an edge whose child is a tree
node. A hybrid edge $e \in E_H$ is an edge whose child is a hybrid
node.  Unrooted phylogenetic networks are typically obtained
by suppressing the root node and the direction of all edges. We
also consider \textit{semi-directed} unrooted networks, where the root
node is suppressed 
and we ignore the direction of all tree edges, but we
maintain the direction of hybrid edges, thus keeping information on
which nodes are hybrids.
The placement of the root is then constrained,
because the direction of the two hybrid edges to a given
hybrid node inform the direction of time at this node:
the third edge must be a tree edge directed away from
the hybrid node and leading to all the hybrid's descendants.
Therefore the root cannot be placed on any descendant of any hybrid
node, although it might be placed on some hybrid edges.

We further assume that the true network is of
\textit{level-1} \cite{Huson2010}, i.e.  any
given edge can be part of at most one cycle. This means that there
is no overlap between any two cycles (but see the Discussion).
Refer to \cite{Huson2010} for other types of evolutionary networks.
Throughout this work, we denote by

\begin{figure}
\includegraphics[scale=0.15]{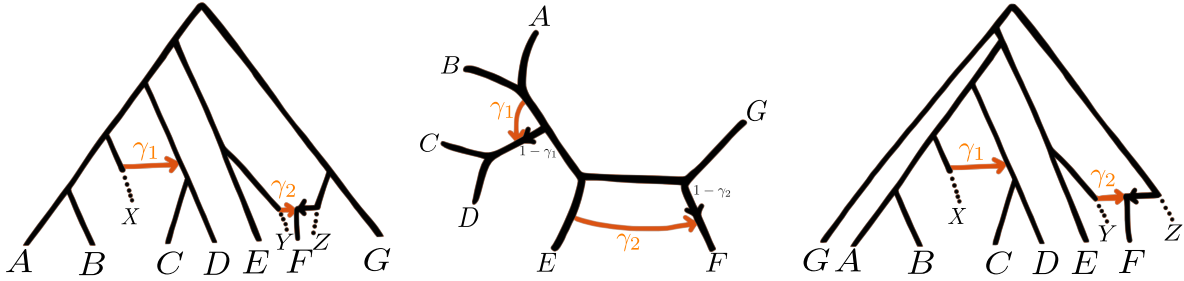}
\caption{\textbf{Example of rooted and semi-directed phylogenetic networks with
 $h=2$ hybridization events and $n=7$ sampled taxa.}
 Inheritance
 probabilities $\gamma$ represent the proportion of genes contributed
 by each parental population to a given hybrid node.
 Left: rooted network modelling several biological processes.
 Taxon F is a
 hybrid between two non-sampled taxa Y and Z with $\gamma_2\approx 0.50$,
 and the lineage ancestral to taxa C and D has received genes
 introgressed from a non-sampled taxon X, for which $\gamma_1\approx 0.10$.
 An alternative process at this event could be the horizontal
 transfer of only a handful of genes, corresponding to a very
 small fraction $\gamma_1\approx 0.001$.
 Center: semi-directed network for the biological scenario
 just described. Although the root location is unknown, its position
 is constrained by the direction of hybrid edges (directed by
 arrows).  For example, C, G or E cannot be outgroups.
 Right: rooted network obtained from the semi-directed network
 (center) by placing the root on the hybrid edge that leads to taxon F
 (labeled by $1-\gamma_2$).} 
\label{exampleNet}
\end{figure}

\begin{itemize}
\item{$n$ the number of taxa,}
\item{$h$ the number of hybridization events and}
\item{$k_i$ the number of nodes in the 
undirected cycle created by the $i$th hybrid node.}
\end{itemize}
For example, in Fig.~\ref{exampleNet} (center) $n=7$, $h=2$, $k_1=3$
and $k_2=4$. The main parameter of interest is the
topology $\mathcal{N}$ of the semi-directed network. Like phylogenetic trees,
this network can be rooted by a known outgroup.
The other parameters of interest are $\mathbf{t}$, the vector of branch
lengths in coalescent units (see below), and a vector of
inheritance probabilities $\boldsymbol{\gamma}$, describing the proportion
of genes inherited by a hybrid node from one of its hybrid parent
(see Fig.~\ref{exampleNet}).
Only identifiable branch lengths are considered in $\mathbf{t}$.
For example, with only one sequenced individual per taxon, the lengths
of external edges are not identifiable and are not estimated.


\paragraph{Pseudolikelihood on a network.}
\label{formula_pseudo}
Pseudolikelihood has already been used to estimate
phylogenetic trees under ILS \cite{Liu2010}, and here we extend
the theory to networks.
The pseudolikelihood of a network is based on the likelihood formulas
of its 4-taxon subnetworks.
These 4-taxon likelihoods are not independent but fast to compute.
A quartet is a 4-taxon unrooted tree. For taxon set
$s=\{a,b,c,d\}$, there are only three possible quartets, represented
by the splits $q_1=ab|cd$, $q_2=ac|bd$ and $q_3=ad|bc$.

The \textit{concordance factor} (CF) of a given quartet (or split) is
the proportion of genes whose true tree displays that quartet (or split)
\cite{Baum2007}.
We use the term `CF' as opposed to `probability'
to emphasize that CFs measure genomic support. Probabilities
(such as posterior probabilities or bootstrap values) are most often
thought to measure statistical uncertainty \cite{ane2010}.
Intuitively, splits between natural evolutionary groups of
organisms are recovered by most or all genes, and thus have
high CFs. On the other hand, the presence of a hybrid
would be captured by intermediate CFs. For example, if $a$ is a hybrid
intermediate between $b$ and $c$, the CFs of $ab|cd$ and $ac|bd$ would
be around 0.5 while the CF of $ad|bc$ would be near 0.

The theoretical CFs $(\cf_{q_1},\cf_{q_2},\cf_{q_3})$
expected under the coalescent model
are already known if the network is a species tree \cite{allmanDegnanRhodes2011}.
Interestingly, these CFs are independent of the root placement
in the species tree, and are given by
$(1-2/3e^{-t},1/3e^{-t},1/3e^{-t})$ if the unrooted species tree is $q_1=ab|cd$
with an internal edge of length $t$ coalescent units.
On a species network with reticulations, the probabilities of
rooted gene trees was fully derived in \cite{Yu2012a} and more efficiently
in \cite{Yu2014}.
But the probabilities of unrooted gene trees was not determined.
We derive the quartet probabilities in the next section.
In particular, they do not depend on the root placement in the
network, which makes them simple and fast to compute.

To calculate the likelihood of a 4-taxon network from gene
trees $\mathcal{G}=\{G_1,G_2,...,G_g \}$ at $g$ loci, we consider
the number of gene trees $\mathbf{X} = (X_{q_1},X_{q_2},X_{q_3})$
that match each of the three quartets.
Assuming unlinked loci, $\mathbf{X}$ follows a multinomial distribution
with probabilities $(\cf_{q_1},\cf_{q_2},\cf_{q_3})$, the quartet CFs
expected under the coalescent on the 4-taxon network.
With a larger network on $n\geq 4$ taxa, we consider all
4-taxon subsets $s$ and combine the likelihood of each 4-taxon subnetworks
to form the full network pseudolikelihood:
\begin{equation}\label{eq:pseudolik}
L=\prod_{s \in{\cal S}} (\cf_{q_1})^{X_{q_1}}(\cf_{q_2})^{X_{q_2}}(\cf_{q_3})^{X_{q_3}}
\end{equation}
where $\cal S$ is the collection of all 4-taxon sets and
$q_i=q_i(s)$ ($i=1,2,3$) are the 3 quartet trees on $s$.
In \eqref{eq:pseudolik}, the data are summarized in the $X$ values,
and the candidate network governs the CF values, which we derive below.

\paragraph{Quartet CF for a 4-taxon network under ILS.}
\label{formula_quartet}
For $h=1$ hybridization, there are 5 different semi-directed
4-taxon networks (up to tip re-labelling).
We describe here the expected quartet CFs (probabilities) for only one case
and refer to S1 Text for the remaining cases. 

\begin{figure}
\centering
\includegraphics[scale=0.2]{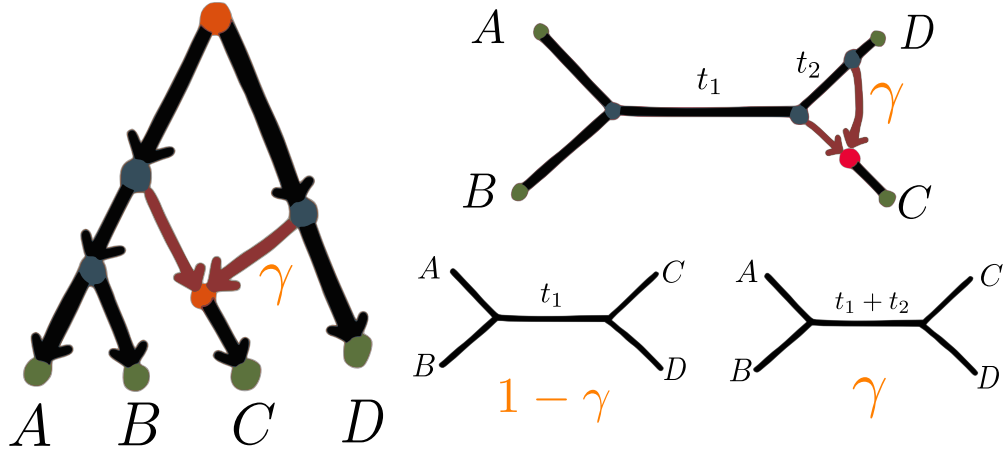}
\caption{
\textbf{Rooted network (left) and its semi-directed version (top right).}
Quartet CFs expected under the network do not depend on the
root placement, and are weighted averages of quartet CFs expected under
the unrooted trees (bottom right).
}
\label{CFhybridformula}
\end{figure}

Under the hybridization model described in \cite{Meng2009,Yu2012a}
and the network in Fig.~\ref{CFhybridformula} (left), each gene from taxon $C$ has
probability $\gamma$ of having descended from the hybridization edge
sister to $D$,
and probability $1-\gamma$ of having descended from the original tree
branch, sister to $(AB)$.
Therefore, the expected CFs are weighted averages of CFs
obtained on 2 species tree with ILS. 
Because the quartet probabilities do not depend on the root placement in each
species tree, they do not depend on the root placement in the original network
either. Fig.~\ref{CFhybridformula} (top right) shows the
corresponding semi-directed network,
and all rooted networks displaying it share the
same quartet CFs, obtained from the coalescent models on the 2 unrooted
species trees shown in Fig.~\ref{CFhybridformula} (bottom right).
These trees have the same topology but different branch
lengths in this case.
Therefore we get that
$\cf_{ab|cd}=\cfnetworksmajor{t_1}{t_1-t_2}$
and the other 2 quartets occur with equal probabilities:
$\cf_{ac|bd}=\cf_{ad|bc}=\cfnetworksminor{t_1}{t_1-t_2}$.


\begin{figure}[ht]
\centering
\includegraphics[scale=0.3]{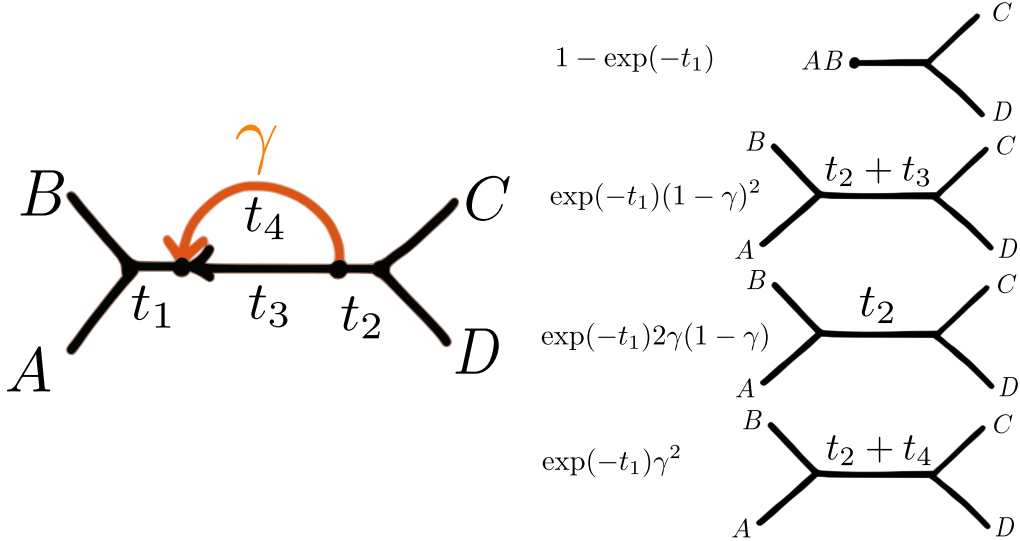}
\caption{\textbf{Example of a 4-taxon semi-directed network (left),
with known direction of both hybrid edges but unspecified
position of the root.} The root can be placed on the internal edges with
length $t_2,t_3,t_4$, or on the external edges to C or D.
The quartet CFs on this network are weighted averages
of CFs under 4 trees with weights as shown (right).}
\label{weighted_average2}
\end{figure}

On other semi-directed networks, more than 2 underlying
unrooted species trees are needed if
a hybrid node has more than one descendent taxon.
In the network in Fig. \ref{weighted_average2}
for instance, the hybrid node has two descendants, $A$ and $B$.
Given this network,
the computation of the CF for the major quartet $AB|CD$ is as follows. First,
$A$ and $B$ can coalesce along the branch of length $t_1$ with probability
$1-e^{-t_1}$. If they do not coalesce (with probability
$e^{-t_1}$) then there are 3 options: both originated
from the minor hybrid edge (probability $\gamma$ each);
both originated from the major hybrid edge
(each with probability $1-\gamma$);
or one ($A$ or $B$) originated from the minor hybrid edge but
the other ($B$ or $A$) from the major.
Assuming that each lineage's origin is independent of the other,
we get 
$ 
\cf_{AB|CD} = 1-e^{-t_1}+e^{-t_1}\Big((1-\gamma)^2(1-2/3 e^{-t_2-t_3})
 +2\gamma(1-\gamma)(1-2/3 e^{-t_2})+\gamma^2(1-2/3 e^{-t_4-t_2})\Big).
$ 
Therefore, this CF is a weighted average of CFs from the 4 species
trees shown in Fig. \ref{weighted_average2} (see S1 Text for all other
cases).

With more than 1 hybridization ($h>1$) there are an infinite number
of semi-directed 4-taxon networks, but we can still calculate the quartet CFs
if we assume that the cycles created by different reticulations
do not share edges. We do so recursively on $h$, by reducing each network
to an equivalent network with $h=0$ or 1 and transformed branch lengths.
For example, the network in Fig.~\ref{CFhybridformula} leads
to equal CFs of the 2 minor quartets $ac|bd$ and $ad|bc$,
so it is equivalent to the unrooted species tree $ab|cd$ with
internal branch length $t_3=-\log((1-\gamma)e^{-t_1}+\gamma e^{-t_1-t_2})$ to ensure
$\cftreeminor{t_3}=\cf_{ac|bd}$ given above.
This new species tree and the original network
have the same expected quartet CFs.
%
The assumption of a level-1 network guarantees non-overlapping reticulation
cycles, such that we can
find an equivalent 4-taxon network with $h=0$ or 1 and
the same expected quartet CFs.
We then just apply the network formulas above.
The transformed branch lengths of the equivalent network
are given in the S1 Text.

\paragraph{Detecting the presence of a hybridization.}
Identifiability is a basic requirement if one seeks to learn about parameters from data.
Here our parameters are
the network topology $\cal N$, branch lengths $\boldsymbol{t}$ and
inheritance values $\boldsymbol{\gamma}$.
We already know that quartet CFs do not depend on the root placement,
so the rooted network is not identifiable and we only
consider semi-directed networks $\cal N$ here.
Our pseudolikelihood model would be identifiable if two different combinations
of parameters $(\mathcal{N},\boldsymbol{t},\boldsymbol{\gamma})$ and
$(\mathcal{N}',\boldsymbol{t}',\boldsymbol{\gamma}')$ yield different
sets of quartet CFs.
We show here and below that
some reticulations and some parameters are impossible (or hard) to detect.
This theory is used later to reduce the parameter
space explored by our heuristic search, to avoid network and parameter
combinations that are not identifiable.

\begin{figure}[ht]
\centering
\includegraphics[scale=0.3]{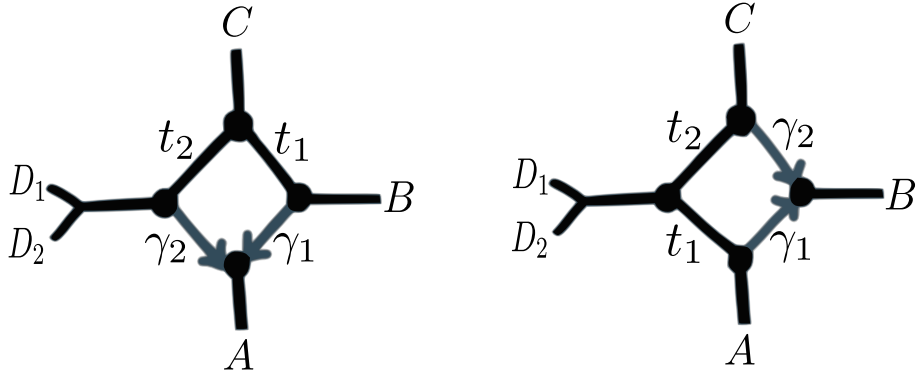}
\caption{\textbf{Networks with $k=4$ nodes in the reticulation cycle and
identical unrooted topologies.} They differ in their hybrid
position (left: good diamond, right: bad diamond I).
If $D_2$ is not sampled ($n=4$),
only $\gamma_i(1-e^{-t_i})$ for $i=1,2$ are identifiable
and the 2 networks are not distinguishable from each other.
}
\label{dir_id}
\end{figure}

On $n=4$ taxa, 
we already showed that the network in Fig.~\ref{CFhybridformula} is
equivalent to a tree with some appropriate internal branch length.  In
fact, the same holds true for all 4-taxon networks with $k=2$ or $3$
nodes in their reticulation cycle: these reticulations cannot be
detected.
If $k=4$ i.e. if the reticulation involves more distantly related
taxa, then the presence of the hybridization can be detected
based on the quartet CFs.
However, networks with the same unrooted topology are unidentifiable
from each other from only 4 taxa,
like the 2 networks in Fig.~\ref{dir_id} if only $D_1$ is sampled ($n=4$).
They only differ in the placement of the hybrid node,
which is therefore not identifiable, even if the unrooted network
and the presence of the reticulation is.


In general, for networks with $n\geq 4$ taxa,
we restrict our focus to the
case when ${\cal N}'$ is the network topology
obtained from $\cal N$ by removing a single hybrid edge of interest.
The assumption of
non-intersecting cycles allows us to study the detectability of this one
hybrid edge given the other hybridizations in the network (see S1 Text).
Assuming that all ${n\choose 4}$ 4-taxon sets are used in the
pseudolikelihood, the network $\cal N$ gives us $3 {n\choose 4}$
quartet CFs expected under the coalescent.
The presence of the hybridization of interest can be detected
if the quartet CFs from $({\cal N},\boldsymbol{t},\boldsymbol{\gamma})$ cannot all be equal
to the quartet CFs from $({\cal N}',\boldsymbol{t}',\boldsymbol{\gamma}')$ simultaneously.
We matched both systems from $\cal N$ and ${\cal N}'$ using Macaulay2
\cite{macaulay2}, and checked the values of
$(\boldsymbol{t},\boldsymbol{\gamma})$ and $(\boldsymbol{t}',\boldsymbol{\gamma}')$
when the two systems of CFs were equal (see S1 Text for full details).
Apart from the obvious case $\gamma=0$ for the hybrid edge absent in
${\cal N}'$, we found that $\cal N$ and ${\cal N}'$ were not
distinguishable when $t_b=0$ or $t_b=\infty$ for some tree
branches $b$, implying either a hard polytomy or a branch with no ILS
and a reduction of the problem to a 4-taxon network.  We can ignore
these cases with the reasonable assumption \centerline{\textbf{A1}: $t
  \in (0,\infty)$ for all tree branches and $\gamma\in (0,1)$.}
\textbf{A1} is not sufficient, however, to ensure that the
  presence of each hybridization in $\cal N$ can be detected.
Increasing taxon sampling helps detect a hybridization only if the
added taxa increase the size of the reticulation cycle.  Namely, if
the cycle only involves $k=2$ nodes (see
Fig.~\ref{hybrid_nodes_cross}), then $\cal N$ is not distinguishable
from ${\cal N}'$, regardless of $n$.  For $k=3$, some
hybridizations are detectable and some are not.  If any
two of the three subtrees defined by the hybridization cycle
(Fig.~\ref{hybrid_nodes_cross}) have only one taxon, then the
hybridization is not detectable.  It is, if
instead at least two subtrees contain more than one taxon.  In
general, hybridizations with $k\geq4$ can be detected if
$n\geq 5$.  Here and below, we use the terms detectable or identifiable
in their \textit{generic} sense \cite{allman2008,allman2009}, which simply
means that some conditions on $(\boldsymbol{t},\boldsymbol{\gamma})$
are required, like \textbf{A1}, but that all these conditions
are met except on a subset of measure zero.


\begin{figure*}
\centering
\includegraphics[scale=0.2]{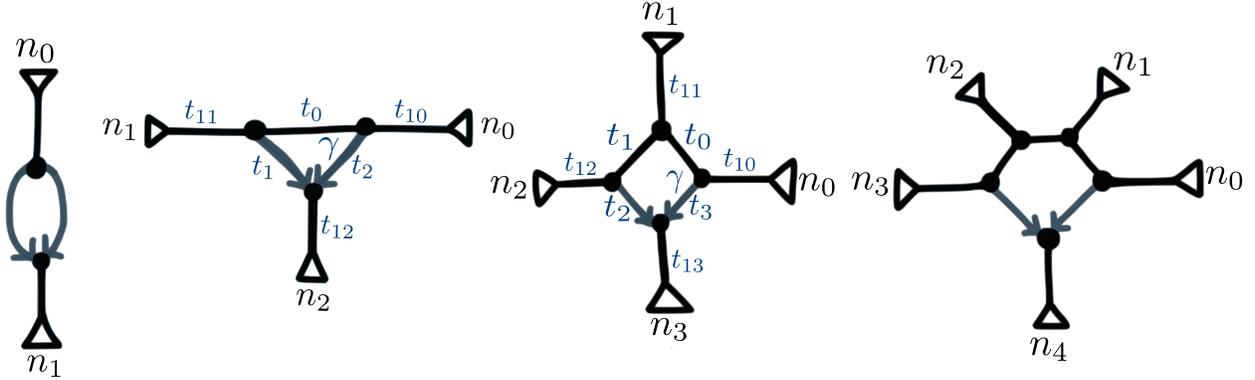}
\caption{\textbf{Networks with $k$ nodes in a hybridization cycle: $k=2,3,4$ and $5$
from left to right.}
 When $k=3$, parameters are not identifiable. A good triangle corresponds
 to $n_1, n_2, n_3\geq 2$, in which case setting $t_{12}=0$ makes the other
 parameters identifiable.
 When $k=4$, parameters are not all identifiable for the bad diamond I
 ($n_0=n_2=n_3=1$ but $n_1\geq 2$) and for the bad diamond II
 ($n_0=n_1=n_2=1$ but $n_3\geq 2$).}
\label{hybrid_nodes_cross}
\end{figure*}

We further determined if the direction of a given
hybrid edge was identifiable (in addition to its presence) when
$n=5$ and $k=4$, in a case when the direction is
not identifiable from 4 taxa.  Fig.~\ref{dir_id} shows two networks that
differ only in the
placement of the hybrid node, but otherwise have the same unrooted topology.
We proved that they yield different
sets of quartet probabilities and therefore are distinguishable from each other,
showing that the direction of the hybridization becomes identifiable when $n\geq 5$.

\paragraph{Identifiability of branch lengths and heritabilities.}
We now turn to the case when ${\cal N}'=\cal N$ to determine if
$\boldsymbol{t}$ and $\boldsymbol{\gamma}$ are identifiable given a
known network topology.  Like before, we used
Macaulay2 to determine under which conditions two different
combinations of parameters $(\boldsymbol{t},\boldsymbol{\gamma})$ and
$(\boldsymbol{t}',\boldsymbol{\gamma}')$ yield different sets of
quartet probabilities for a fixed network $\cal N$
(see S1 Text for details). 

Just as before, the identifiability depends on the type of network
(Fig.~\ref{hybrid_nodes_cross}). With only $4$ taxa, there are more
parameters than equations (3 quartet CFs), so $\boldsymbol{t}$ and
$\boldsymbol{\gamma}$ are not separately identifiable, so we
focus first on the case with $n \geq 5$.

If $n\geq 5$, parameter identifiability is again easier if the reticulation
involves more distantly related taxa.
If $k \geq 5$, all the parameters are identifiable. 
If $k \leq 3$, parameters are not identifiable.
If $k = 4$, 
parameters are identifiable if
either $n_0\geq 2$ (or $n_2$, symmetrically), or if both $n_1$ and
$n_3\geq 2$ (see Fig.~\ref{hybrid_nodes_cross}).
We call this a \textit{good diamond}.
Parameters are not all identifiable in the remaining 2 cases,
which we call \textit{bad diamonds} I and II (see Fig.~\ref{hybrid_nodes_cross}).
The bad diamond I already lacked identifiability
under a different model in \cite{Pickrell2012}.

\paragraph{Practical consequences.}
A naive search for the most likely network would get stuck alternating
between non-distinguishable networks or parameter sets. Hence we
reduced the searchable space to only consider networks whose
reticulations involve enough nodes. 
Indeed, all reticulations with $k=2$ and most with
$k=3$ are either not detectable at all, or their parameters are not
all identifiable.
For hybridizations with $k=3$, we only kept those with $n_i\geq 2$
for all $i=0,1,2$ (see Fig.~\ref{hybrid_nodes_cross}) 
and we enforced $t_{12}=0$ to make the other 6 parameters identifiable.
We denote this case as a \textit{good triangle}.
For bad diamonds I ($k=4$), we reparametrized the 3 non-identifable values
$(\gamma,t_1,t_0)$ into 2 identifiable ones
$(\gamma(1-e^{-t_0}),(1-\gamma)(1-e^{-t_1}))$ (see S1 Text). 
For bad diamonds II, we set $t_{13}=0$ and
kept the other 5 parameters $(\gamma,t_0,t_1,t_2,t_3)$.

\subsection*{Network estimation procedure}
\label{compu}
The input for our method is a table of quartet CFs observed from multi-locus data
(the $X$ values in \eqref{eq:pseudolik}),
across many or all 4-taxon subsets from the $n$ taxa of interest.

\paragraph{Pseudolikelihood optimization.}
The maximum pseudolikelihood (MPL) estimate is the network, branch lengths $\boldsymbol t$
and $\boldsymbol\gamma$ heritabilities that maximize the pseudolikelihood
\eqref{eq:pseudolik}.
This MPL optimization was fully implemented in SNaQ (Species Networks
applying Quartets) and is part of our open source package PhyloNetworks in
Julia \cite{julia}. 
The numerical optimization of branch lengths and $\gamma$ parameters
for a fixed topology 
is performed with a
derivative-free methodology in the NLopt package for Julia.
The heuristic optimization of the network topology uses a strategy
similar to that in \cite{Yu2014}.  Given a fixed maximum number of
hybridizations ($h_m$), we search for the MPL network with at most
$h_m$ hybridizations. Since the pseudolikelihood can only improve when
hybridizations are added, we expect the final network to have $h=h_m$
exactly.
A network is estimated for various values of $h_m$, followed by
a model selection procedure to select the appropriate number
of hybridizations (see below). For a given $h_m$,
the search is initialized with a tree from a very fast
quartet-based tree estimation method like ASTRAL \cite{Mirarab2014} or
Quartet MaxCut \cite{Snir2012-qmc,Avni2015-qmc}.  The length of each
branch is initialized using the average observed CF of the quartets
that span that branch exactly, $\overline{\cf}$, transformed to
coalescent units by $t=-\log(1-3/2\,\overline{\cf})$.  The search then
navigates the network space by altering the current network using one
of 5 proposals, chosen at random: 1) move the origin of an existing
hybrid edge, 2) move the target of an existing hybrid edge, 3) change
the direction of an existing hybrid edge, 4) perform a
nearest-neighbor interchange move (NNI) on a tree edge, and 5) add a
hybridization if the current topology has $h<h_m$.  Any new proposed
network is checked to verify that it is a semi-directed level-1
network with $h\leq h_m$ and with at least one valid placement for the
root.
More details on these moves are provided in S1 Text. 
Although the deletion of a current hybridrization is not proposed
(because the MPL network should have $h=h_m$), this deletion is still
performed when suggested by the data, if the numerical optimization
of parameters returns a $\hat\gamma=0$. 
In this case, the corresponding hybrid edge is removed and the search
attempds to add it back at random in the neighborhood of the original
hybrid edge.
If this attempt fails for all neighbors, the hybridization is deleted
entirely and the search continues from a network with 1 fewer hybridization.
Similarly, if the numerical optimization returns a branch of length
$t=0$, an NNI move is proposed immediately on that branch.
The search continues until the pseudolikelihood
converges or until the number of consecutive failed proposals reaches a limit.

In \cite{Huber2015}, Huber et al. proved that the space of unrooted level-1 networks is
connected by local subnetwork transfers,
which generalize the NNI operations on trees
and which are similar to our moves 1, 2 and 4.
Although we do not have a formal proof that the MPL network
can be reached from the starting tree using our proposals, the
results in \cite{Huber2015} suggest that it is the case.

\paragraph{Statistical uncertainty.}
There are two sources of uncertainty when we estimate CFs from
sequence data.  Gene trees are not observed directly but estimated,
and only a finite number of genes can be sampled.  Our preferred
estimation of quartet CFs integrates over both sources of noise using
BUCKy \cite{Ane2007,Larget2010}, to estimate true gene tree conflict
and discard conflict due to gene tree error. BUCKy
returns estimated genome-wide CFs and their 95\% credibility intervals.
These CFs were shown to be most influenced by highly informative genes
and least influenced by genes with large tree uncertainty \cite{ane2010}.
Very briefly, MrBayes
\cite{mrbayes3} is run on each gene separately and the full tree
samples from all genes serve as input for BUCKy, which is run
separately on each 4-taxon set. BUCKy has a prior
probability of $(1+\alpha/3)/(1+\alpha)$ that 2 genes share the same
quartet tree. For example, choosing $\alpha=1$ amounts to assuming a
prior concordance probability of 0.667, compared to 0.333 if gene
trees matched just by chance.

A faster way to estimate CFs from sequences would be to use maximum likelihood
with RAxML \cite{raxml8} (or maximum parsimony for faster estimation)
on each gene separately, and then to simply count the number of genes
that display each quartet tree.
To account for tree uncertainty, one may drop any gene, for a given 4-taxon set,
that does not have bootstrap support above some threshold (like 70\%)
for one of the 3 quartets. This method would not account for the
uncertainty due to having a limited number of genes.
With only 10 genes, for instance, the estimated CFs would necessarily
be of low precision, of the form $i/10$ in the best case when all
10 gene trees are strongly supported.

If some genes are missing some taxa, the quartet CFs obtained on a
given 4-taxon set can simply use the subset of genes that have sequences
for the 4 taxa of interest. In most cases, a large number of genes
can be included for each given 4-taxon sets,
even if none of the genes have data across the full taxon set.
Furthermore, the collection of 4-taxon sets with available CF data
does not need to be exhaustive, as the sum in \eqref{eq:pseudolik}
simply involves the sampled 4-taxon sets.

To measure uncertainty in the network, one may re-do the network analysis on
bootstrap data sets. If we estimated credibility intervals for CFs with BUCKy,
then 100 credible sets of quartet CFs can be obtained by sampling each CF
from its posterior distribution, approximated by its credibility interval.
If CFs were obtained using RAxML and observed quartet frequencies in gene trees,
then bootstrap sets of quartet CFs could be obtained by sampling one
bootstrap tree from each gene. 
To summarize the networks estimated from these bootstrap sets,
we first calculated the support for edges being in the major tree:
the tree obtained by suppressing the minor hybrid edge (with $\gamma<0.5$) at
each reticulation. We then summarized the support for the placement of each minor 
hybrid edge on that tree,
considering 2 edges as equivalent if they are of the same type 
(hybrid or tree edges) and define the same clusters \cite{Yu2014}.

Uncertainty in the number of hybridizations $h$ is more
difficult to capture (see Discussion). We used here a slope heuristic 
to find where the network score changes from a sharp
to a slow linear decrease as $h$ increases. We also looked to see if the 
bootstrap support for successive reticulations dropped at the same $h$ value.

\section*{Results}
\label{results}
\subsection*{Simulated data}
We carried out simulations to compare the speed and accuracy of
SNaQ and PhyloNet.
Given that PhyloNet uses the rooted and full gene trees, 
SNaQ can only be expected to perform as
accurately as PhyloNet at best. Our simulations show that a pseudolikelihood
approach does not compromise too much accuracy, but greatly improves speed.

We simulated $g$ gene trees with
ms \cite{Hudson2002} under four different networks:
$(n,h)=(6,1)$, 
$(6,2)$, 
$(10,1)$ 
and $(15,3)$, 
with $\gamma$ values set to $0.2$ or $0.3$ on each minor hybrid edge
(see S1 Text) 
These network topologies were chosen at random by simulating a tree
with $n$ taxa under the coalescent, then 
choosing two edges at random for the origin and target of each hybridization
and rejecting networks of level $>1$.
On 6 taxa all reticulations were hard to reconstruct with $k=4$,
including a bad diamond I in the case $h=2$.
On 10 and 15 taxa, both networks also had a diamond,
of the bad type II for $n=10$.
We varied the number of genes between 10 and 3000. 
All analyses were run on 2.7-3.5 GHz processors. 


We first used the true simulated gene trees for inference.
The rooted gene trees served as input for PhyloNet
and the unrooted quartet CFs as observed in the $g$ gene trees
served as input for SNaQ.
The semi-directed network returned by SNaQ was rooted by the
outgroup species, when compatible with the estimated hybrid edges.
Next, we used
Seq-Gen \cite{seqgen} to simulate sequences of length $500$
under HKY, $\kappa=2$, 
A,C,G and T frequencies of
$0.300414,0.191363,0.196748,0.311475$
and population mutation rate $\theta=0.036$, as in \cite{Yu2014}.
Gene trees were estimated with MrBayes \cite{mrbayes3} using
$10^6$ generations sampled every 200, 25\% burnin and an HKY model.
The consensus trees (one per gene) served as input for
PhyloNet. The posterior tree samples were then used in BUCKy
\cite{Ane2007,Larget2010} for each 4-taxon set, to estimate quartet CFs
and use them as input for SNaQ.
For this pipeline, we used the tools implemened by \cite{Stenz2015}
and available at \verb+https://github.com/nstenz/TICR+. 
This procedure was replicated 30 times.
The accuracy of each method was measured as the proportion of times
that the estimated network matched the true network.
To compare rooted networks we used the distance in \cite{Nakhleh2010},
which is a metric on reduced networks (including
level-1 networks) and is implemented in PhyloNet.
We used it to detect equality between rooted networks,
but not to measure how ``close'' networks were, because this distance
is very sensitive to small differences such as a change in the direction of
a hybrid edge.

\begin{figure}
\centering
\includegraphics[scale=0.6]{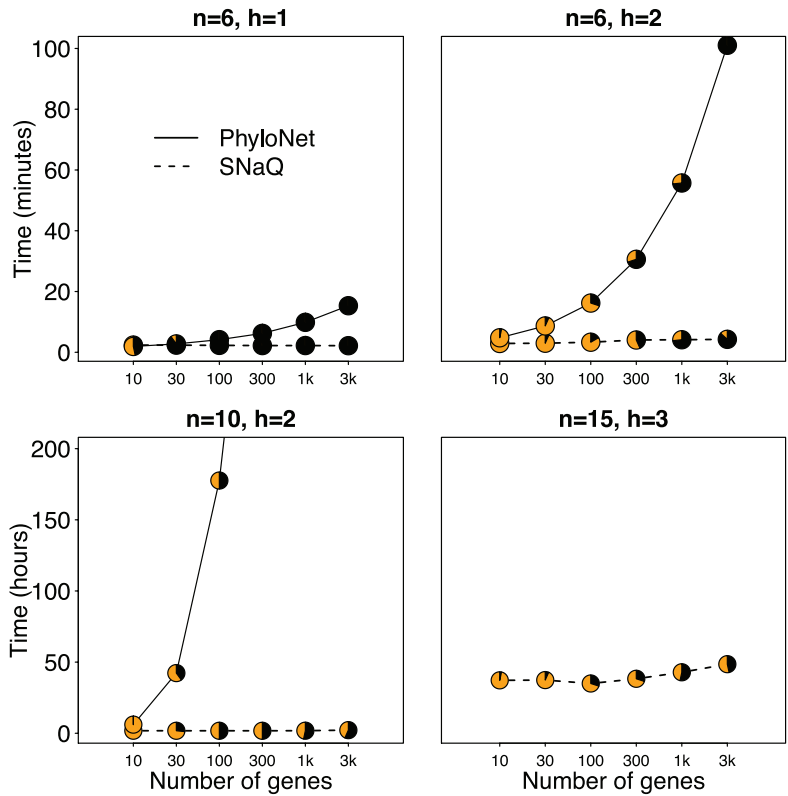}
\caption{\textbf{Performance (average computing time per replicate)
of SNaQ and PhyloNet:} in simulations using true gene trees
on networks with $n=6$, 10 or 15 taxa and $h=1$, 2 or 3. Each
replicate consisted of 10 independent runs with full optimization of
branch lengths and inheritance probabilities for each run.
Pie charts display accuracy (black: probability of recovering the true network).
With $n=10$ and 300 or more loci, or with $n=15$, PhyloNet was too slow to run.
}
\label{fig:knownGT}
\end{figure}

Fig.~\ref{fig:knownGT} summarizes the accuracy and
speed of SNaQ and PhyloNet. On 10 or 15 taxa PhyloNet was too slow to run
(a single replicate with 10 taxa and 300 loci required over 400 hours),
so we cannot provide a comparison of accuracy on these 2 larger networks.

\begin{figure}
\centering
\includegraphics[scale=0.6]{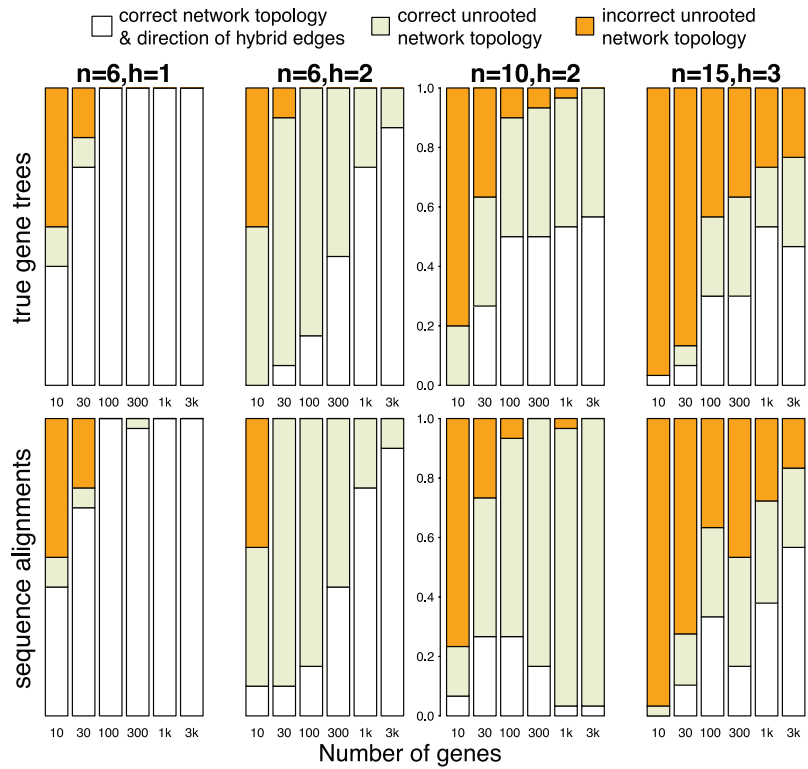}
\caption{\textbf{Accuracy of SNaQ in simulations using true gene trees
  or sequence alignments.}
Even when the semi-directed topology was not recovered, the
unrooted topology was estimated correctly for most replicates
using 30 loci or more and $h\leq 2$.}
\label{snaqAcc}
\end{figure}

For networks with $h=2$ or more, the accuracy of SNaQ decreased.
So, for each semi-directed network estimated by SNaQ, we determined if
its unrooted topology matched that of the true network.
Fig.~\ref{snaqAcc} shows that in the vast majority of cases when the
directed network was incorrectly estimated, its unrooted topology was
still correctly inferred from true gene trees and for $n=6$ with
estimated gene trees. For $n\geq 10$, the inferred direction
of hybrid edges degraded when gene trees were estimated.
In most replicates on 10 taxa, this was because the bad diamond II
near the root in the true network had a wrong estimated
placement of the hybrid node.

\begin{figure}
\centering
\includegraphics[scale=0.6]{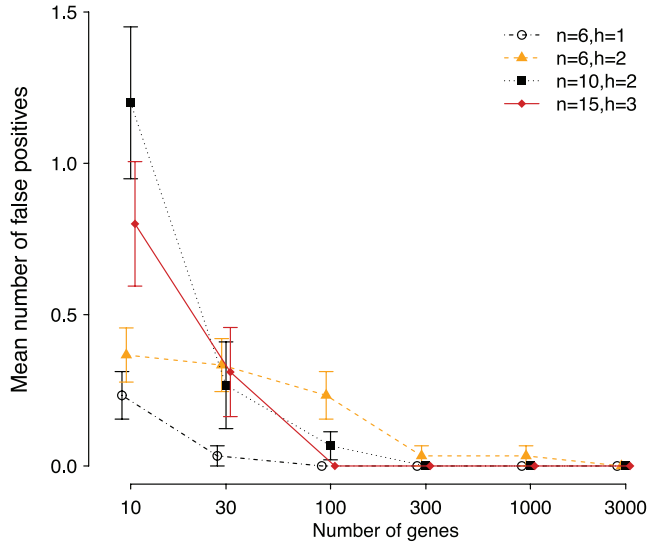}
\caption{\textbf{Accuracy of SNaQ to recover the major tree
in the species network, from sequence alignments.}
The major tree is obtained by suppressing all minor hybrid edges ($\gamma<0.5$)
to capture the major vertical inheritance pattern.
Accuracy is measured as half the Robinson-Foulds distance between
the true and estimated tree, i.e. the number of incorrect edges
in the estimated tree.
A lot fewer genes are needed to accurately estimate the major vertical
pattern, compared to the horizontal pattern.}
\label{RFdist}
\end{figure}

To detemine which features in the network were correctly
estimated, we extracted the major tree from each network, that is,
the tree obtained by keeping the major hybrid edge and suppressing the
minor hybrid edge at each hybrid node.
We then compared the true major tree (from the true network)
to the estimated major tree using the Robinson-Foulds distance (see
Fig. \ref{RFdist}).
The major tree was correctly estimated from 300 or more genes
in all scenarios, except when $n=6,h=2$ and
300 genes (1 replicate out of 30) and 1000 genes (1 replicate out of 30).
In both cases, the true major tree was displayed
in the estimated network but the major hybrid edge was estimated
as a minor edge with $\gamma<0.5$.
Therefore, the network's ``backbone", i.e. the major vertical inheritance pattern,
can still be estimated accurately even when the full network
and hybrid edges are not (Fig. \ref{snaqAcc}).

\begin{figure}
\centering
\includegraphics[scale=0.7]{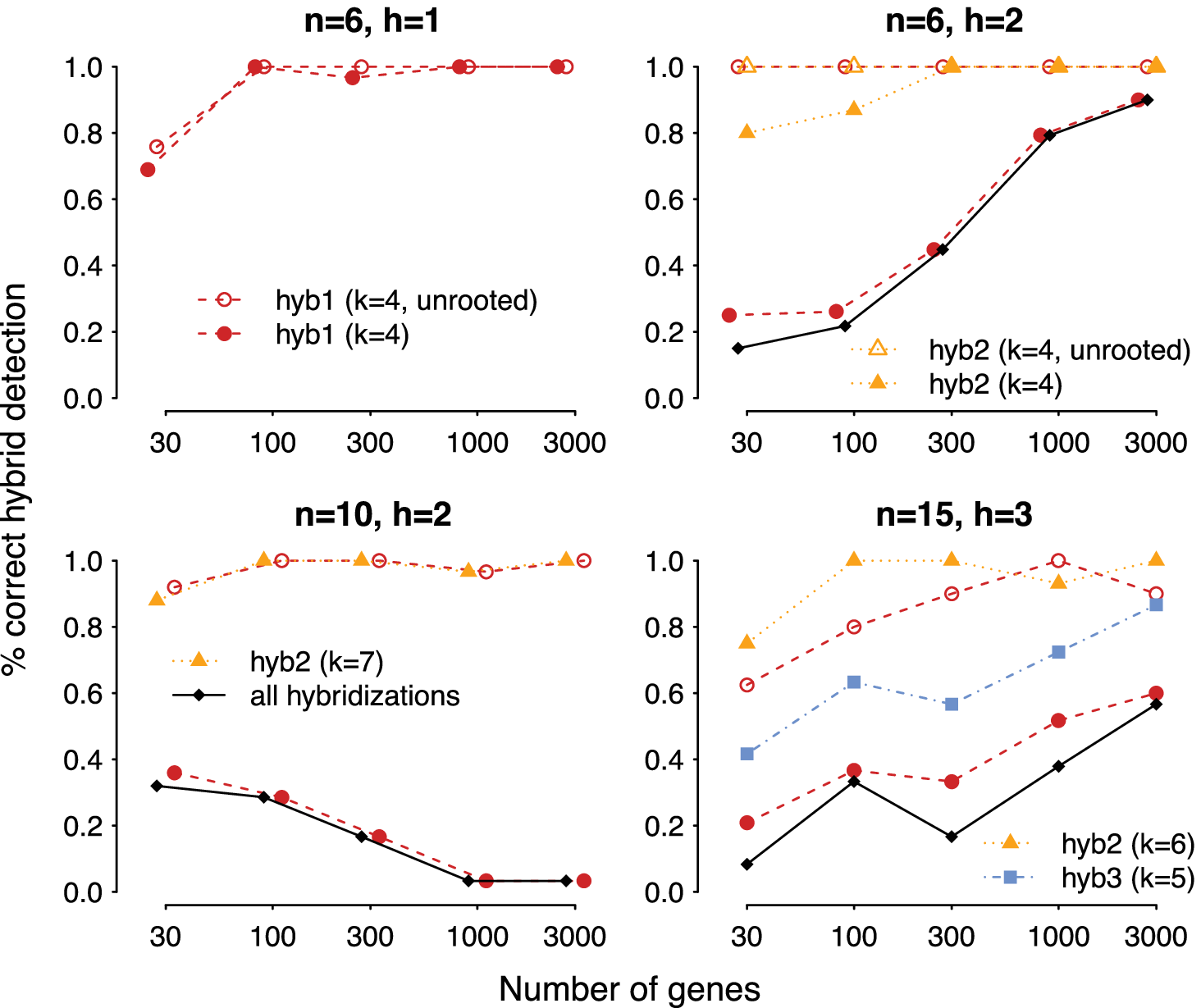}
\caption{\textbf{Accuracy of SNaQ to recover each hybridization:}
proportion of times each reticulation was correctly inferred (connecting 
the correct donor edge to the correct recipient edge in the major tree), 
among sequence alignments in which the major tree was recovered.
Minor hybrid edges are numbered and colored as in S1 Text. 
For reticulations creating a cycle of $k=4$ nodes, we also calculated
the proportion of times that this undirected (or ``unrooted") cycle was 
correctly inferred, even though the identity and direction of hybrid edges
in this cycle might be incorrect (empty symbols).
The proportion of times that all hybridizations were correctly inferred 
(black lines) was low when a single hybridization with $k=4$ was hard to 
recover (bad diamond I in case $n=6$, $h=2$, and bad diamond II in case $n=10$).
}
\label{hybridDet}
\end{figure}

Among cases when the major tree was correctly estimated, we
  determined the detection accuracy of each true hybridization
  event. To do so, we compared each estimated hybridization with the
  true hybridization of interest. In each network (true and
  estimated), we removed the other hybridizations by suppressing their
  minor hybrid edges and used the known outgroup to root both
  networks. We then calculated the hardwired cluster distance between
  the two resulting networks
  to determine if the estimated hybridization event matched the true
  hybridization of interest: connecting the same donor edge to the
  same recipient edge in the major tree (Fig.~\ref{hybridDet}).
  For $n=6$, the hybridizations forming a good diamond were recovered 
  with high accuracy from 100 genes, but the hybridization forming a 
  bad diamond I (case $h=2$) was very hard to recover, needing more than 1000 
  genes for an accurate inference of the hybrid edges' direction. 
  Still, the unrooted cycle was correctly estimated from 100 genes or more. 
  For $n=10$ and $n=15$ taxa,
  the hybridization creating a cycle of $k=4$ nodes was also very hard
  to detect with its correct direction, although its undirected cycle was
  accurately recovered from a few hundred genes.
  Hybridizations were recovered more accurately as their cycles 
  spanned more nodes, with a high recovery rate
  for the hybridizations with $k=6$ and $k=7$ from 100 genes or more.

\subsection*{\textit{Xiphophorus} fishes evolution}
\begin{figure*}
\centering
\includegraphics[scale=0.4]{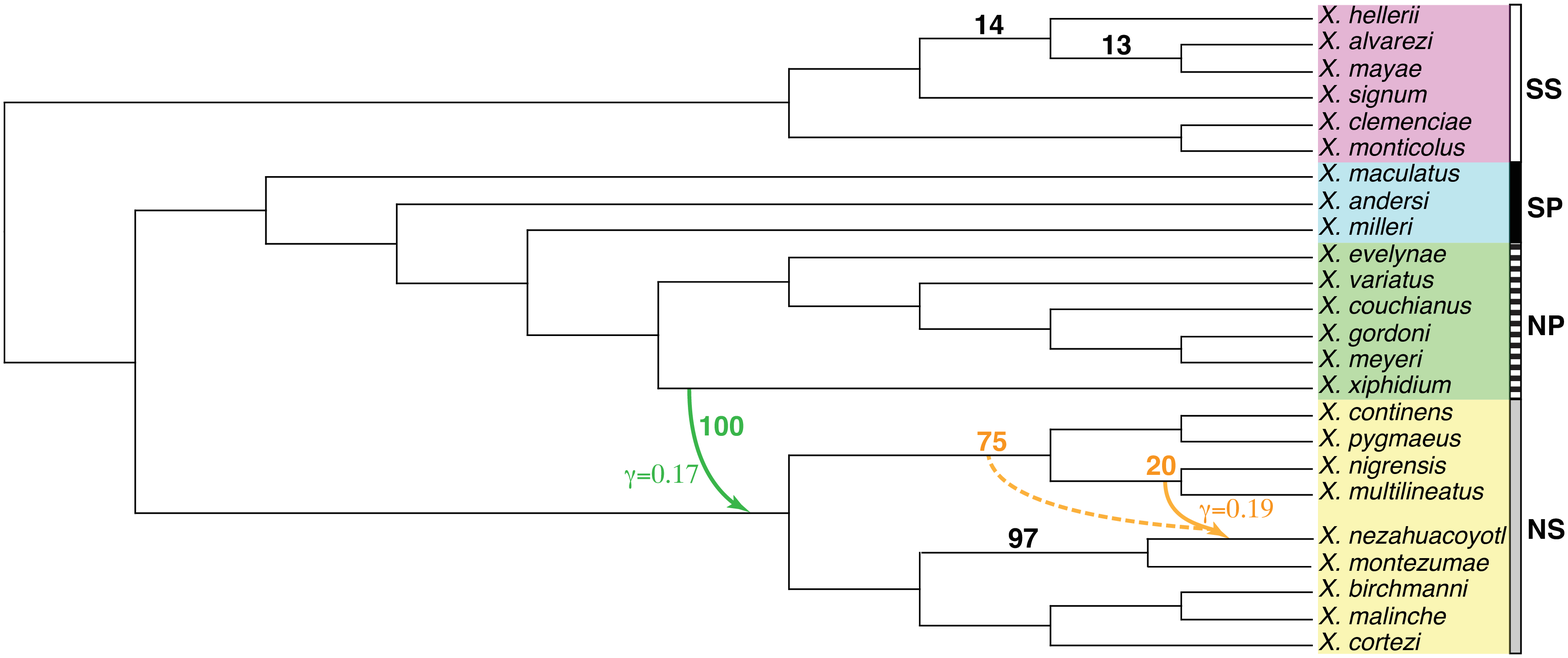}
\caption{\textbf{\textit{Xiphophorus} reticulate evolution estimated with SNaQ:}
  from 1183 genes, $h=2$, rooted with the southern swordtails
  outgroup clade (SS).  NS: northern swordtails, SP: southern
  platyfishes, NP: northern platyfishes.
  Black edges: major tree (including hybrid edges with $\gamma>0.5$).
  Colored solid arrows: minor hybrid edges, annotated by their estimated $\gamma$.
  Black numbers: bootstrap support for edges in the major tree, if different
  from 100\%.
  Colored numbers: bootstrap support for the placement of minor hybrid edges.
  One reticulation had 75\% support for a different donor lineage
  (dotted arrow) than inferred from the original data.
}
\label{fig:cui}
\end{figure*}

We re-analyzed transcriptome data from \cite{Cui2013} to reconstruct
the evolutionary history of 24 swordtails and platyfishes
(\textit{Xiphophorus}: Poeciliidae).  Based on high CFs of splits in
conflict with their species tree followed by a series of
ABBA-BABA tests \cite{durand-etal-2011}, \cite{Cui2013} concluded that
hybridization or gene flow was widespread in the history of these
tropical fishes.  We re-analyzed their first set of 1183 transcripts.
BUCKy was performed on each of the 10,626 4-taxon sets.
The resulting quartet CFs were used in SNaQ, using  $h_m=0$ to $5$
and 10 runs each.
The network with $h=0$ and the major tree in the network with $h=1$
were identical to the total evidence tree in \cite{Cui2013}, 
with \textit{X.~xiphidium} 
placed within the grade of southern platyfishes (SP),
making the northern platyfishes (NP) paraphyletic (see S1 Text). 
With $h\geq 2$ the major tree was almost identical but with NP monophyletic 
(Fig.~\ref{fig:cui}) because \textit{X.~xiphidium} was found 
sister to the rest of the NP species, but involved in a reticulation (see below).
With $h\geq 3$, a reticulation within the southern swordtails (SS)
was found consistently ($\gamma=0.43$),
but with a direction in conflict with SS being an outgroup clade.
Its cycle had only $k=5$ nodes, 4 of them leading to a single taxon (
see S1 Text) 
so we suspect an error in the inferred hybrid node and gene flow direction.
The extra 2 reticulations found with $h=4$ and 5 had low $\gamma$ values
(in $[0.006-0.16]$).

The network scores (negative log-pseudolikelihood) 
decreased sharply from $h=0$ to $h=2$ then
slightly and somewhat linearly (see S1 Text), 
suggesting that $h=2$
best fits the fish data using a slope heuristic
\cite{birgeMassart2006,baudry2012}.  The network estimated with
$h=2$ (Fig.~\ref{fig:cui}) found \textit{X.~xiphidium} involved 
in an ancient reticulation, contributing a proportion $\gamma=0.17$ 
of genes to the lineage ancestral to northern swordtails (NS).
This reticulation might explain the placement of {\it X.~xiphidium} 
closer to the root in \cite{Cui2013}, from tree-based methods that do
not account for potential gene flow.
The second hybridization ($\gamma=0.20$) was found from the
population ancestral to \textit{X.~multilineatus} and \textit{X.~nigrensis}
into \textit{X.~nezahuacoyotl}, and relates to a high CF found by \cite{Cui2013} 
for a clade uniting \textit{X.~nezahuacoyotl} and the \textit{nigrensis} group.

Bootstrap data sets were simulated by sampling
each quartet CF from a uniform distribution on its 95\% credibility interval
(conservatively) then normalizing the sampled CFs across the 3 quartets 
on each 4-taxon set. For each bootstrap data set we estimated a network using 3 runs,
and $h=3$
(instead of 2) because the third inferred reticulation had a high
$\gamma$ (see S1 Text) 
and to assess the ability of the bootstrap procedure to identify 
the best $h$ value. If the bootstrap was consistent with the slope heuristic,
we expected high bootstrap support for the placement of the first 2 reticulations 
and lower support for the third. 
As expected, this third reticulation and network topology within the SS clade
was variable among bootstrap networks (see S1 Text), 
suggesting uncertainty in the major tree
within this clade (Fig.~\ref{fig:cui}). The rest of the tree was highly supported,
as was the placement of the reticulation involving \textit{X.~xiphidium}.
The reticulation involving \textit{X.~nezahuacoyotl} had split support for
its donor lineage, with 75\% support for a more ancestral lineage (Fig.~\ref{fig:cui}).

\section*{Discussion}
\label{conc}
Many methods are being developped
to understand organisms whose evolution behaves more net-like rather
than tree-like. There is evidence of reticulation at all levels
in the tree of life: deep among early prokaryotic and eukaryotic groups,
to shallow among recently diverged species
(e.g. \cite{clarkMesser2015,fontaine2015,jonsson2015})
or even among populations of the same species.
Our new and fast statistical
method to infer phylogenetic networks from multi-locus data
could be used at these various levels in the tree of life.

\paragraph{Network model and assumptions.}
Network inference is theoretically and computationally challenging.
Split networks can be estimated rapidly, yet
lack an evolutionary model and biological interpretability.
\cite{camaraLevineRabadan} proposed a very fast distance-based approach
to reconstruct topological ancestral recombination graphs (tARGs) from
a long alignment, but the biological interpretability of tARGs is still
limited.
The evolution model in \cite{Yu2012a} uses an explicit network and
satisfyingly accounts for various processes: reticulation events,
deep coalescences, and substitutions.
Yet a full likelihood estimation of large network (as in \cite{Yu2014})
seems beyond computational reach.
Our pseudolikelihood method offers an alternative, allowing the
estimation of bigger and more complex networks while maintaining
biological interpretability and a flexible evolutionary model.

We assumed a level-1 network throughout, where each
hybrid node is part of a single cycle. This assumption is quite restrictive,
but \cite{Pardi2015} showed that
sequence data and gene trees on present-day species
do not contain enough information
to reconstruct complex networks, even from many loci. Therefore,
some assumption has to be made to limit the network complexity.
Extending our method to networks with intersecting cycles
will need further work to restrict the search to candidate networks
that are distinguishable from each other. Indeed, \cite{Pardi2015}
show that different level-2 networks can have the exact
same likelihood, and hence pseudolikelihood. So no method based
on gene trees can ever decide which of these level-2 networks
is true.
Under a model without ILS, using full gene trees
and branch length in substitutions per sites comparable across genes,
\cite{Pardi2015} showed that level-1 networks are distinguishable but
level-2 networks are not necessarily.
Extending our approach to higher level networks,
with or without ILS, will require
extensive theory to work around this lack of identifiability.


Our approach allows for multiple individuals per species.
All alleles from the same species simply need to be treated as a known
and fixed polytomy in the network.
Future work could include this and other topology constraints on the network,
to reduce the computational burden when there are known phylogenetic
relationships.

\paragraph{Branch lengths.}
We allow hybrid edge lengths to be 0, but we do not constrain
them to be 0 (unlike in \cite{Meng2009,Yu2012a})
even though each gene flow event has to occur between contemporary
populations. If one parental population went extinct or
has no sampled descendants, the hybrid edge from this parent has a
positive length in the observable network.
A second reason is that a long branch can fit a population
bottleneck, as might be expected in the formation of a new hybrid species.
Not constraining hybrid branch lengths to 0 has a computational burden,
however. Future implementations might enforce this constraint,
when taxon sampling is thorough and extinction of parental populations
can be ruled out. 

By considering quartet topologies only, we ignored branch lengths
in gene trees. 
This choice frees us from various assumptions.
Using gene tree branch lengths, which are in substitutions per site,
would require some assumption on gene rates to make branch lengths
comparable across trees, and a molecular clock on gene trees.
Other assumptions would also be needed on
population sizes, 
shared or not across lineages.
The recent approach in \cite{YuNakhleh2015} should scale well to
many taxa, but makes these strong assumptions because it requires
accurate distances obtained from branch lengths in gene trees.
On the contrary, our approach should be robust to rate variation
across genes and across lineages, and does not require
any assumption on population sizes.


\paragraph{Identifiability of the topology.}
Yu et al. \cite{Yu2012a} already noted a lack of identifiability from rooted gene trees
for reticulations with $k=3$ from only 4 taxa (including the outgroup).
We found a similar lack of identifiability from unrooted quartets if $n<5$.
In practice, some reticulations are hard to detect even with 5 or more taxa,
if some branches are long with no ILS (close to violating \textbf{A1}).
However, in these cases the unrooted topology of the network can still
be recovered, even if the direction of gene flow
and the placement of the hybrid node is not.
Therefore, heuristic strategies that keep the unrooted network
unchanged, or that just slightly modify it, may improve the search for the
best network.

More tools are needed to study unrooted and semi-directed phylogenetic networks.
For instance, no distance measure has been developed for such networks,
that we know of. Distances between rooted networks would also be needed,
that would be less sensitive to small changes in the unrooted or semi-directed
topologies than the distance proposed in \cite{Nakhleh2010}.
New notions of edge equivalence would also be needed on unrooted and
semi-directed networks. It would help summarize a bootstrap sample
of networks for instance, with no need for an outgroup.

We propose here a tree-based but informative summary by
extracting the major tree from each network, obtained by dropping
any minor hybrid edge (with inheritance $\gamma<0.5$).
Because this tree summarizes the major vertical inheritance pattern
at each node, it can be considered an estimate of the species tree.
We found that recovering the underlying species tree can be much
easier (requiring fewer genes) than recovering the horizontal signal.
Even if the species tree is the main purpose of a study, \cite{Solis-Lemus2015}
showed that species-tree methods can be inconsistent in recovering the
vertical signal if there is gene flow, so using a network can be beneficial
to avoid the possible inconsistency of tree-based coalescent methods.

\paragraph{Missing data.}
All data analyzed here had full taxon sampling from each gene, and we were
able to use all 4-taxon sets. Future work could assess the impact
of missing data (gene sequences, or 4-taxon sets) on the method's accuracy.
Missing 4-taxon sets will be necessary for large networks, because the
number of 4-taxon sets grows very rapidly with the number of taxa ($\sim n^4/24$).
With many taxa, one may randomly select a collection of 4-taxon sets
and/or choose them specifically.
SNaQ calculates the number of quartets involving each taxon and
provides information about under-represented taxa, if any.
With many individuals per species, one may greatly reduce the
collection of 4-taxon sets to be analyzed by randomly sampling from those
containing at most one individual per species.
If the assignment of individuals to species is correct, any 4-taxon set
containing 2 individuals from the same species
would be non-informative about the species-level relationships.
This strategy is used in \cite{chifmanKubatko2014} to infer species trees under ILS.

\paragraph{Uncertainty in the number of hybridizations}
Model selection is necessary to estimate the number of hybridizations $h$,
because the pseudolikelihood is bound to improve as $h$ increases,
like the likelihood or parsimony score in \cite{Yu2013}.
We used here the log pseudolikelihood profile with $h$.
A sharp improvement is expected until $h$ reaches the
best value and a slower, linear improvement thereafter.
Such data-driven slope heuristics can indeed be used with contrast functions
(like pseudolikelihoods) for model selection in regression frameworks
\cite{birgeMassart2006,baudry2012}. 

Information criteria have already been used to select $h$ (e.g. \cite{kubatko2009}),
but these criteria are inappropriate if the full likelihood
is replaced by a pseudolikelihood. Theory is missing to compare the
pseudolikelihoods of different networks, because of the possible
correlation between quartets from different 4-taxon sets.
It can be shown, however, that quartets from two 4-taxon sets $s_1$ and $s_2$
are independent if $s_1$ and $s_2$ overlap by at most one taxon and if the
true 4-taxon subnetworks share no internal edges.
Future work could exploit this partial independence to construct hypothesis tests.

Cross validation has been proposed by \cite{Yu2014}, and was shown to have
good performance.
In our framework, the cross-valication error could
be measured from the difference between the quartet CFs observed in the
validation subset and the quartet CFs expected from the network estimated
on the training set.
Because $K$-fold cross-validation requires partitioning the loci into $K$ subsets
and re-estimating a network $K$ times at each $h$ value,
this approach can be computationally heavy.

Finally, \cite{Stenz2015} proposed a goodness-of-fit test, also based on
quartet CFs, to determine if a tree with ILS fits the observed data
or if a network is needed instead. This test could be extended to networks,
to decide if a given $h$ provides an adequate fit.
One advantage to this approach
is that testing the adequacy of a given $h$ does not require to estimate a larger
network with $h+1$ hybridizations, whereas other approaches above would
require estimation of both networks in order to decide
that the simpler network is sufficient.

\paragraph{Pseudolikelihood with rooted triples.}
After submission, we learned about similar work 
using subnetworks and a pseudolikelihood approach \cite{yu2015},
which scales to many taxa.
In \cite{yu2015}, the pseudolikelihood is based on rooted triples
whereas we use unrooted quartets.
There are fewer triples, so the method in \cite{yu2015} is potentially
faster. However, fewer triples means less
information. For example, the networks $\Psi_1$ and $\Psi_2$ shown in Fig. 2 of
\cite{yu2015}, which are not distinguishable from triplets,
are in fact distinguishable from quartets
(see S1 Text).
Our thorough study of the network identifiability allowed us to
implement a search that avoids jumping between networks that are not
distinguishable, which facilitates convergence.
The downside of our approach is the assumption of a level-1 network.
Instead, \cite{yu2015} do not assume any restriction on the network.
Finally, our method
does not require rooted gene trees as input, which we view as a major
advantage because rooting errors are avoided.

\section*{Acknowledgments}


We are very grateful to Douglas Bates for encouraging us to use Julia.
We thank Noah Stenz for helping with the CF estimation pipeline, Ray
Cui for providing MrBayes output of their original analysis,
Sarah Friedrich for her help with artwork, David Baum for many
  meaningful discussions and two anonymous
  reviewers whose constructive feedback greatly improved the
  manuscript.


%
%
%


\newpage
\section*{\LARGE S1 Text: Supplementary Material}
\appendix
\beginsupplement
\section{Quartet CFs under the coalescent  with hybridization}

\subsection{Quartet CFs for a 4-taxon network with one hybridization}
\label{formulas_quartet}
A argument similar to what is described in the main text was applied
to all 4-taxon networks with $h=1$.  The results are summarized below.

\noindent\begin{minipage}{0.35\textwidth}
\centering
\includegraphics[scale=0.15]{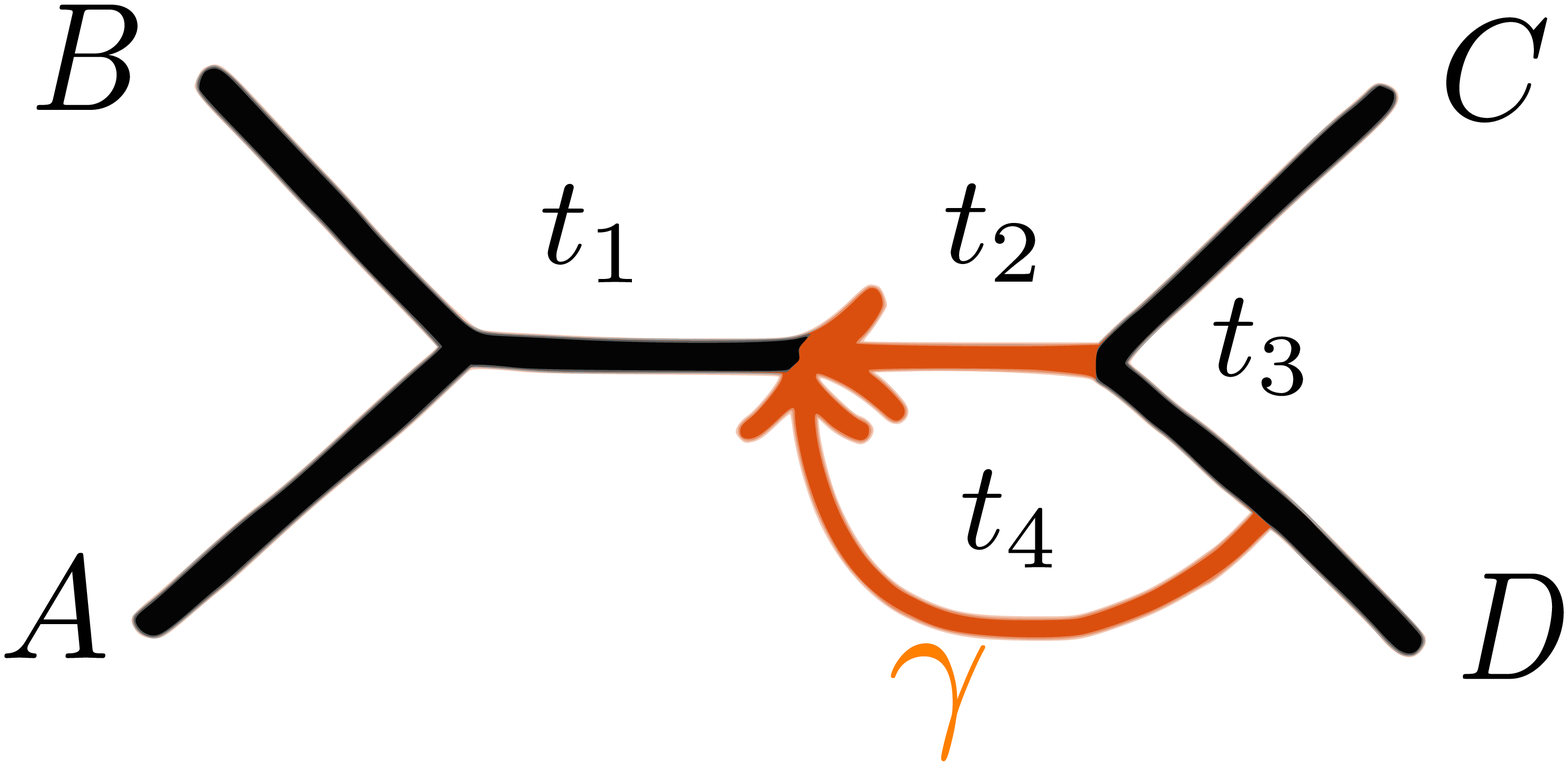}
\\Network 1
\end{minipage}
\begin{minipage}{0.55\textwidth}
\footnotesize
\begin{align*}
CF_{AB|CD}&=(1-\gamma)^2(1-2/3\exp(-t_1-t_2))\\
&+2\gamma(1-\gamma)(1-\exp(-t_1)+1/3\exp(-t_1-t_3))\\
&+\gamma^2(1-2/3\exp(-t_1-t_4))\\
CF_{AC|BD}&=(1-\gamma)^2(1/3\exp(-t_1-t_2))\\
&+\gamma(1-\gamma)\exp(-t_1)(1-1/3\exp(-t_3))\\
&+\gamma^2(1/3\exp(-t_1-t_4))\\
CF_{AD|BC}&=(1-\gamma)^2(1/3\exp(-t_1-t_2))\\
&+\gamma(1-\gamma)\exp(-t_1)(1-1/3\exp(-t_3))\\
&+\gamma^2(1/3\exp(-t_1-t_4))
\end{align*}
\end{minipage}

\noindent\begin{minipage}{0.35\textwidth}
\centering
\includegraphics[scale=0.15]{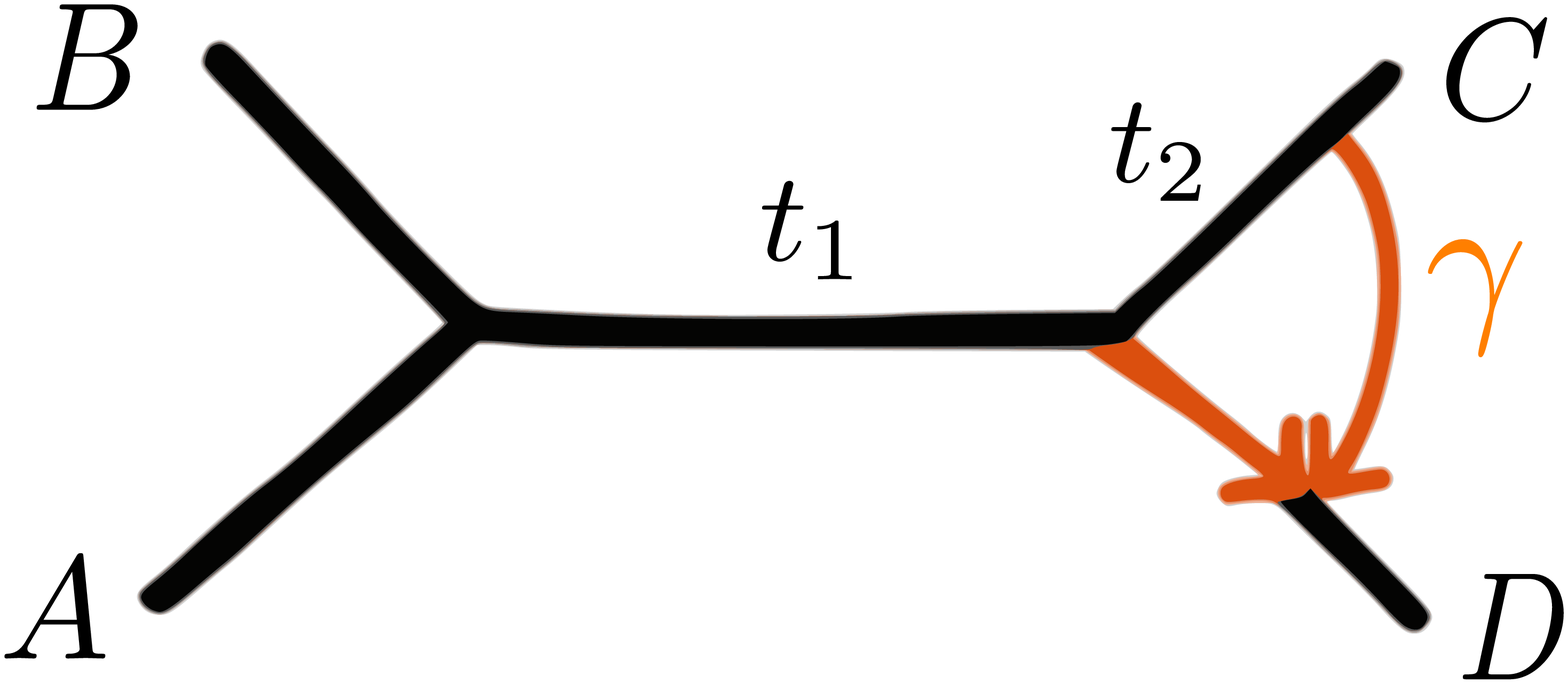}
\\Network 2
\end{minipage}
\begin{minipage}{0.55\textwidth}
\footnotesize
\begin{align*}
CF_{AB|CD}&=(1-\gamma)(1-2/3\exp(-t_1))+\gamma(1-2/3\exp(-t_1-t_2))\\
CF_{AC|BD}&=(1-\gamma)1/3\exp(-t_1)+\gamma1/3\exp(-t_1-t_2)\\
CF_{AD|BC}&=(1-\gamma)1/3\exp(-t_1)+\gamma1/3\exp(-t_1-t_2)
\end{align*}
\end{minipage}

\noindent\begin{minipage}{0.35\textwidth}
\centering
\includegraphics[scale=0.15]{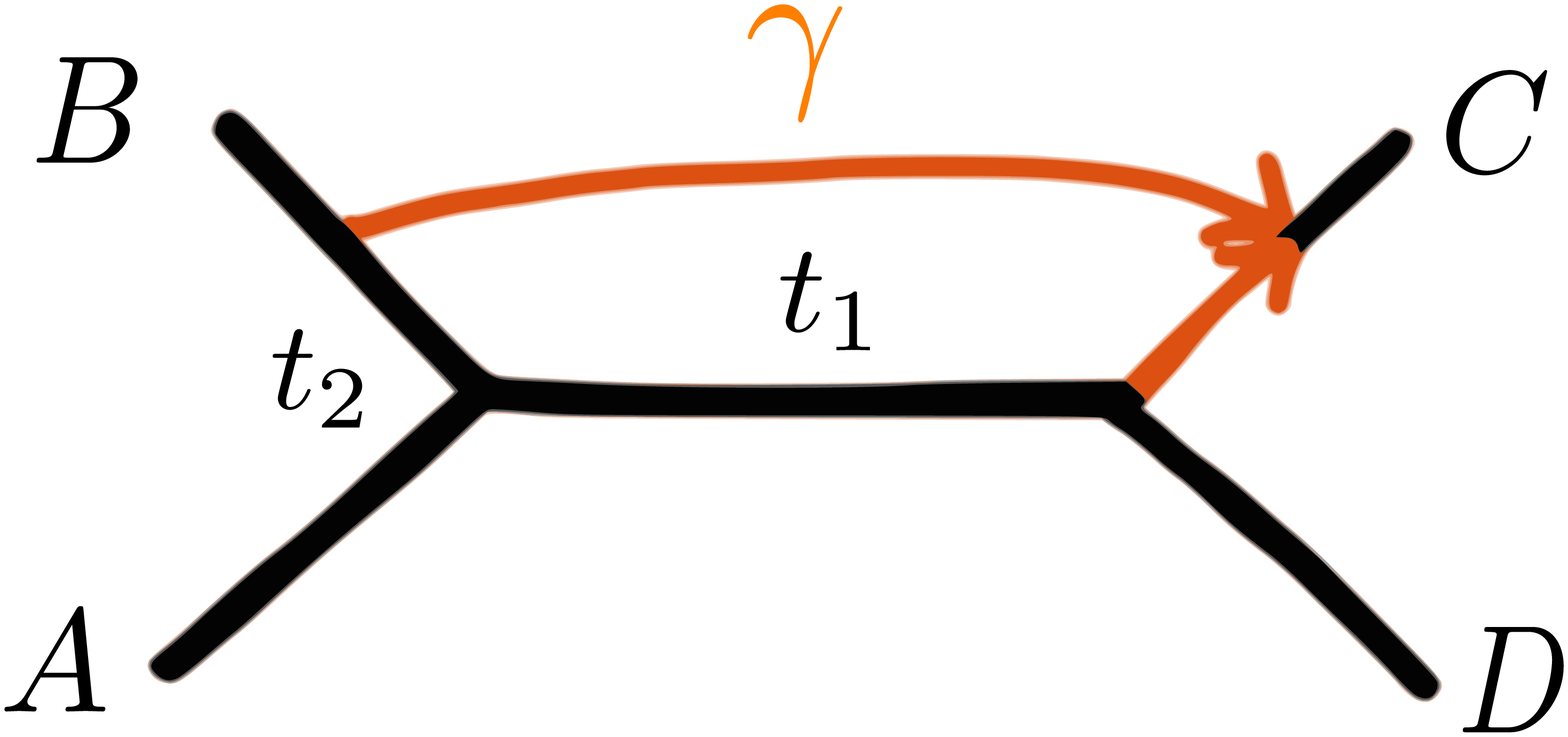}
\\Network 3
\end{minipage}
\begin{minipage}{0.55\textwidth}
\footnotesize
\begin{align*}
CF_{AB|CD}&=(1-\gamma)(1-2/3\exp(-t_1))+\gamma(1/3\exp(-t_2))\\
CF_{AC|BD}&=(1-\gamma)1/3\exp(-t_1)+\gamma(1-2/3\exp(-t_2))\\
CF_{AD|BC}&=(1-\gamma)1/3\exp(-t_1)+\gamma1/3\exp(-t_2)
\end{align*}
\end{minipage}

\noindent\begin{minipage}{0.35\textwidth}
\centering
\includegraphics[scale=0.15]{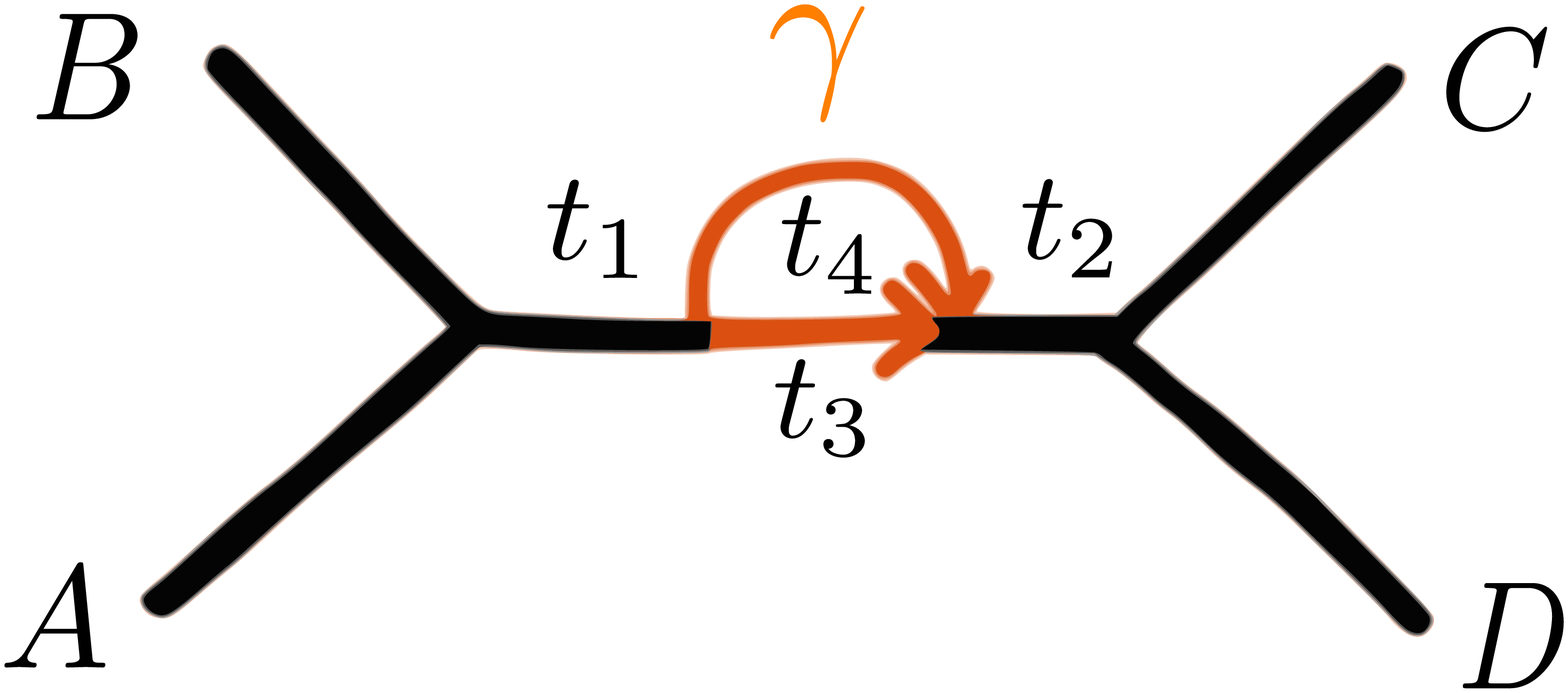}
\\Network 4
\end{minipage}
\begin{minipage}{0.55\textwidth}
\footnotesize
\begin{align*}
CF_{AB|CD}&=(1-\gamma)^2(1-2/3\exp(-t_1-t_2-t_3))\\
&+2\gamma(1-\gamma)(1-2/3\exp(-t_1-t_2))\\
&+\gamma^2(1-2/3\exp(-t_1-t_2-t_4))\\
CF_{AC|BD}&=(1-\gamma)^2(1/3\exp(-t_1-t_2-t_3))\\
&+2\gamma(1-\gamma)(1/3\exp(-t_1-t_2))\\
&+\gamma^2(1/3\exp(-t_1-t_2-t_4))\\
CF_{AD|BC}&=(1-\gamma)^2(1/3\exp(-t_1-t_2-t_3))\\
&+2\gamma(1-\gamma)(1/3\exp(-t_1-t_2))\\
&+\gamma^2(1/3\exp(-t_1-t_2-t_4))
\end{align*}
\end{minipage}

\noindent\begin{minipage}{0.35\textwidth}
\centering
\includegraphics[scale=0.15]{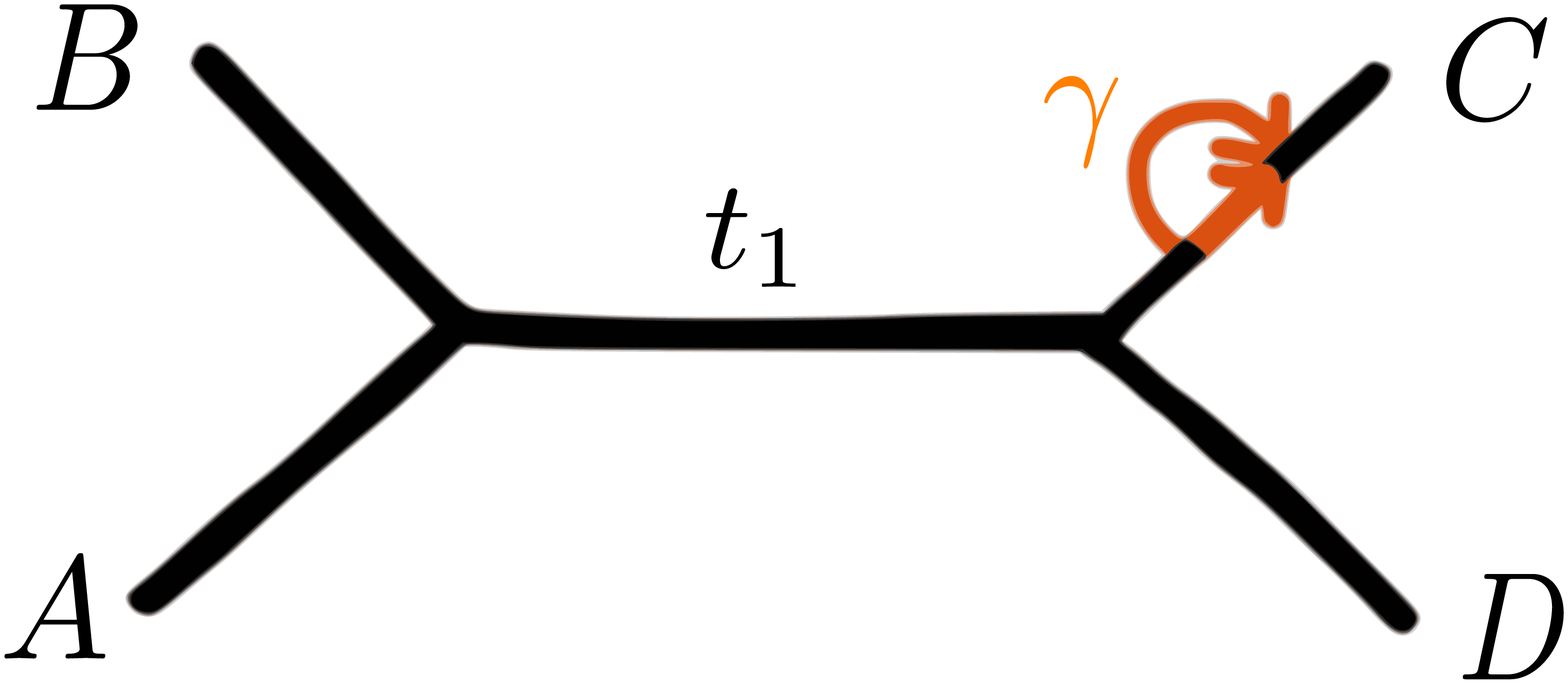}
\\Network 5
\end{minipage}
\begin{minipage}{0.55\textwidth}
\footnotesize
\begin{align*}
CF_{AB|CD}&=1-2/3\exp(-t_1)\\
CF_{AC|BD}&=1/3\exp(-t_1)\\
CF_{AD|BC}&=1/3\exp(-t_1)
\end{align*}
\end{minipage}

\subsection{Subnetwork equivalence for quartet CFs}
\label{reparam_quartets}

We show here how a level-1 four-taxon network with any number of hybridizations
can be reduced to an equivalent network with $h=0$ or 1 hybridizations only,
where equivalence means that both networks give the same quartet CFs.
The network is reduced by induction on $h$, by replacing each ``blob'' in the network
(a set of nodes and edges along a given cycle \cite{Huber2015}) by
a simpler network. To simplify notations we use $z_i=1-\exp(-t_i)$.
A subnetwork is of \textit{type~1} if it leads to two equal minor CFs,
and is therefore equivalent to a subnetwork with no hybridization.
A subnetwork is of \textit{type~2} if it leads to three different quartet CFs,
and is not equivalent to a tree.
We summarize below all the possible subnetwork configurations on 4 taxa
(on the left in each figure) and the equivalent form (on the right),
with the formula giving the branch length in the equivalent network
to obtain the same CFs from both networks.


\noindent\begin{minipage}{0.35\textwidth}
\centering
\includegraphics[scale=0.15]{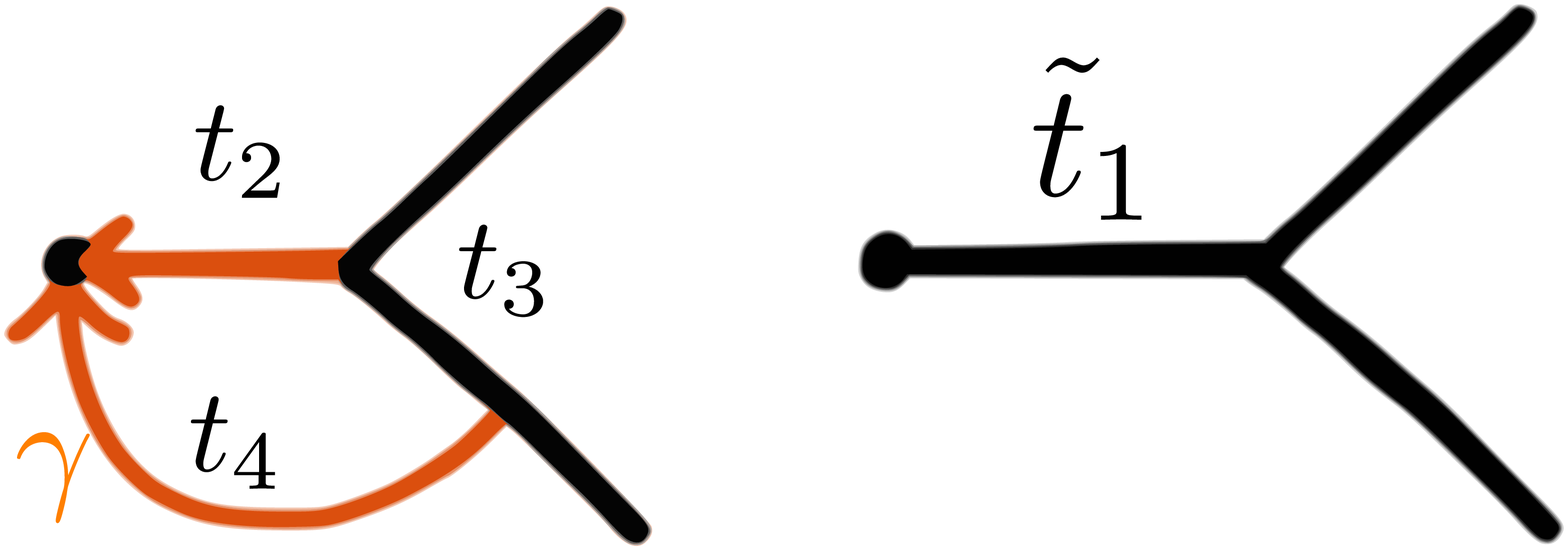}
\\Subnetwork 1 (type 1)
\end{minipage}
\begin{minipage}{0.55\textwidth}
\centering{\footnotesize
$\exp(-\tilde{t}_1):=1+\gamma z_3-\gamma^2 z_4-\gamma^2 z_3-(1-\gamma)^2 z_2$}
\end{minipage}

\noindent\begin{minipage}{0.35\textwidth}
\centering
\includegraphics[scale=0.15]{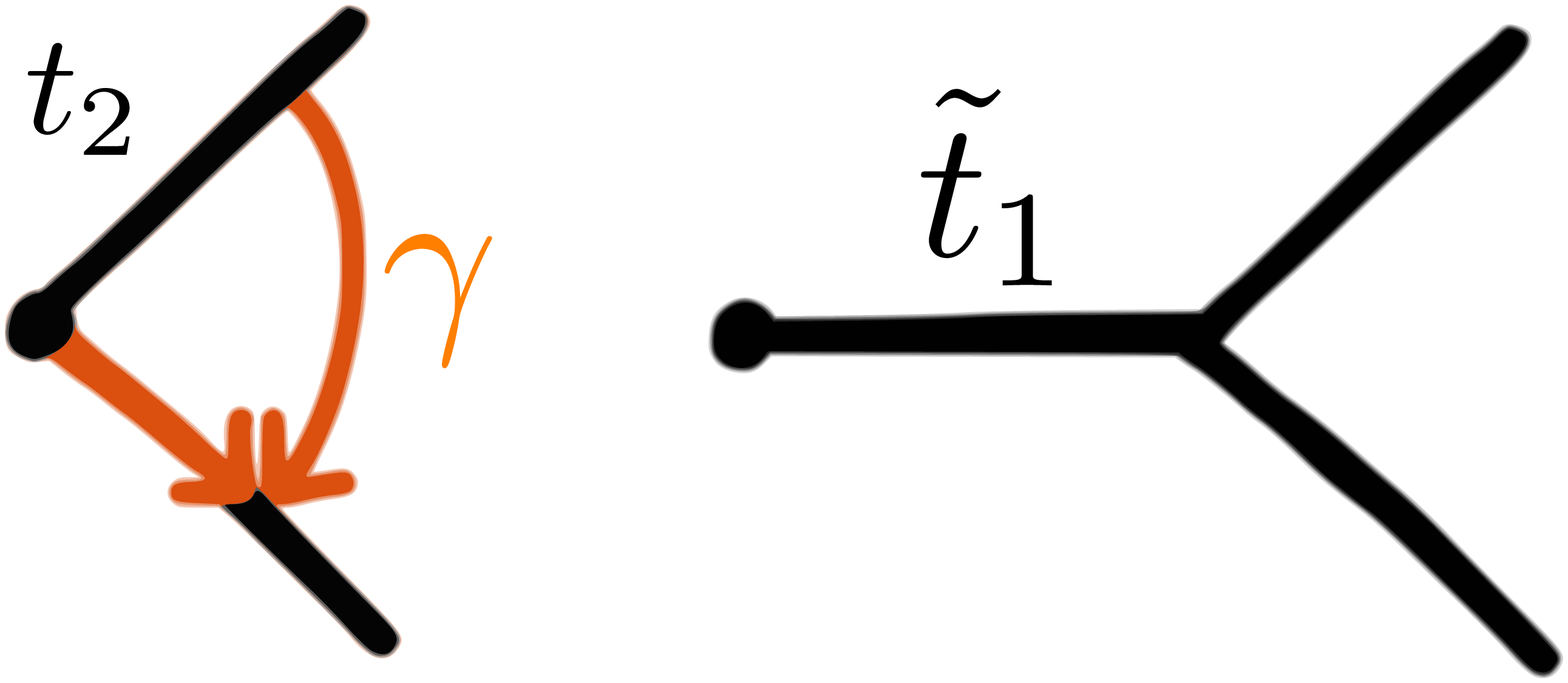}
\\Subnetwork 2 (type  1)
\end{minipage}
\begin{minipage}{0.55\textwidth}
\centering{\footnotesize
$\exp(-\tilde{t}_1):=1-\gamma z_2$}
\end{minipage}

\noindent\begin{minipage}{0.35\textwidth}
\centering
\includegraphics[scale=0.15]{quartet4.eps}
\\Subnetwork 3 (type 2)
\end{minipage}
\begin{minipage}{0.55\textwidth}
\footnotesize
\begin{align*}
3CF_{AB|CD}&=1-2(1-\gamma)z_1-\gamma z_2\\
3CF_{AC|BD}&=1+2\gamma z_2-(1-\gamma)z_1\\
3CF_{AD|BC}&=1-(1-\gamma)z_1-\gamma z_2
\end{align*}
\end{minipage}
\normalsize

\noindent\begin{minipage}{0.35\textwidth}
\centering
\includegraphics[scale=0.15]{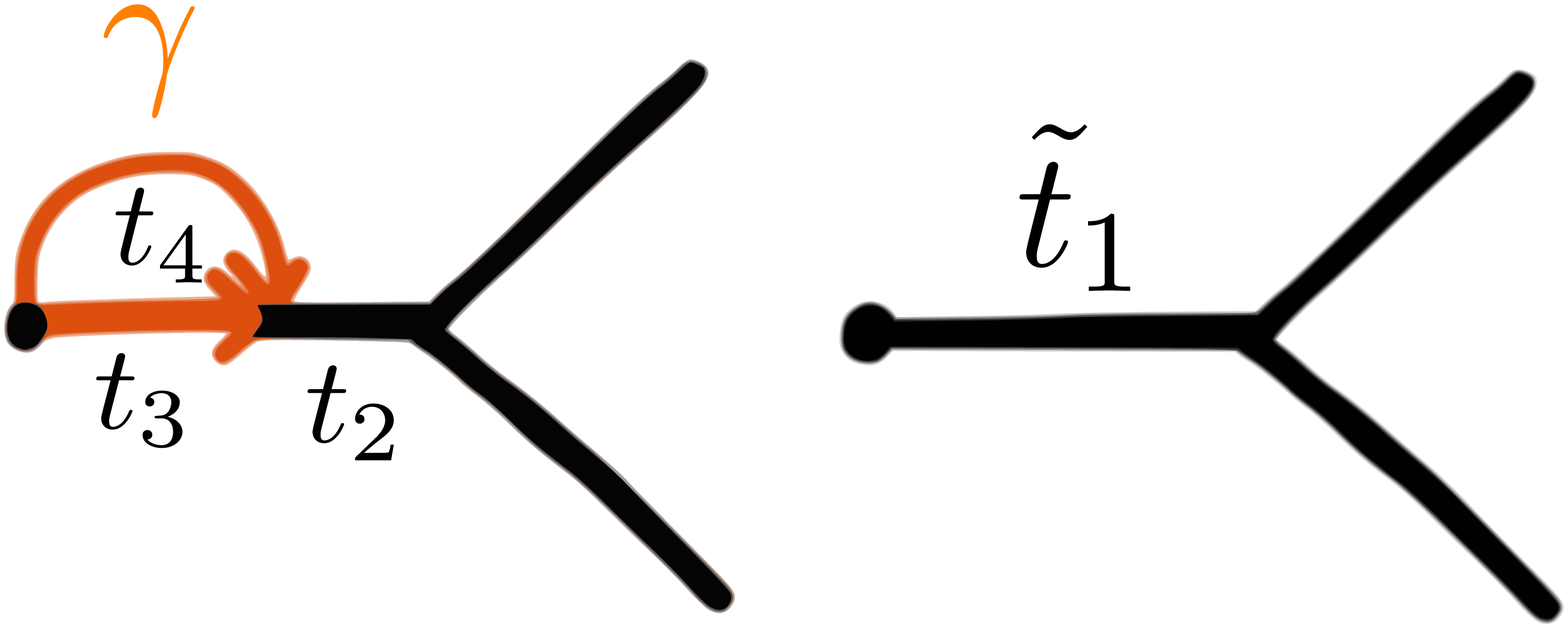}
\\Subnetwork 4 (type 1)
\end{minipage}
\begin{minipage}{0.55\textwidth}
\centering{\footnotesize
$\exp(-\tilde{t}_1):=\exp(-t_2)[1-\gamma^2 z_4 -(1-\gamma)^2 z_3]$}
\end{minipage}

\subsection{Four-taxon networks with more than one hybridization}
\begin{figure}[ht]
\centering
\includegraphics[scale=0.2]{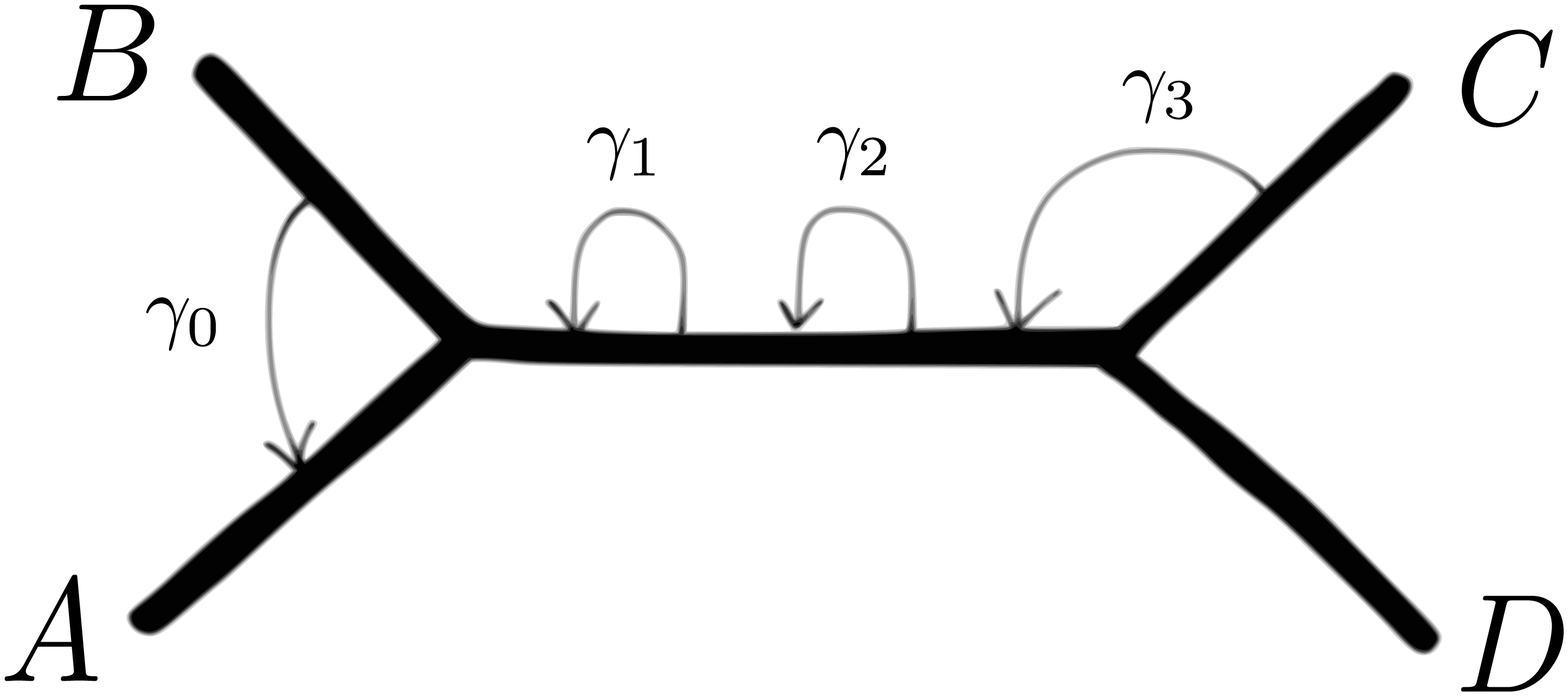}
\hspace{1cm}
\includegraphics[scale=0.25]{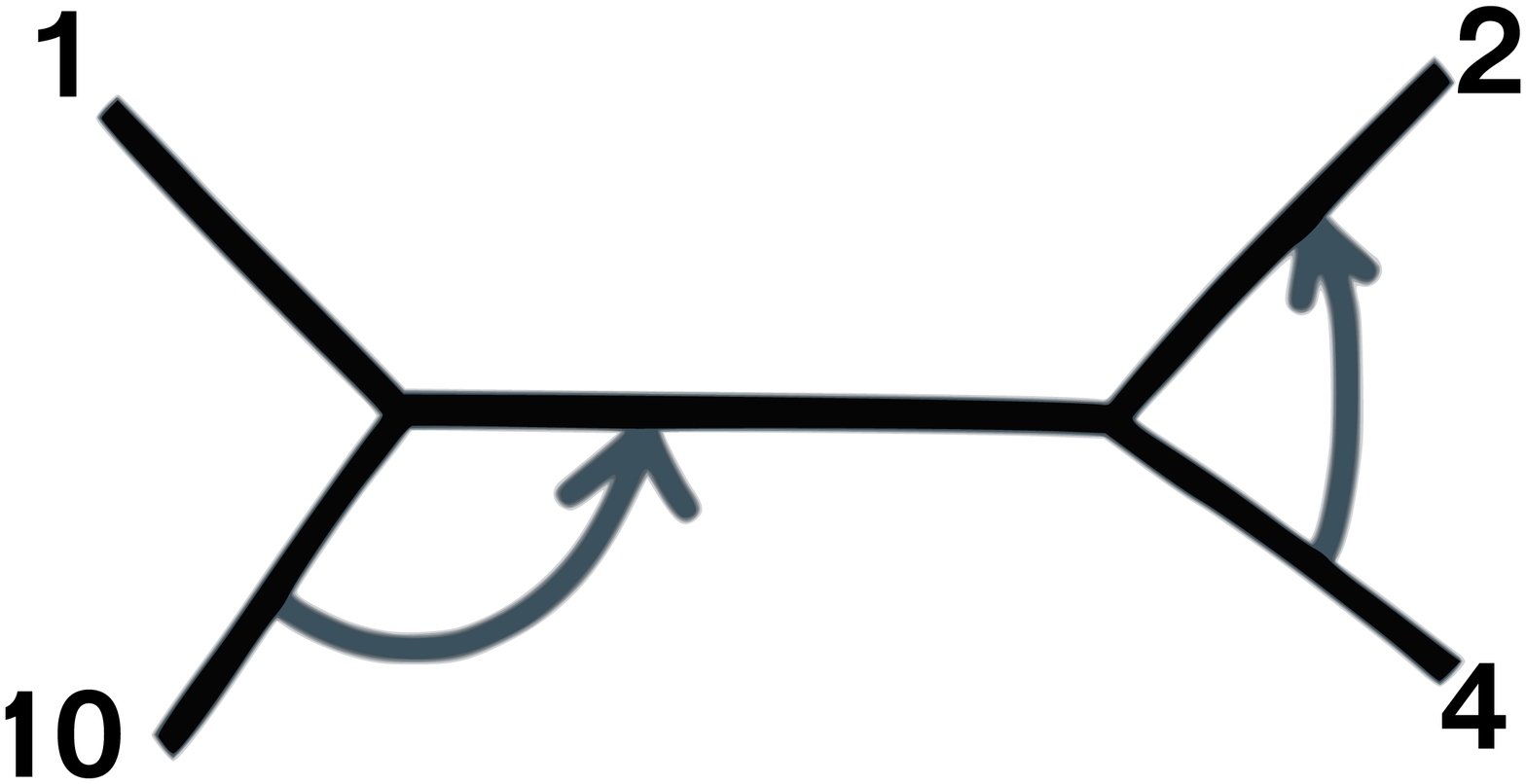}
\caption{Left: Network on 4 taxa with multiple hybridizations of
  type~1. This network is equivalent to a 4-taxon tree (without
  hybridizations) in terms of expected CFs. 
  Right: 4-taxon network extracted from the network in Fig.~\ref{simNet}~(d)
  with 2 hybridizations of type~1. This network is equivalent to a
  4-taxon tree (without hybridizations) in terms of expected CFs.
}
\label{morehyb}
\end{figure}

We can use the subnetwork equivalence just described to account for
multiple hybridizations on the same network (see example on
Fig. \ref{morehyb} left).  
Even when none of the hybridizations are of type 1 on the full
network, some hybridizations may become of type 1 after pruning taxa.
Fig.~\ref{morehyb} (right) shows such an example: 
when reduced to the four taxa 1, 2, 4 and 10, 
the network in Fig.~\ref{simNet}~(d)
has two hybridizations of type~1. Both can be
removed with the subnetwork equivalence described in the previous section.

\section{Topology identifiability}

\subsection{5-taxon network: topology identifiability}
\label{5top_id}

Assume first $h=1$. A network with $n=5$ taxa has $5 \choose 4$
$=5$ four-taxon subsets, and each one has three possible quartets.
Each quartet CF expected on this network is given in previous section
\ref{formulas_quartet}.  Thus, a 5-taxon network defines $3*5=15$
quartet CF equations.  The same is true for a 5-taxon species tree
obtained by removing the hybridization event in the network, but with
different CF formulas.

The hybridization in the network is identifiable if the same set of
quartet CF values cannot solve simultaneously the system of equations
for the network and the system of equations for the tree. If there
exists a set of CF values that can solve both systems of equations,
then we cannot identify the network from the tree based on quartet
CFs. Thus, the conditions under which $CF_{network}=CF_{tree}$ have a
solution allow us to identify the conditions under which the network
is not identifiable from the tree.
\begin{figure}[ht]
\centering
\includegraphics[scale=0.2]{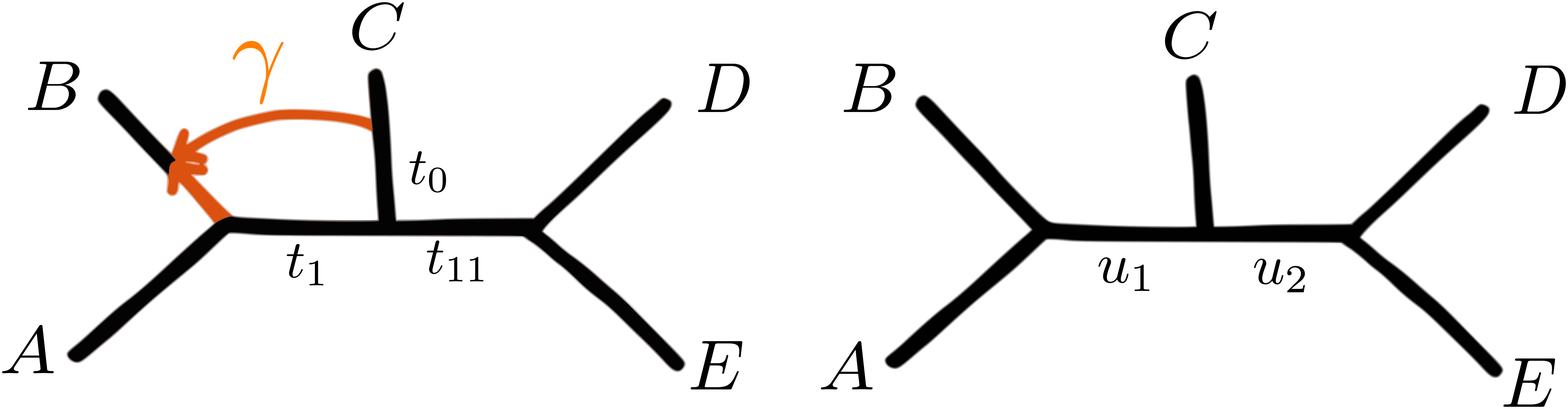}
\caption{The network (left) is identifiable from the tree (right)
based on quartet CFs,
provided that $t_0 > 0$, $t_1 > 0$, $t_{11}<\infty$ and $\gamma \in (0,1)$,
for any value of $u_1$ and $u_2$.}
\label{network5}
\end{figure}
Take, for example, the 5-taxon network in Fig. \ref{network5} (left)
with system of CF equations in table \ref{networkCF}. To prove that it
is identifiable from the tree in Fig. \ref{network5} (right), we
computed the 15 quartet CFs $c_1,c_2,...,c_{15}$ for the tree with the
formulas in table \ref{treeCF} for given values of branch
lengths. Then, with aid of Macaulay2, we investigated if the set of
equations for the network would have a solution when equal to those CF
values.  It turns out that regardless of the values of the tree branch
lengths $u_1,u_2$, the 5-taxon network is identifiable from the tree
as long as $t_0 > 0$, $t_1 > 0$, $t_{11}<\infty$ and $\gamma \in (0,1)$.

\begin{table}
\caption{System of concordance factors equations for 5-taxon network in Fig. \ref{network5} (left)}
\label{networkCF}
\centering
\begin{tabular}{cl}
$ABCD$ & $\begin{array}{l} CF_{AB|CD}=(1-\gamma)(1-2/3\exp(-t_1))+\gamma(1/3\exp(-t_0)) \\
CF_{AD|BC}=(1-\gamma)(1/3\exp(-t_1))+\gamma(1-2/3\exp(-t_0)) \\
CF_{AC|BD}=(1-\gamma)(1/3\exp(-t_1))+\gamma(1/3)\exp(-t_0)
\end{array}$ \\
\hline
$ABDE$ & $\begin{array}{l} CF_{AB|CE}=(1-\gamma)(1-2/3\exp(-t_1))+\gamma(1/3\exp(-t_0)) \\
CF_{AE|BC}=(1-\gamma)(1/3\exp(-t_1))+\gamma(1-2/3\exp(-t_0)) \\
CF_{AC|BE}=(1-\gamma)(1/3\exp(-t_1))+\gamma(1/3)\exp(-t_0)
\end{array}$ \\
\hline
$ABDE$ & $\begin{array}{l} CF_{AC|DE}=(1-\gamma)(1-2/3\exp(-t_1-t_{11}))+\gamma(1-2/3\exp(-t_{11})) \\
CF_{AD|CE}=(1-\gamma)(1/3\exp(-t_1-t_{11}))+\gamma(1/3\exp(-t_{11})) \\
CF_{AE|CD}=(1-\gamma)(1/3\exp(-t_1-t_{11}))+\gamma(1/3\exp(-t_{11}))
\end{array}$ \\
\hline
$ACDE$ & $\begin{array}{l} CF_{AC|DE}= 1-2/3\exp(-t_{11})\\
CF_{AD|CE}= 1/3\exp(-t_{11})\\
CF_{AE|CD}= 1/3\exp(-t_{11}) \end{array}$ \\
\hline
$BCDE$ & $\begin{array}{l} CF_{AC|DE}=(1-\gamma)(1-2/3\exp(-t_{11}))+\gamma(1-2/3\exp(-t_0-t_{11}))\\
CF_{AD|CE}=(1-\gamma)(1/3\exp(-t_{11}))+\gamma(1/3\exp(-t_0-t_{11}))\\
CF_{AE|CD}=(1-\gamma)(1/3\exp(-t_{11}))+\gamma(1/3\exp(-t_0-t_{11}))
\end{array}$
\end{tabular}
\end{table}

\begin{table}
\caption{System of concordance factors equations for 5-taxon unrooted tree in Fig. \ref{network5} (right)}
\label{treeCF}
\centering
\begin{tabular}{cl}
$ABCD$ & $\begin{array}{l} CF_{AB|CD}=\cftreemajor{u_1} \\ CF_{AD|BC}=\cftreeminor{u_1} \\ CF_{AC|BD}=\cftreeminor{u_1}
\end{array}$ \\
\hline
$ABDE$ & $\begin{array}{l} CF_{AB|CE}=\cftreemajor{u_1} \\ CF_{AE|BC}=\cftreeminor{u_1} \\ CF_{AC|BE}=\cftreeminor{u_1}
\end{array}$ \\
\hline
$ACDE$ & $\begin{array}{l} CF_{AC|DE}=\cftreemajor{u_2} \\ CF_{AD|CE}=\cftreeminor{u_2} \\ CF_{AE|CD}=\cftreeminor{u_2}
\end{array}$ \\
\hline
$BCDE$ & $\begin{array}{l} CF_{BC|DE}=\cftreemajor{u_2} \\ CF_{BD|CE}=\cftreeminor{u_2} \\ CF_{BE|CD}=\cftreeminor{u_2}
\end{array}$ \\
\hline
$ABDE$ & $\begin{array}{l}  CF_{AB|DE}=\cftreemajor{u_1-u_2} \\ CF_{AD|BE}=\cftreeminor{u_1-u_2} \\ CF_{AE|BD}=\cftreeminor{u_1-u_2} \end{array}$ \\
\end{tabular}
\end{table}

A similar study was done to every possible 5-taxon network with one
hybridization to conclude that if $t_i > 0$ for all tree edge $i$
and if $\gamma \in (0,1)$,
then the network is generically identifiable from the tree
(as long as the number of nodes in the hybridization cycle is $k\geq3$).

\subsection{\textit{n}-taxon network: topology identifiability}
\label{top_id}

To generalize to $n \geq 5$ taxa, not all $n \choose 4$ $\times 3$
quartet CF equations are needed.  In Fig. \ref{hybrid_nodes_cross},
the networks are ordered by the number of nodes in the hybridization
cycle. Ignoring the case with $k=2$ which is not identifiable from a
tree, and assuming that the network is level-1, the hybridization
cycle in the network is attached to $k$ subnetworks, each represented
by a small triangle with a given number of taxa $n_i$.  With $k=3$
nodes in the hybridization cycle for example, the network has three
subgraphs, each with $n_0,n_1,n_2$ taxa respectively.  If $n_0\geq3$
in the top-right subnetwork, one can form a 4-taxon subset by taking
three taxa from this subnetwork and one taxon from any of the other 2
subnetworks. However, the three CF formulas for such a 4-taxon subset
do not involve any of the parameters near the hybridization cycle of
interest. That is, the CFs resulting from choosing three taxa from one
subtree and one taxon from another subtree yield no information about
the hybridization event of interest. In other words, we can ignore all
the equations that involve three taxa from the same subgraph.

Thus, to study the identifiability of an $n$-taxon network, it suffices
to consider only the subset of quartets involving
$n_i \leq 2$ taxa for any subnetwork $i$. If only one tip is taken
for subnetwork $i$, then hybridizations in this subnetwork do
not affect the quartet CFs. If the 4-taxon set has $n_i=2$ tips from subnetwork $i$,
then there could be a hybridization involving them within this subnetwork.
Because the network is of level 1, however, this hybridization has
to be of type~1 (see Section \ref{reparam_quartets}) and can be
removed without affecting the CFs, using a transformed branch length leading to
subnetwork $i$. Thus, we can study the
identifiability of a given hybridization of interest ignoring
hybridizations in the subnetworks.
Using Macaulay2, we found
the same sufficient conditions as with $n=5$ for generic
identifiability of the topology: $t\in (0,\infty)$ for all tree branch
lengths and $\gamma \in (0,1)$.
We give below weaker, but still sufficient conditions for each value of $k$.
The $n$-taxon network with one hybridization and $k$ nodes in this
hybridization cycle is generically identifiable from a tree
with the same topology but the hybridization removed if: 
\begin{align*}
\boldsymbol{k=3}:\;\;\;
&t_{10},t_{11},t_{12}<\infty;\;\; t_0>0;\;\; \gamma \in (0,1) \\
&(1-\gamma)(1-\exp(-t_1)) \neq \gamma((1-\exp(-t_2))+(1-\exp(-t_0)))\\
&\gamma(1-\exp(-t_2)) \neq (1-\gamma)((1-\exp(-t_0))+(1-\exp(-t_1)))
\end{align*}
\begin{align*}
\boldsymbol{k=4}:\;\;\;
&t_{12},t_{13}<\infty;\;\; t_0,t_1>0;\;\; \gamma \in (0,1) \\
&\gamma(1-\exp(-t_3)) \neq (1-\gamma)((1-\exp(-t_1))+(1-\exp(-t_2)))
\end{align*}
\begin{align*}
\boldsymbol{k=5}:\;\;\;
&t_{13},t_{14}<\infty;\;\; t_0,t_1,t_2>0;\;\; \gamma \in (0,1) \\
&\gamma(1-\exp(-t_4)) \neq (1-\gamma)((1-\exp(-t_2))+(1-\exp(-t_3)))
\end{align*}

\begin{figure}
\centering
\includegraphics[scale=0.2]{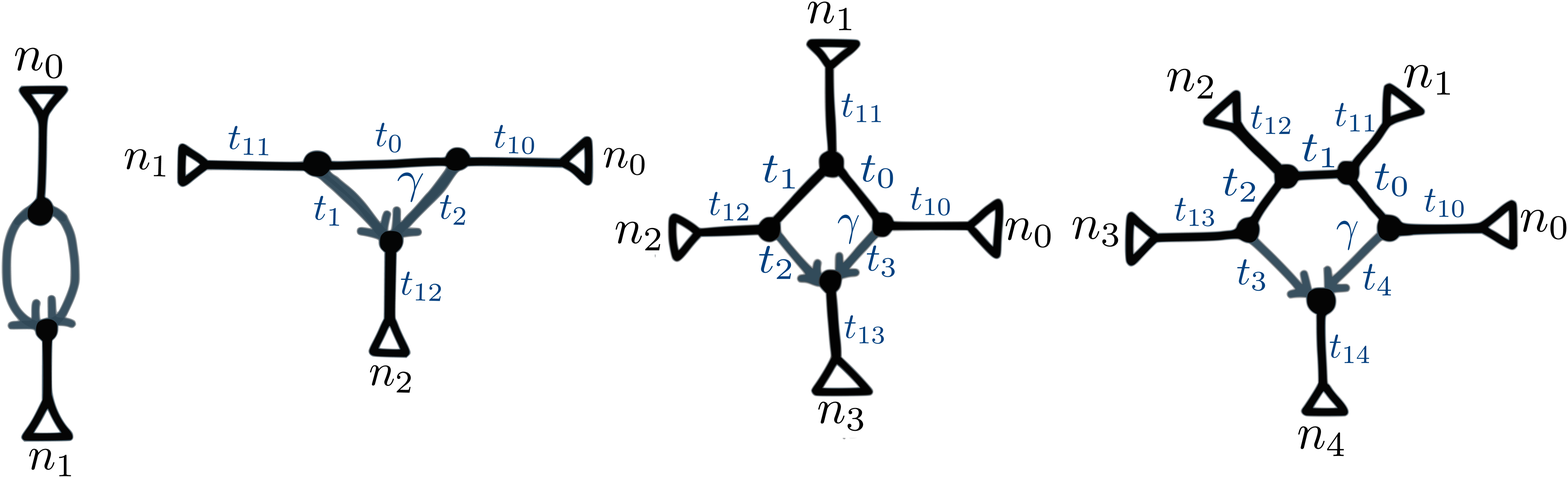}
\caption{Characterization of $n$-taxon networks depending on the
  number of nodes in the hybridization cycle: $k=2,3,4,5$ in order
  from left to right.}
\label{hybrid_nodes_cross}
\end{figure}

\section{Parameter identifiability}
We study here whether branch lengths and heritabilities are identifiable
from quartet CFs, given a fixed network topology.

\subsection{Identifiability of parameters in a 5-taxon network}
\label{5par_id}

As mentioned before, a network with $n=5$ taxa has 15 quartets,
with quartet CFs given in section \ref{formulas_quartet}.
Thus, a 5-taxon network defines
15 CF equations in the unknown parameters of branch lengths and
$\gamma$.
If this system has a unique solution for all the unknown parameters
given a set of CF values $c_1,c_2,...,c_{15}$, then we say that the
parameters are identifiable.  For example, recall the 5-taxon network
in Fig. \ref{network5} (left) with system of CF equations in table
\ref{networkCF}.  From a quick inspection, it is evident that not all
vectors of CF values can yield a solution. For instance, the quartet CFs
are identical for the 4-taxon sets ABCD and ABCE. Thus, if we consider
the first 6 equations:

\begin{align*}
CF_{AB|CD}&=(1-\gamma)(1-2/3\exp(-t_1))+\gamma(1/3\exp(-t_0)) =c_1\\
CF_{AD|BC}&=(1-\gamma)(1/3\exp(-t_1))+\gamma(1-2/3\exp(-t_0)) =c_2\\
CF_{AC|BD}&=(1-\gamma)(1/3\exp(-t_1))+\gamma(1/3)\exp(-t_0) =c_3\\
CF_{AB|CE}&=(1-\gamma)(1-2/3\exp(-t_1))+\gamma(1/3\exp(-t_0)) =c_4\\
CF_{AE|BC}&=(1-\gamma)(1/3\exp(-t_1))+\gamma(1-2/3\exp(-t_0)) =c_5\\
CF_{AC|BE}&=(1-\gamma)(1/3\exp(-t_1))+\gamma(1/3)\exp(-t_0)=c_6,
\end{align*}
it is obvious that a solution can exist only if
\[c_1=c_4,\quad c_2=c_5,\quad\mbox{and }c_3=c_6.\]
In addition to these three conditions, there are many others that the
CF values $c_1,c_2,...,c_{15}$ need to fulfill for a solution to exist.
In other words, the structure of the network imposes conditions on the
CF values that need to be satisfied for the system of equations to be
consistent.  We will call these conditions \textit{invariants}, which
are like \textit{consistency checks} for the system.
On a tree, the invariants are easy to list.
For the tree in Fig. \ref{network5}, for example, equating the
expected quartet CFs from table \ref{treeCF} to a set of CF values $c_i$
gives the following system of equations for the 2 unknown branch lengths $u_1$ and $u_2$:
\begin{align*}
CF_{AB|CD}&=\cftreemajor{u_1} =c_1\\
CF_{AD|BC}&=\cftreeminor{u_1} =c_2\\
CF_{AC|BD}&=\cftreeminor{u_1} =c_3\\
CF_{AB|CE}&=\cftreemajor{u_1} =c_4\\
CF_{AE|BC}&=\cftreeminor{u_1} =c_5\\
CF_{AC|BE}&=\cftreeminor{u_1} =c_6\\
CF_{AC|DE}&=\cftreemajor{u_2} =c_7\\
CF_{AD|CE}&=\cftreeminor{u_2} =c_8\\
CF_{AE|CD}&=\cftreeminor{u_2} =c_9\\
CF_{BC|DE}&=\cftreemajor{u_2} =c_{10}\\
CF_{BD|CE}&=\cftreeminor{u_2} =c_{11}\\
CF_{BE|CD}&=\cftreeminor{u_2} =c_{12}\\
CF_{AB|DE}&=\cftreemajor{u_1-u_2} =c_{13}\\
CF_{AD|BE}&=\cftreeminor{u_1-u_2} =c_{14}\\
CF_{AE|BD}&=\cftreeminor{u_1-u_2}=c_{15}.
\end{align*}
Several equations are obviously identical. Hence, these conditions on the $c_i$ values are
necessary for the system to have a solution $(u_1,u_2)$:
\[c_1=c_4,\quad c_2=c_3=c_5=c_6,\quad c_7=c_{10},\quad c_8=c_9=c_{11}=c_{12}\quad
\mbox{and }c_{14}=c_{15}.\]
In addition, the three CF from each 4-taxon set must add up to 1:
\[c_1+c_2+c_3 = c_4+c_5+c_6 = c_7+c_8+c_9 = c_{10}+c_{11}+c_{12} = c_{13}+c_{14}+c_{15} = 1.\]
Finally, three times the minor CF of
ABCD multiplied by three times the minor CF of ACDE should be equal to
three times the minor CF of ABDE. This is because
$\exp(-u_1)\exp(-u_2)$ should be equal to $\exp(-u_1-u_2)$.
We thus obtain the following additional invariant
\begin{align*}
(3c_2)(3c_8)&=3c_{14}.
\end{align*}
For this species tree, we obtained 13 \emph{independent} invariants
that the $c_i$ values must satisfy for the system to have a solution.
If the values $c_1,c_2,...,c_{15}$ are obtained from a tree with
branch lengths $u_1^*,u_2^*$ (using table \ref{treeCF}),
then these $c_i$ values automatically satisfy the invariants of the system,
and $(u_1,u_2)=(u_1^*,u_2^*)$ is necessarily one solution of the
system of equations. Our hope is that this solution is unique.

The question of identifiability can now be restated in terms of the
number of algebraically independent equations that the system has.
For the tree above, there are 15 original equations but 13 independent
invariants, therefore two algebraically \emph{independent} equations.
Since we only have 2 unknown parameters $u_1$ and $u_2$,
we can solve for them.
We can do a similar analysis with the network example in Fig.
\ref{network5} with CF equations in table \ref{networkCF}. We have a
system of 15 equations, and we would like to know how many independent
invariants are defined by this system in order to determine how many
algebraically independent equations we have.  We know from algebraic
geometry \cite{Cox2007} that a system with the same number of
algebraically independent equations as unknown parameters has finitely
many solutions. This does not imply that the parameters are
identifiable, but it does imply that the parameters are generically
indentifiable. It also implies that, given perfect data
(infinitly many genes, correctly reconstructed gene trees),
the pseudolikelihood has finitely many maxima.
Proving uniqueness of solution is a
hard problem in algebraic geometry and is beyond the scope of the
present work.

Therefore, to study the generic identifiability of parameters,
we obtained the system of quartet CF equations for each network,
and we verified whether the number of algebraically
independent equations was equal to the number of parameters. We
automated this using Macaulay2.

The network in Fig. \ref{network5} needs four
parameters ($\gamma,t_0,t_1,t_{11}$) and 15 equations. In this example,
the 15 equations have 12 independent invariants. Thus, we only have three
algebraically independent equations and four parameters. This means
that the system has an infinite number of solutions and we cannot solve
for $\gamma,t_0,t_1,t_{11}$.

In this case, which is a bad diamond I, we decided to reparametrize
$(\gamma,t_0,t_1,t_{11})$.
The quartet CFs can be expressed in terms of the following 3 parameters
only, which are identifiable from the 3 algebraically independent equations:
$x=\gamma(1-\exp(-t_0))$, $y=(1-\gamma)(1-\exp(-t_1))$ and $t_{11}$.

They have the following interesting interpretation: the
lineage from species B in Fig. \ref{network5} (left) either originated
from the hybridization edge and coalesced with C along the edge of length $t_0$
with probability $x=\gamma(1-\exp(-t_0))$, or it originated from the other hybridization
edge and coalesced with A along the edge of length $t_1$,
with probability $y=(1-\gamma)(1-\exp(-t_1))$.

\subsection{\textit{n}-taxon network: parameter identifiability when $h=1$}
\label{par_id}
In this section we assume that the network has a single hybridization
(assumption relaxed in the next section)
and we seek to determine if the parameters around the cycle created by this
hybridization are identifiable (Fig.~\ref{hybrid_nodes_cross}),
given the network topology.
Parameter identifiability depends on
the number of nodes in the cycle created by the hybridization
event. We summarize the results below.

For $\boldsymbol{k=3}$, the parameters are not identifiable if
  $n\leq5$. If $n\geq6$ and $n_i\geq2$ for all $i=0,1,2$ (see
  Fig. \ref{hybrid_nodes_cross}), we have 6 algebraically
  independent equations and 7 parameters. Thus, the 7 parameters
  are not identifiable, but 6 of them are identifiable if the remaining parameter
  if known. We decided to set $t_{12}=0$ and estimate the other 6 paramerers. We call this
  case a \textit{good triangle}.

  For $\boldsymbol{k=4}$, all parameters are identifiable if either
  $n_0\geq 2$ (or $n_2$, symmetrically), or if both $n_1$ and $n_3\geq
  2$ (see Fig.~\ref{hybrid_nodes_cross}).  Parameters are not all
  identifiable in the remaining 2 cases, which we call \textit{bad
    diamonds} I and II (see Fig.~\ref{hybrid_nodes_cross}).  The bad
  diamond I was described in the previous section. For a bad diamond
  II, 6 parameters are needed around the hybridization and we found
  only 5 algebraically independent equations, with no simple
  reparametrization. So, we set $t_{13}=0$ and solve for the remaining
  parameters.

For $\boldsymbol{k=5}$ all parameters are identifiable. To show it, we first
 considered the case when $n_0=\cdots=n_4=1$, that is, we considered information from
 quartets with at most one taxon from each subnetwork. Using Macaulay2 we
 found that $\gamma$ and the 3 tree edge lengths in the cycle ($t_0,t_1,t_2$) were
 identifiable. Next, we used previous results from $k=4$.
 If $n_i\geq 2$ and $i\neq 4$, the length $t_{1i}$ of the edge attaching subnetwork $i$
 is identifiable because we can extract a good diamond from which
 we can identify $t_{1i}$. If $n_4 \geq 2$, the hybrid branch lengths $t_3,t_4$
 and the subnetwork branch length $t_{14}$ are needed. To identify these parameters,
 we can extract a bad diamond II, which originally provided 6 algebraically
 independent equations. These are enough to identify the remaining
 3 unknown branch lengths.

For $\boldsymbol{k>5}$ we can prove that all needed parameters are identifiable
 by extracting subnetworks with $k=5$, each identifying a different subset
 parameters, together spanning all parameters.

\smallskip
Note that the branch lengths labelled in Fig. \ref{hybrid_nodes_cross} are
not all needed for all networks. In particular, if $n_i=1$ then the branch
leading to subnetwork $i$ is an external branch, and its length $t_{1i}$
is irrelevant for any CF.
For a bad diamond I for example ($n_1\geq 2$ but other $n_i=1$),
$t_{10}$, $t_{12}$ and $t_{13}$ are irrelevant and obviously non-identifiable.
In this bad diamond I, the hybrid branch lengths $t_2$ and $t_3$ are also
irrelevant, as they only have 1 descendant ($n_3=1$) like external edges.
These branch lengths were omitted from our study of identifiability:
we did not include them in the list of parameters to be studied
and we excluded them from the list of parameters during the
pseudolikelihood optimization search.

\subsection{Parameter identifiability in level-1 networks}
We now extend the results from the previous section to networks with
$h>1$, provided that the network is of level 1.
As already noted in section~\ref{top_id}, only a subset of
quartets bear information on the hybridization of interest and on
parameters around this hybridization: those involving $n_i\leq 2$
taxa from any given subnetwork $i$.
Because the network is of level 1, any hybridization in subnetwork $i$
does not affect the quartet CFs if $n_i=0$ or $n_i=1$.
For 4-taxon sets that need to involve $n_i=2$ taxa from subnetwork $i$,
we noted in section~\ref{top_id} that the subnetwork reduced to these 2 taxa was equivalent
(in terms of quartet CFs) to a subnetwork with no hybridizations
where the 2 taxa are sister
(subnetwork of type~1 in section~\ref{reparam_quartets}).
This may come at the cost of having to transform the length $t_{1i}$
of the branch linking the cycle from the hybridization of interest to
subnetwork $i$ (Fig.~\ref{hybrid_nodes_cross}).
If there is at least one pair of taxa from subnetwork $i$ such that
the equivalent subnetwork reduced to this pair
is separated from the cycle by a branch of untransformed length $t_{1i}$,
then the results stated in the previous section apply directly.
If there is no such pair of taxa from subnetwork $i$, then the results
stated in the previous section apply to the transformed branch length
$\tilde{t}_{1i}$ instead.

\section{Heuristic search in the space of networks}
\label{search}
The search is initialized with a user-specified network, which could
be a tree obtained with a very fast quartet-based tree estimation
method like ASTRAL \cite{Mirarab2014}. If a tree topology is given
with no branch lengths, those are initialized using the average
observed CF of the quartets that span that branch exactly,
$\overline{\cf}$, transformed to coalescent units by
$t=-\log(1-3/2\,\overline{\cf})$.
By default, SNaQ performs 10 independent searches, each using
a starting topology based on the user-specified network or tree.
For each search, an NNI is performed on the user-specified topology with
probability 0.7 by default, so that independent runs
have different starting topologies. If the starting topology is
  a network, then with probability 0.7 by default the origin (or
  target) of a hybrid edge is moved.
Each search then navigates the network space by altering the current
network using one of 5 proposals, chosen at random:
\begin{enumerate}
\item Move the origin of an
existing hybrid edge. A hybrid node is chosen uniformly at
random, then one of the 2 parent edges of that node is chosen
according to the inheritance probabilities: the edge with inheritance $\gamma$
is chosen with probability  $1-\gamma$.
A new edge is then chosen at random
from the vicinity of the current origin.
This edge is cut into 2 smaller edges to insert the new origin.
The two branches at the old origin are merged along with their
branch lengths, whose proportions are used for the 2 new branches
created around the new origin.
\item Move the target of an existing hybrid edge, similarly to
the previous move.
\item Change the direction of an existing hybrid edge.
 A hybrid node is chosen uniformly at
 random, then one of the 2 parent edges of that node is chosen
 according to the inheritance probabilities: the edge with inheritance $\gamma$
 is chosen with probability  $1-\gamma$. The direction of the chosen hybrid edge
 is flipped. The former hybrid node becomes a tree node and the other node
 attached to the hybrid edge becomes a hybrid node.
 Branch lengths are left unchanged.
\item Perform a
 nearest-neighbor interchange (NNI) on a tree edge, around a tree edge
 chosen uniformly at random. \cite{Huber2015} describe a similar type of NNI
 that yields a level-1 (unrooted) network, and showed that the resulting network
 is level-1 if the chosen tree edge is not a link.
 Branch lengths are left unchanged.
\item Add a hybridization if the current topology has $h<h_m$.
 Two tree edges are chosen at random and a new hybrid edge is created
 between these 2 edges. The new nodes (origin and target) are placed
 uniformly along each chosen edge. $\gamma$ is drawn uniformly in $(0,0.5)$.
 The new edge length is initialized at 0.
 If the new cycle intersects a previously existing cycle,
 the origin or the target is chosen at random, to be moved to
 a neighboring edge immediately.
\end{enumerate}
Any new proposed network is checked to verify that it is of level 1,
with $h\leq h_m$ and with at least one valid placement for the root.
If not, the move fails immediately and a new move is proposed at random.
The search continues until one the following criterion is reached:
\begin{itemize}
\item the absolute difference between the pseudolikelihood value of
the newly accepted and the current network is smaller than
a tolerance threshold, 0.001 by default.
\item the pseudo-deviance (difference between the theoretical maximum pseudolikelihood
and the network pseudolikelihood) is below the tolerance threshold.
\item the number of failed moves reaches a limit, 100 by default.
\item 
  for all move types, the number of failed attempts reaches a limit, which is
  specific to the proposal type, $n$ and $h$.  This is
  to avoid repeated proposals of the same new network.  For example, on a
  network with $h=1$, there are at most $N_1=8$ ways to move the
  origin of either hybrid edge to a neighboring location.
  For this network, we chose a limit of 28 failed moves of type 1
  to ensure that all of the 8 distinct proposals were attempted with
  high probability, based on
  the theory of the coupon's collector \cite{Feller1950}.
  In general, for $N=N_i$ distinct ways to attempt a move of type $i$,
  the upper threshold was set to
$\displaystyle
  N\sum_{i=1}^N\frac{1}{i}+\sqrt{\frac{\pi}{6}} N.$
We calculated $N_i$ for each move type.
In general there are $N_1=N_2=8h$ ways to move a hybrid origin or target
and $N_3=2h$ ways to change a hybrid edge direction.
There are $2n-3$ internal tree edges on $n$ taxa, so there are
$N_4=2n-3$ ways to propose an NNI and
$N_5=$ $2n-3\choose2$ ways to choose two tree edges to add a new
hybridization.
\end{itemize}

\section{Identifiability from quartets versus triples}

\begin{figure}[ht]
\centering
\subfigure[{\large $\Psi_1$}]{\includegraphics[scale=0.2]{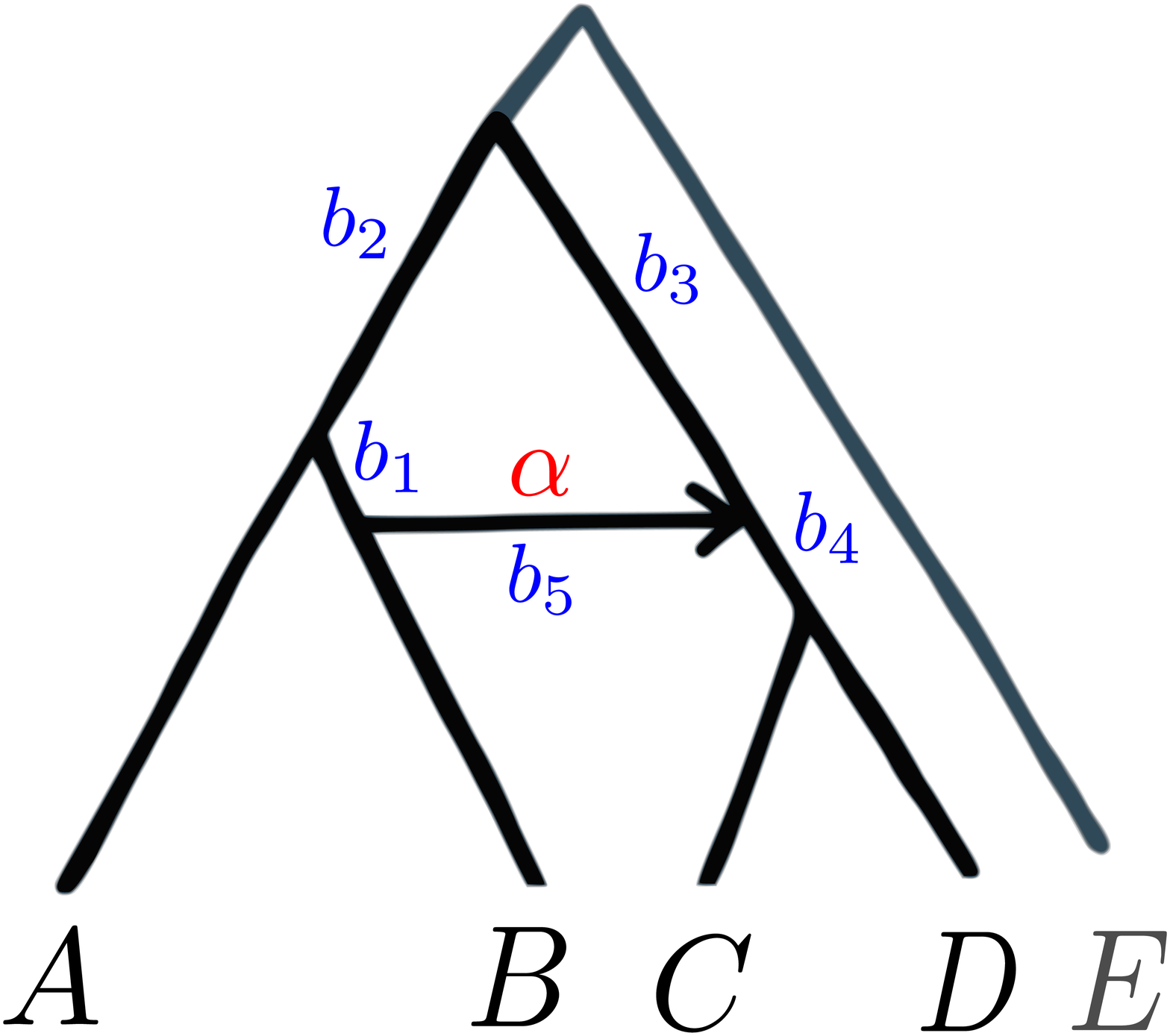}}\hspace{1cm}
\subfigure[{\large $\Psi_2$}]{\includegraphics[scale=0.2]{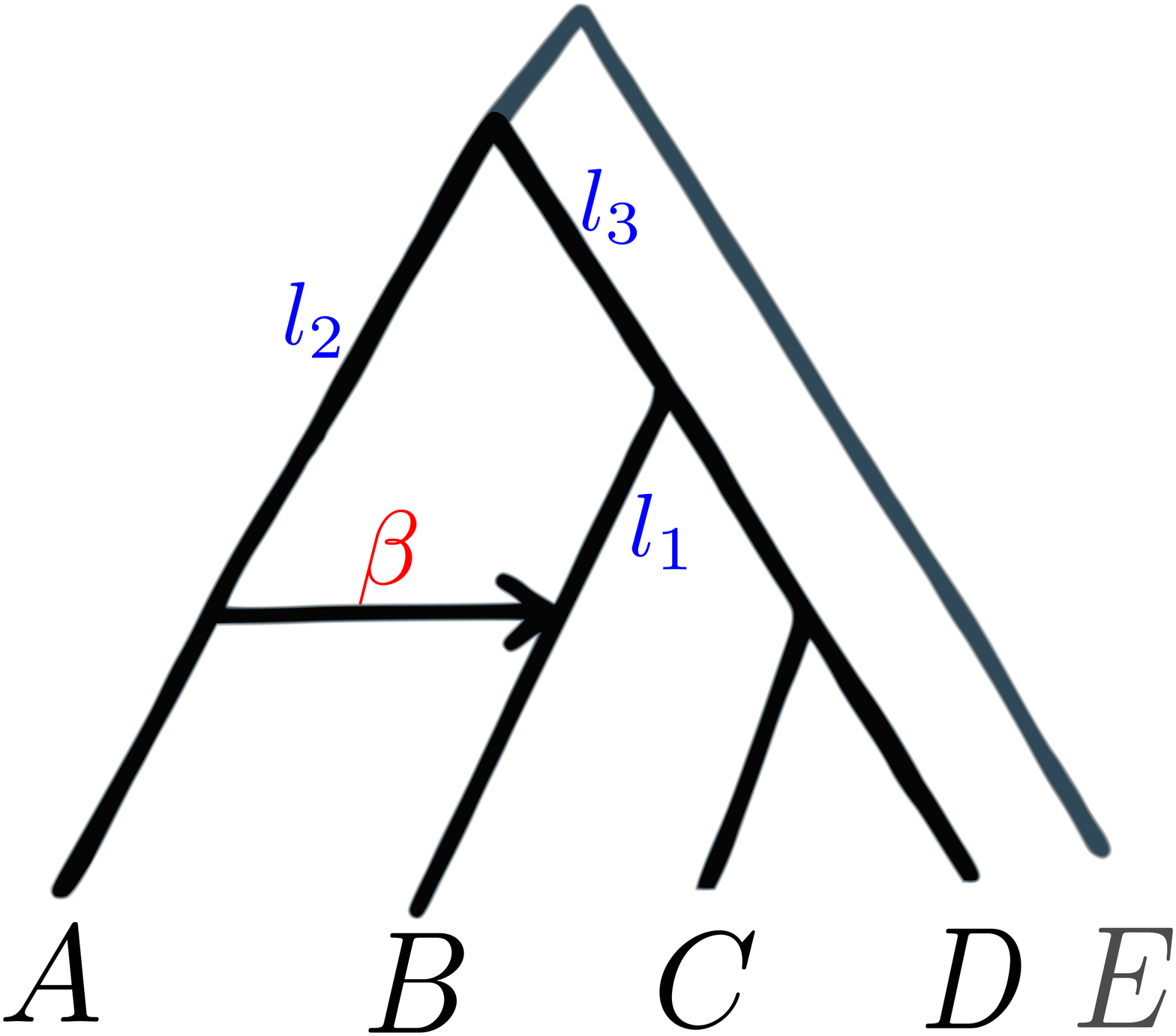}}
\caption{Example from \cite{yu2015} of networks non-identifiable with rooted triples
from the 4 ingroup taxa $(A,B,C,D)$. These two networks are identifiable from each other
based on unrooted quartets from all 5 taxa.}
\label{netYu}
\end{figure}

\begin{figure}[ht]
\centering
\subfigure[$n=6$ taxa,  $h=1$ hybridization ($k=4$)]{\includegraphics[scale=0.4]{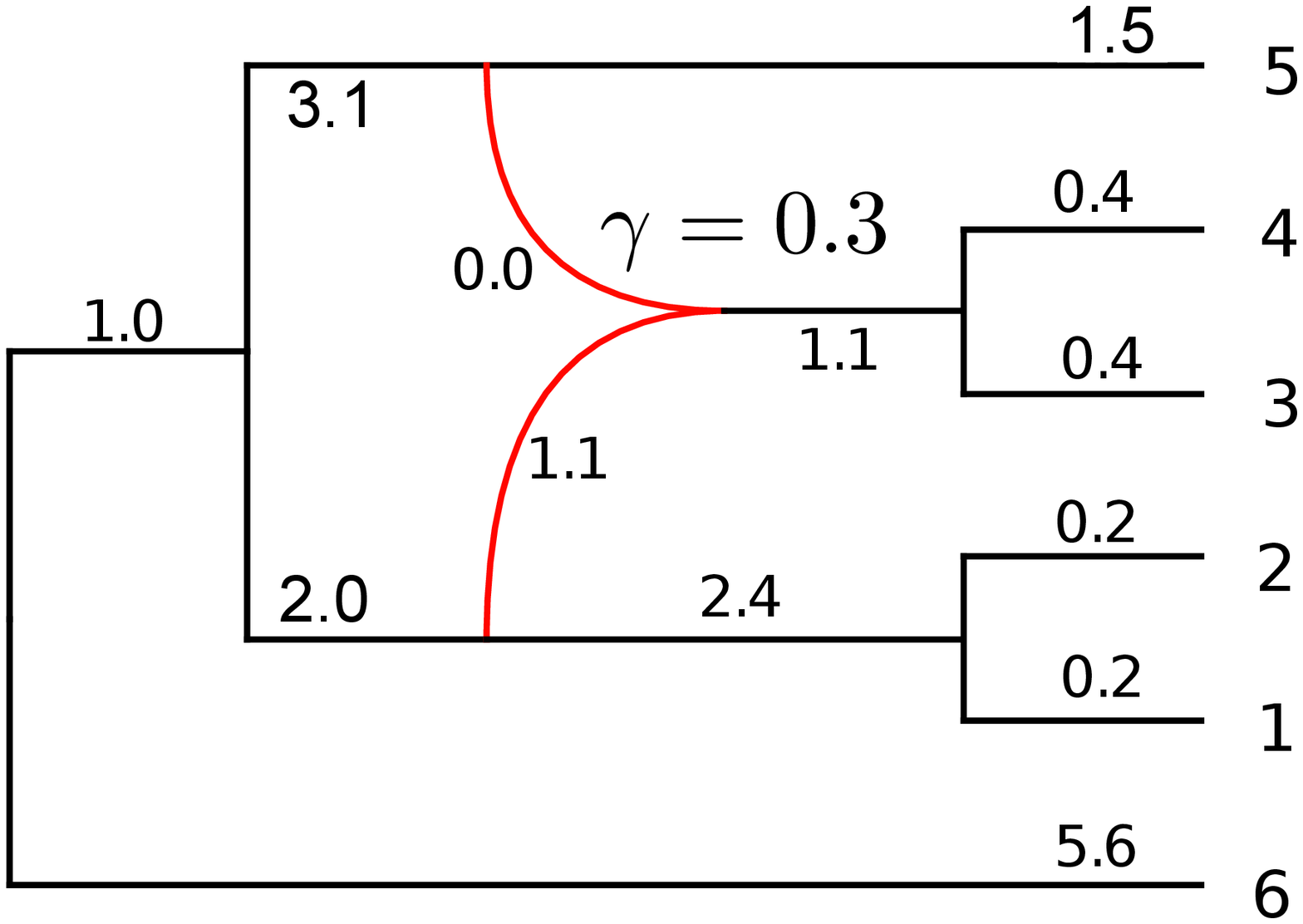}}
\subfigure[$n=6$ taxa,  $h=2$ hybridizations ($k=4,4$)]{\includegraphics[scale=0.4]{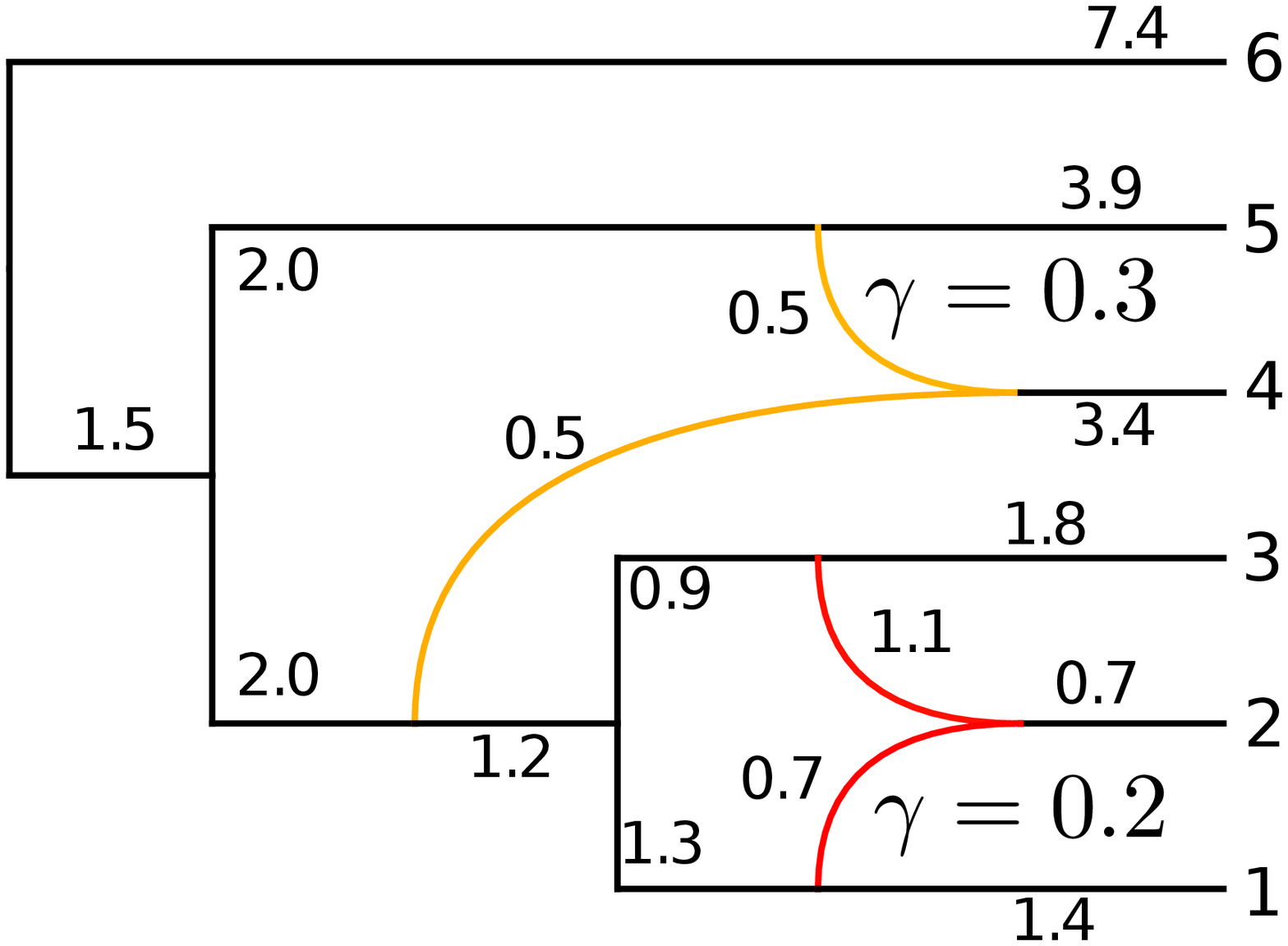}}
\subfigure[$n=10$ taxa, $h=2$ hybridizations ($k=4,7$)]{\includegraphics[scale=0.4]{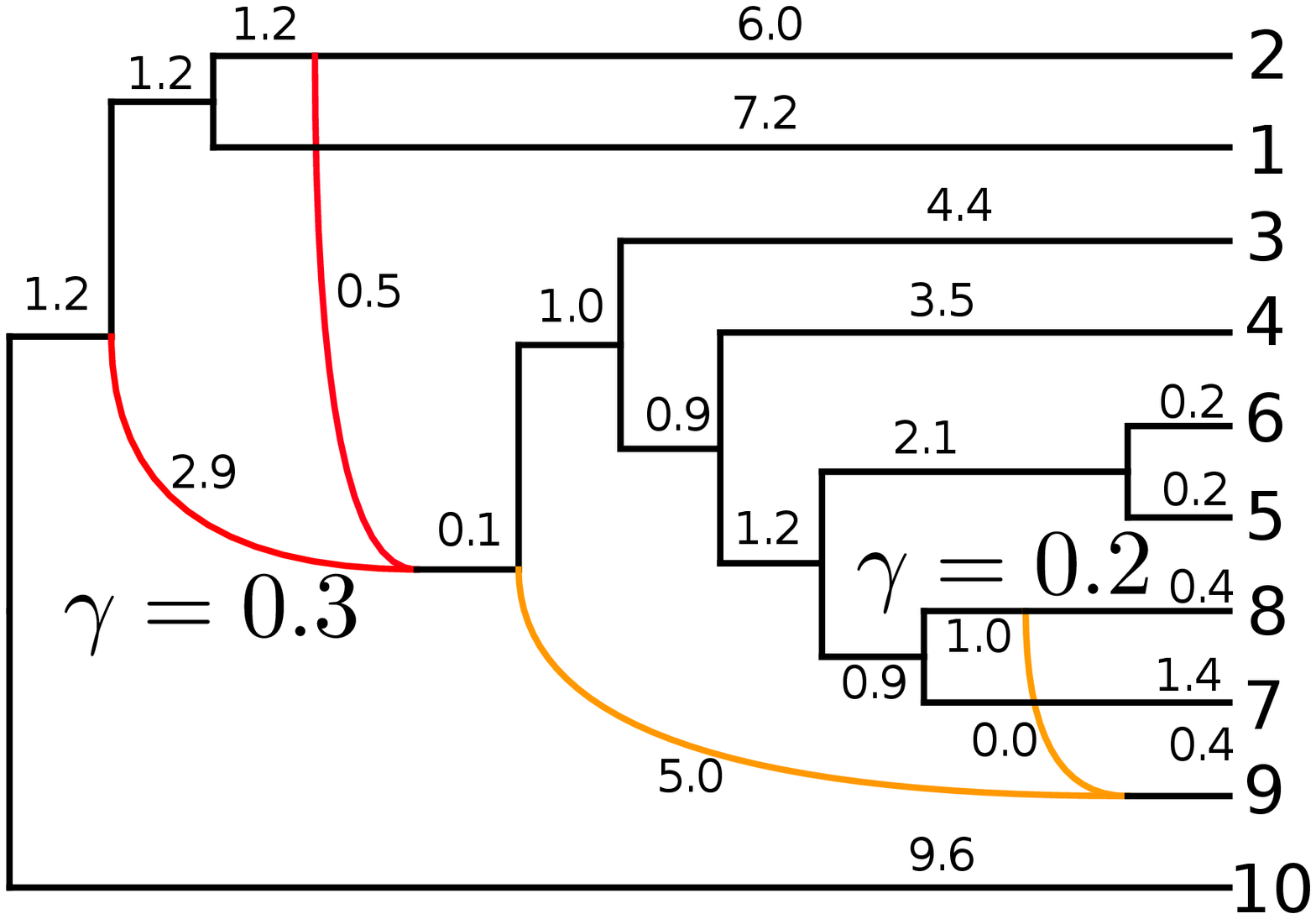}}
\subfigure[$n=15$ taxa, $h=3$ hybridizations ($k=4,5,6$)]{\includegraphics[scale=0.4]{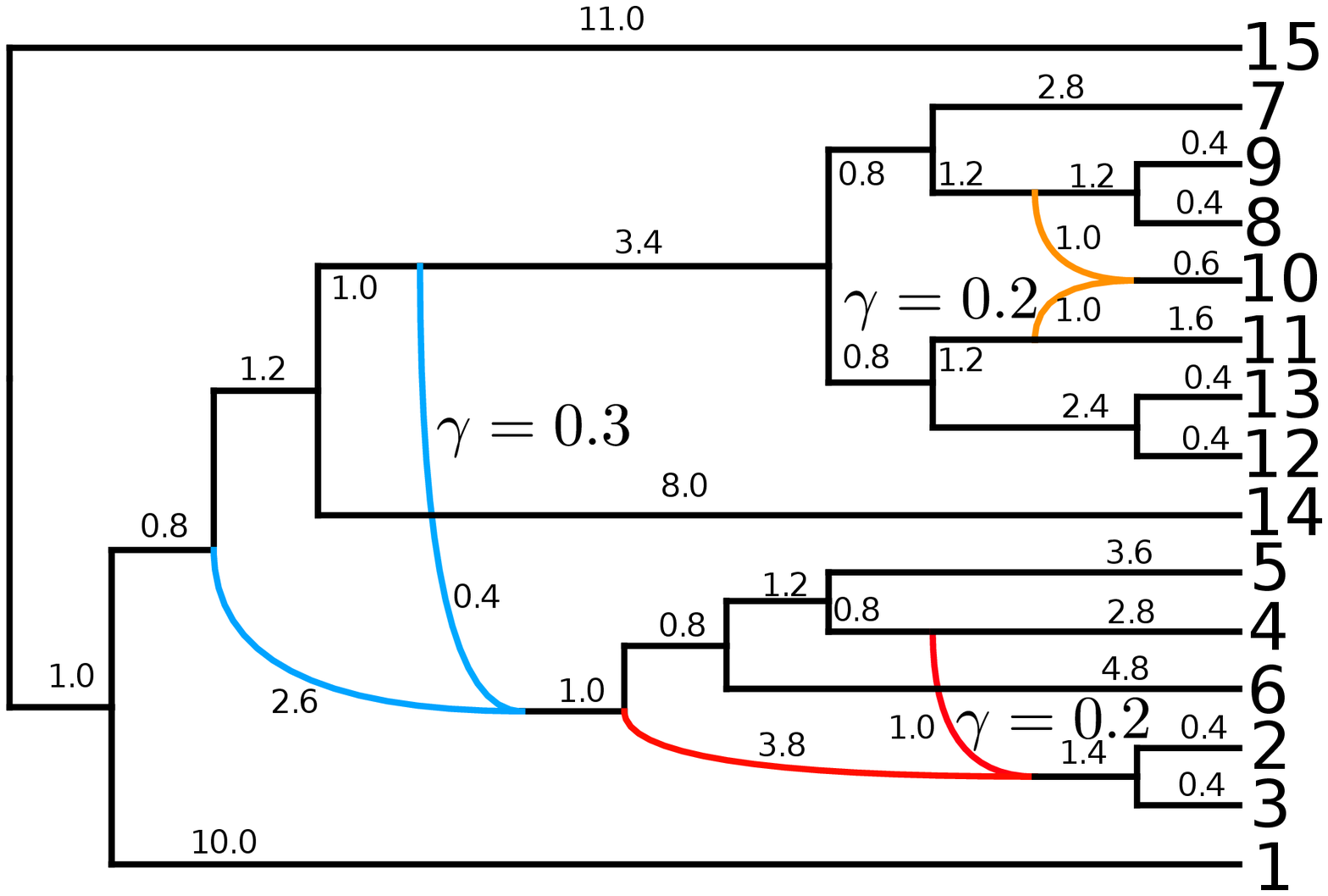}}
\caption{True networks used for the simulations, including
a bad diamond I (top right, $n=6$, with $\gamma=0.2$)
and a bad diamond II (bottom left, $n=10$, with $\gamma=0.3$). Branch
lengths are in coalescent units.}
\label{simNet}
\end{figure}

A pseudolikelihood based on rooted triples is used in \cite{yu2015}
to estimate a rooted network, whereas we use unrooted quartets to estimate
a semi-directed network.  We show here that unrooted quartets provide more
information to identify networks, that may not be identifiable with
rooted triples. For example, \cite{yu2015} present two networks $\Psi_1$ and $\Psi_2$
that are not identifiable by their set of triples (see Fig. \ref{netYu}). We
show here that these networks are identifiable from the set of
quartets.

\begin{table}[ht]
\centering
\begin{tabular}{c||lllll}
Rooted triples    & ABC  & ACD  & ABD  & BCD  & \\
Unrooted quartets & ABCE & ACDE & ABDE & BCDE & ABCD
\end{tabular}
\caption{The collection of 3-taxon subsets from a 4-taxon rooted network
  corresponds in part to the collection of 4-taxon subsets from an unrooted network on
  the same 4 taxa (A,B,C,D) plus an outgroup (E).
  Extra subsets are obtained by sampling ingroup taxa only, like ABCD.}
\label{triquart}
\end{table}

A rooted triple is equivalent to an unrooted quartet by adding the
outgroup used to root the triple, if gene trees were initially rooted
using an outgroup. If gene trees were rooted using a molecular clock assumption
or midpoint rooting, for example, the root still implies an outgroup shared by
all gene trees. Thus, the full set of rooted triples correspond to the
unrooted quartets on any 3 ingroup taxa and the outgroup
(see table \ref{triquart}). With 4 ingroup taxa, the collection of 4-taxon
subsets has one extra set that provides enough information to
distinguish between the two networks in Fig. \ref{netYu}.
Fig. \ref{netYu} corresponds to the networks $\Psi_1$ and $\Psi_2$ in
\cite{yu2015} with an added outgroup $E$.
As mentioned in
  previous sections, the 15 CF equations are not independent as they
  need to add up to one for any given 4-taxon subset (or 3-taxon subset)
  and some subsets are of type~1 so the two minor CFs are
  equal. The independent formulas
  for the CFs from $\Psi_1$ with added outgroup E are:
\begin{align*}
CF_{BC|AE}&=(1-\beta)(\cftreemajor{l_3})+\beta\cftreeminor{l_2}\\
CF_{BA|CE}&=(1-\beta)\cftreeminor{l_3}+\beta(\cftreemajor{l_2})\\
CF_{BE|CD}&=(1-\beta)(\cftreemajor{l_1})+\beta(\cftreemajor{l_3-l_{1}})\\
CF_{AE|CD}&=\cftreemajor{l_{1}-l_3}\\
CF_{AB|CD}&=(1-\beta)(\cftreemajor{l_{1}})+\beta(\cftreemajor{l_3-l_2-l_{1}})
\end{align*} whereas the formulas from $\Psi_2$ are
\begin{align*}
CF_{BC|AE}&=\alpha(\cftreemajor{b_1})+(1-\alpha)\cftreeminor{b_2}\\
CF_{BA|CE}&=\alpha\cftreeminor{b_1}+(1-\alpha)(\cftreemajor{b_2})\\
CF_{BE|CD}&=\alpha^2(\cftreemajor{b_4-b_5})+2\alpha(1-\alpha)(1-e^{-b_4}+\cftreeminor{b_2-b_4-b_1})
+(1-\alpha)^2(\cftreemajor{b_4-b_3})\\
CF_{AE|CD}&=\alpha^2(\cftreemajor{b_4-b_5-b_1})+2\alpha(1-\alpha)(1-e^{-b_4}+\cftreeminor{b_2-b_4})
+(1-\alpha)^2(\cftreemajor{b_4-b_3})\\
CF_{AB|CD}&=\alpha^2(\cftreemajor{b_4-b_5})+2\alpha(1-\alpha)(1-e^{-b_4}+\cftreeminor{b_4-b_1}) 
+(1-\alpha)^2(\cftreemajor{b_4-b_3-b_2}).
\end{align*}
If we set $b_2=1, b_3=2, b_4=1,b_5=0,b_1=1, \alpha=0.1,
  \beta=0.663163, l_2=1.951019, l_3=0.207841, l_1=1.841435$ as in
  \cite{yu2015}, we see that the first 4 CFs are identical between $\Psi_1$
and $\Psi_2$, but the
  fifth one (corresponding to the subset without the outgroup, not
  present in the triples) differs between $\Psi_1$ and $\Psi_2$,
  allowing us to distinguish
  between both networks.
\begin{align*}
    CF_{BC|AE}(\Psi_1) &= CF_{BC|AE}(\Psi_2) = 0.18584\\
    CF_{BA|CE}(\Psi_1) &= CF_{BA|CE}(\Psi_2) = 0.69153\\
    CF_{BE|CD}(\Psi_1) &= CF_{BE|CD}(\Psi_2) = 0.90743 \\
    CF_{AE|CD}(\Psi_1) &= CF_{AE|CD}(\Psi_2) = 0.914114\\
    CF_{AB|CD}(\Psi_1) = 0.92956 &\neq CF_{AB|CD}(\Psi_2) = 0.956292.
\end{align*}

\pagebreak
\section{Simulated data}
Gene trees were simulated on the networks shown below.

\pagebreak
\section{\textit{Xiphophorus} fish network analysis}

\begin{figure}[ht]
\centering
\includegraphics[scale=0.5]{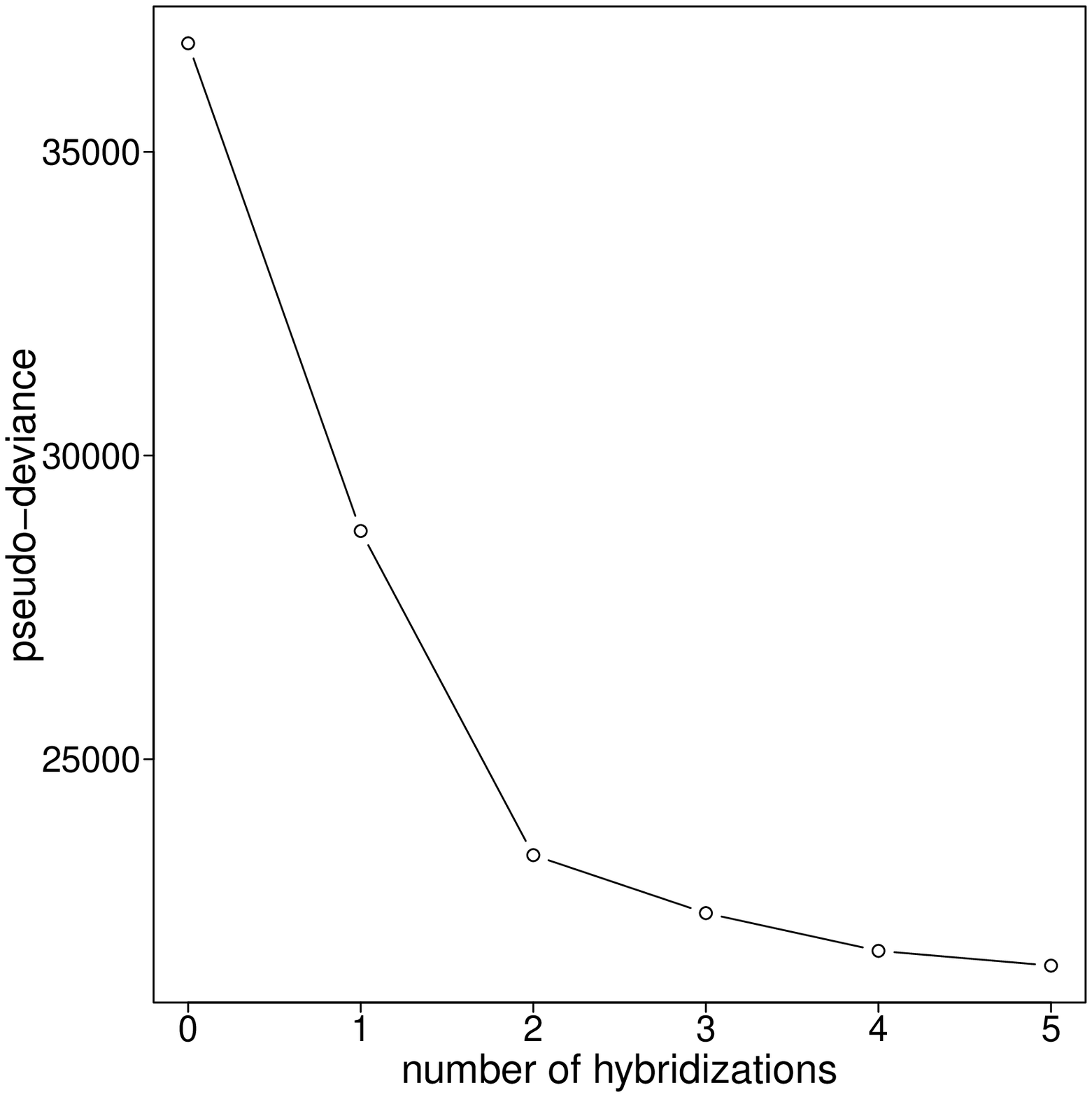}
\caption{Pseudo-deviance score vs number of hybridizations
    for the \textit{Xiphophorus} fish data} 
\label{slope}
\end{figure}


\begin{figure}
\centering
\subfigure[$h=0$]{\includegraphics[scale=0.4]{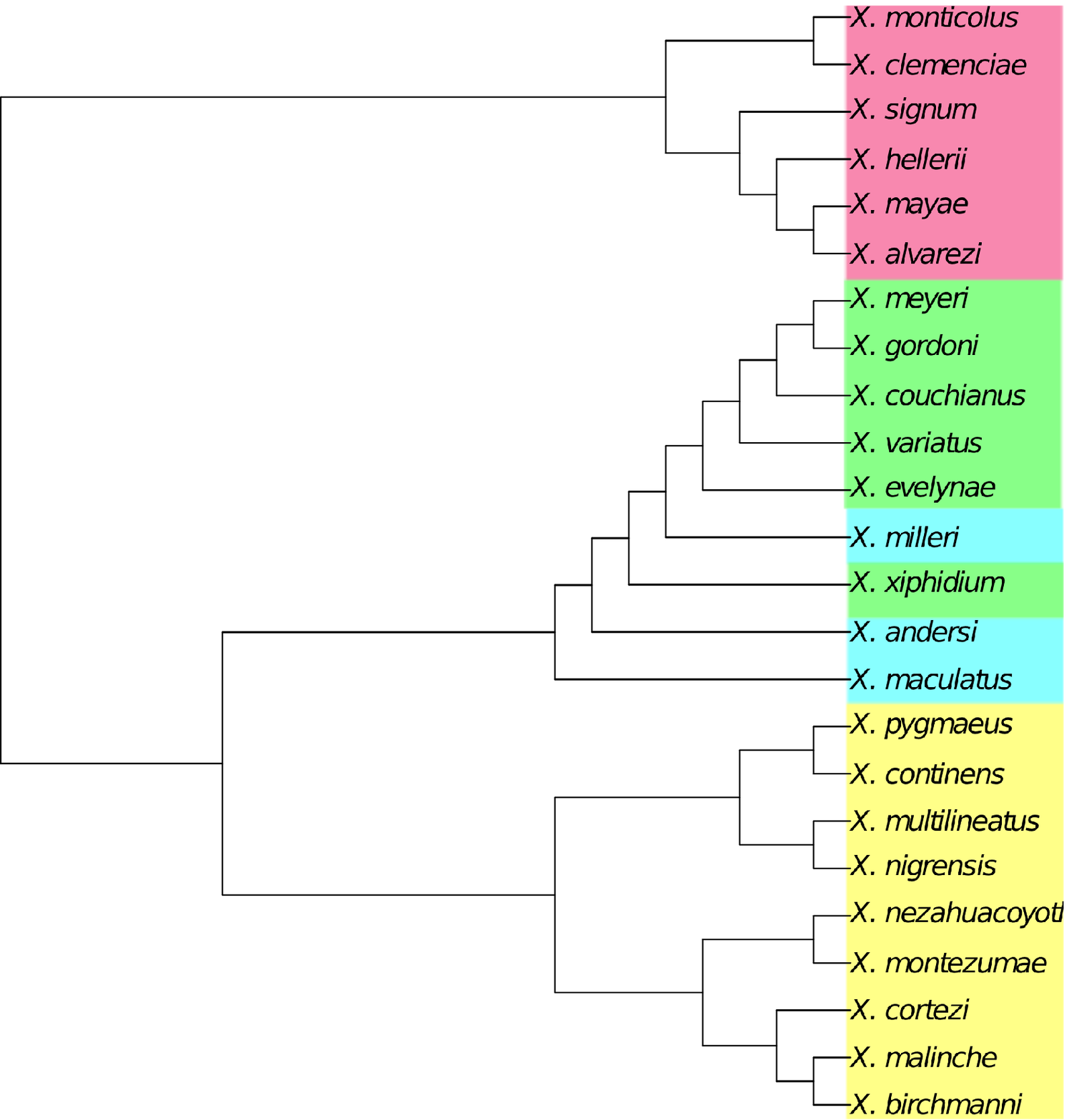}}
\subfigure[$h=1$]{\includegraphics[scale=0.45]{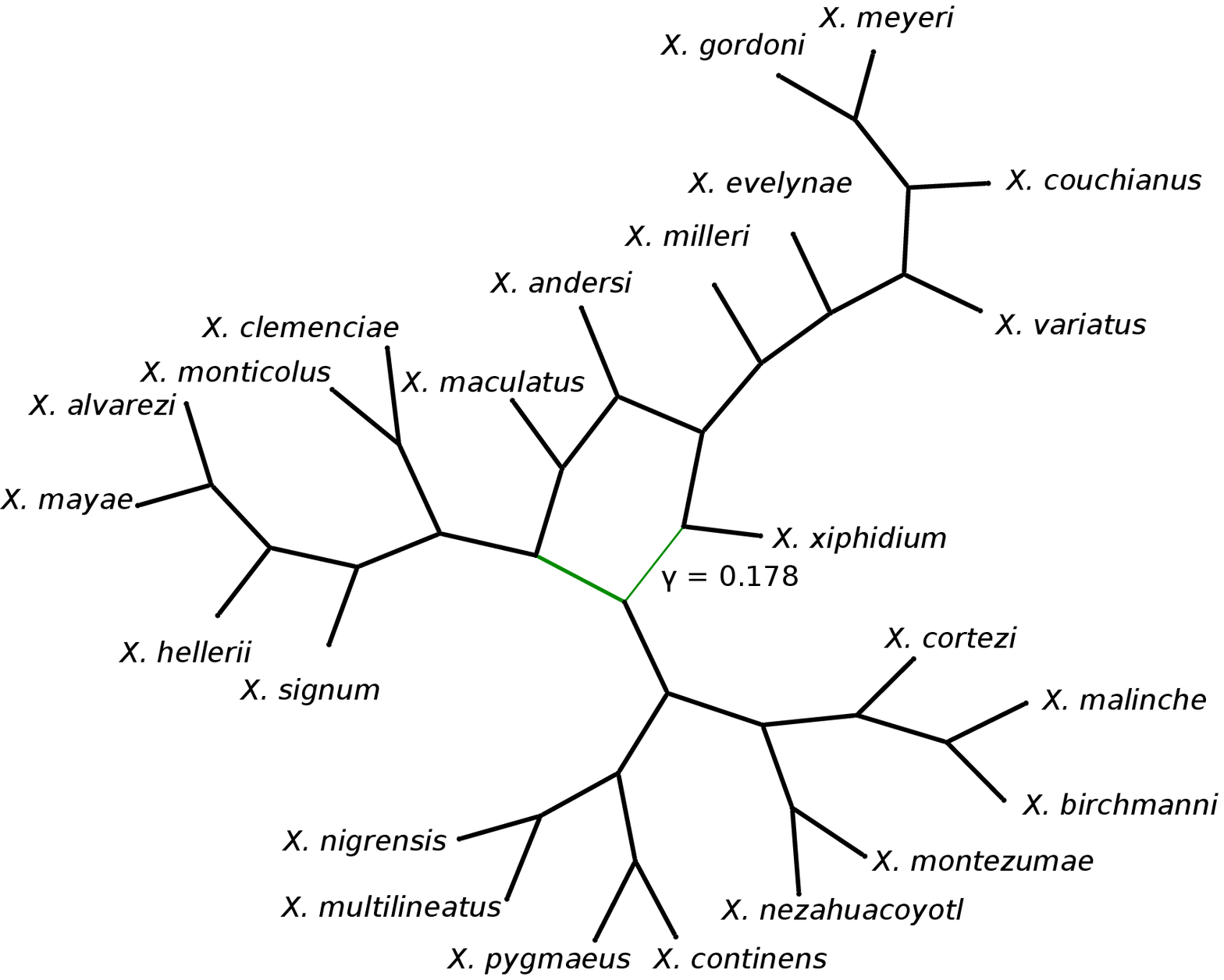}}
\subfigure[$h=2$]{\includegraphics[scale=0.45]{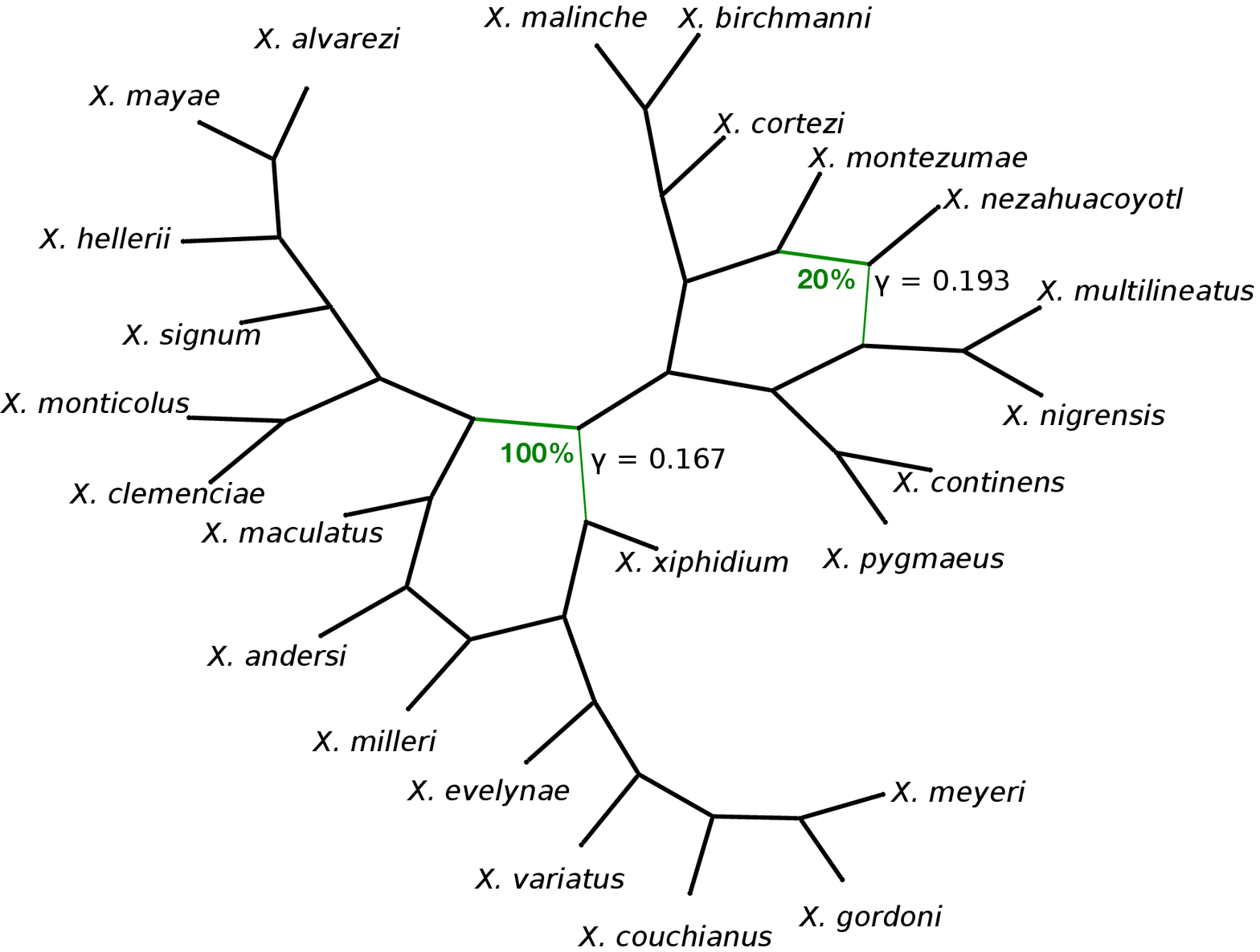}}
\subfigure[$h=3$]{\includegraphics[scale=0.5]{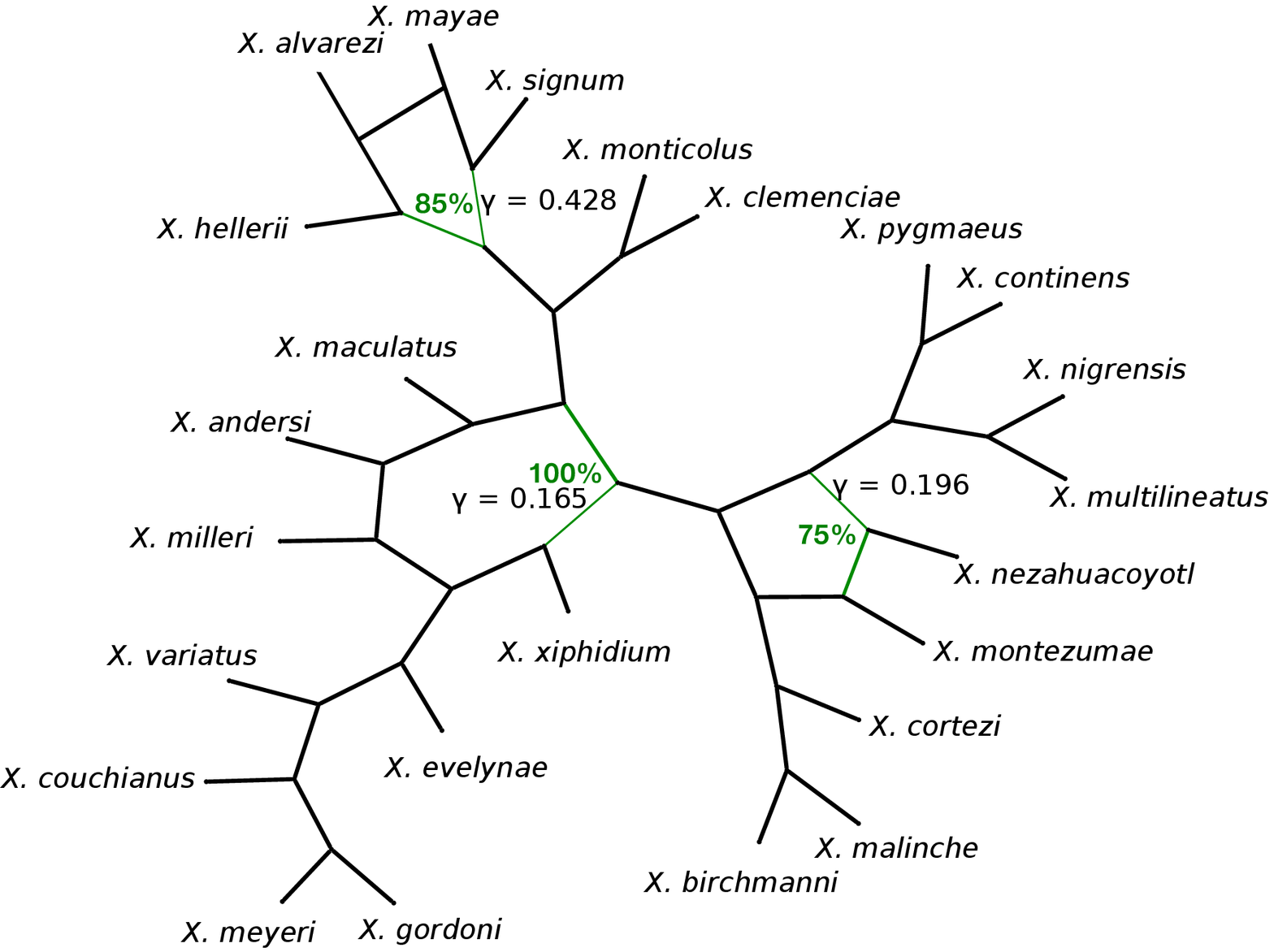}}
\label{cuiPlots}
\end{figure}

\begin{figure}
\centering
\subfigure[$h=4$]{\includegraphics[scale=0.45]{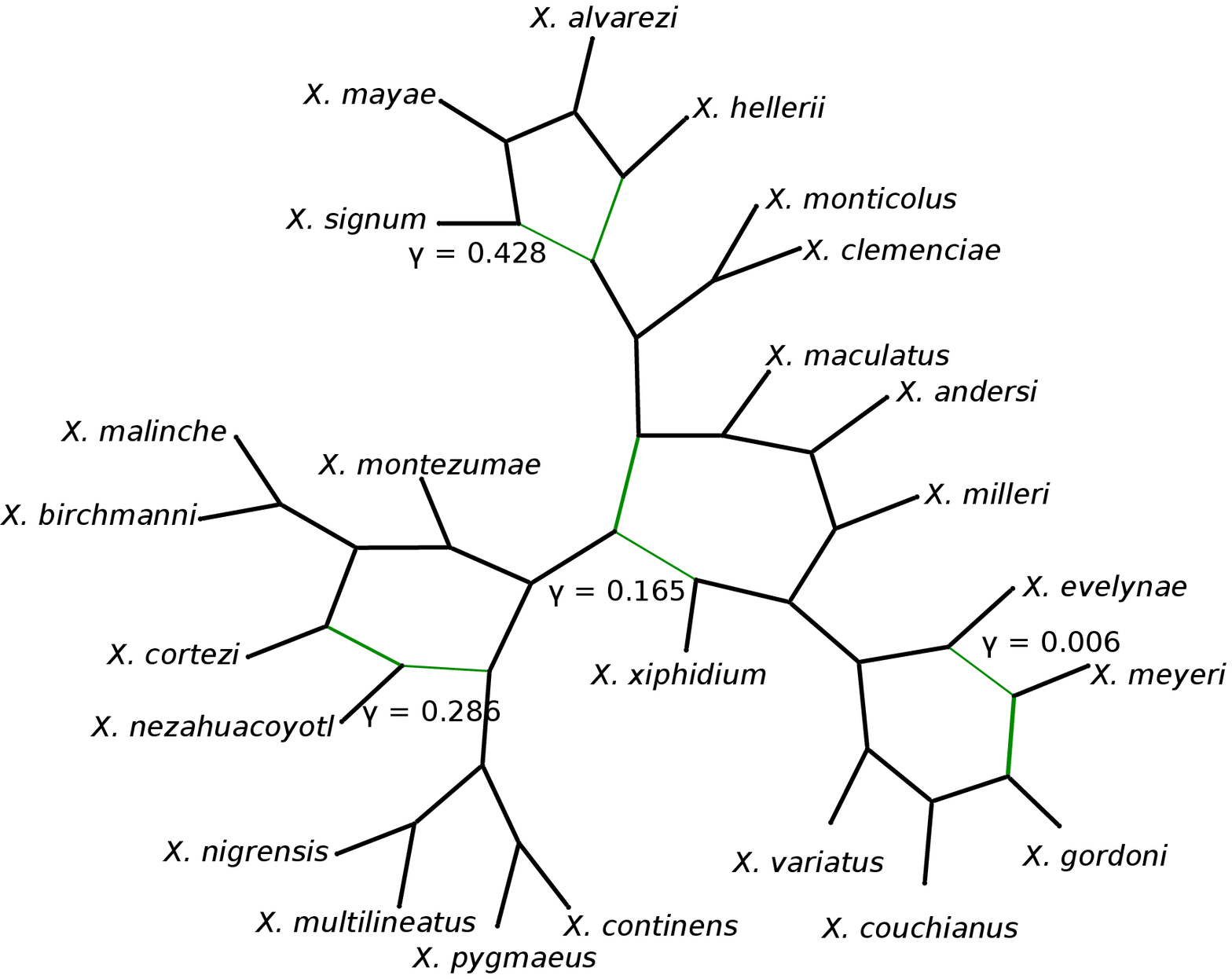}}
\subfigure[$h=5$]{\includegraphics[scale=0.45]{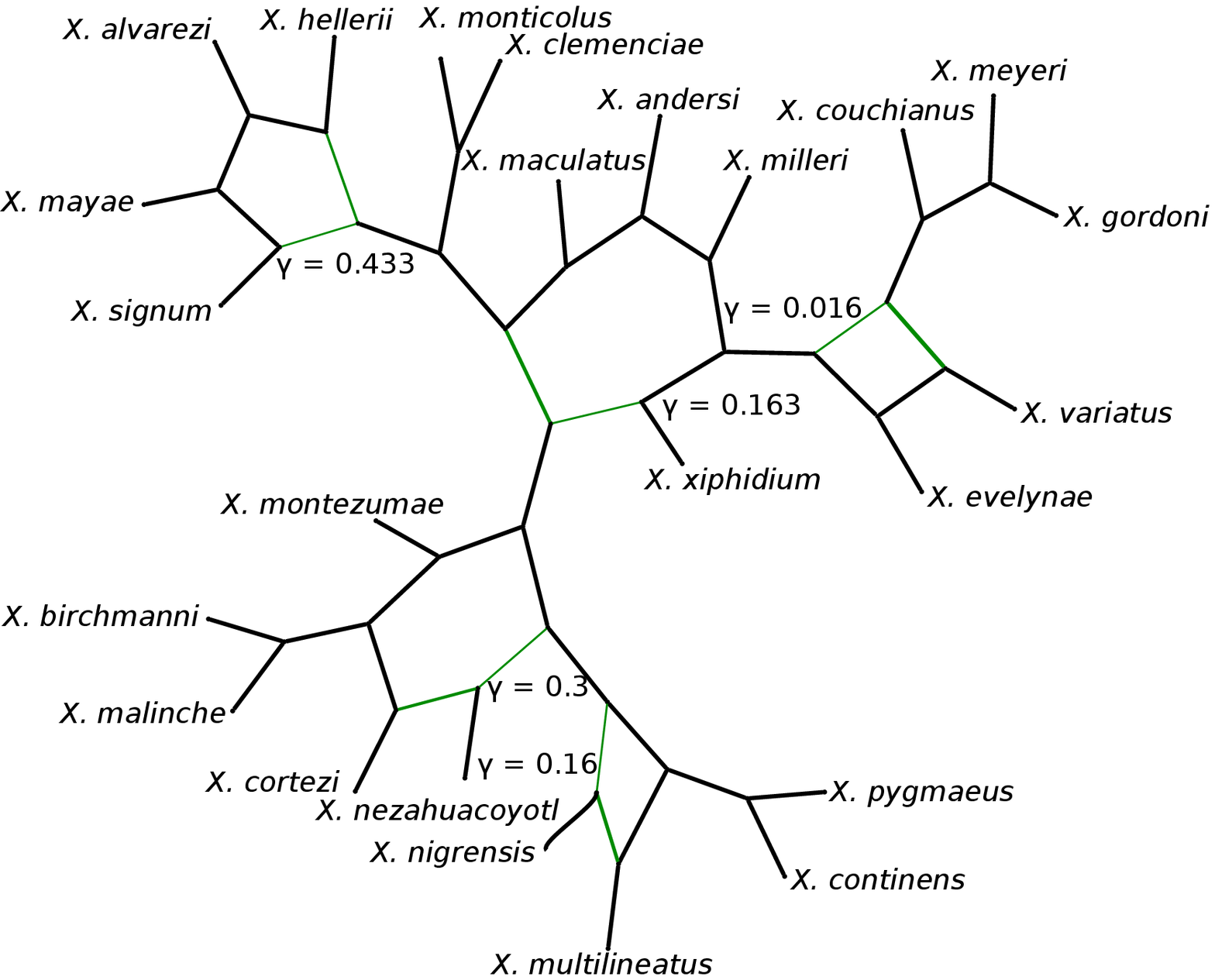}}
\caption{Estimated networks for the \textit{Xiphophorus} fish
    data for $h=0$ to $5$. The estimated tree ($h=0$) is rooted with
    the southern swordtail (SS) outgroup clade (pink). Networks with
    $h\geq1$ are shown as semi-directed networks, as estimated. 
    Hybrid edges, shown in green, are directed toward their hybrid node,
    which is the node where the two hybrid edges meet and represents an 
    ancestral species of mixed parental origins. The minor hybrid edge is
    thinner and annotated with the inheritance probability ($\gamma$)
    whereas the major hybrid edge is thicker and its inheritance
    probability is given by $1-\gamma$ (not shown).    
    With $h\geq 3$, the direction of one hybridization in the SS clade
    conflicts with placing the root at the base of this clade
    (see main text). 
    In the estimated network with $h=2$ and $3$, numbers in green 
    represent bootstrap support for a given
    hybridization. In particular, the 75\% for $h=3$ (or 20\% for
    $h=2$) support value refers to the full hybridization cycle,
    including the placement of \textit{X.~nezahuacoyotl} as sister
    to \textit{X.~montezumae} in the major tree. 
    \textit{X.~nezahuacoyotl} was placed differently (sister to the clade
    \textit{X.~cortezi+X.~birchmanni+X.~malinche}) in 3 bootstrap networks,
    for which the inferred reticulation still had \textit{X.~nezahuacoyotl}
    as the recipient lineage. It also had the same donor lineage as shown 
    in (c) or in (d).} 
\label{cuiPlots2}
\end{figure}












\begin{thebibliography}{100}
\providecommand{\natexlab}[1]{#1}


\bibitem{Huson2010}
Huson D, Rupp R, Scornavacca C. 2010.
\newblock {Phylogenetic Networks}.
\newblock New York, NY: Cambridge University Press, 1st edition.


\bibitem{Spillner2013}
Spillner A, Bastkowski S, Moulton V, Gr\"{u}newald S, B\"{o}gershausen A. 2013.
\newblock {SuperQ: computing supernetworks from quartets.}
\newblock \emph{IEEE/ACM transactions on computational biology and
  bioinformatics / IEEE, ACM}. 10:151--60.


\bibitem{Than2008}
Than C, Ruths D, Nakhleh L. 2008.
\newblock {PhyloNet: a software package for analyzing and reconstructing
  reticulate evolutionary relationships.}
\newblock \emph{BMC bioinformatics}. 9:322.


\bibitem{Yang2014}
Yang J, Gr\"{u}newald S, Xu Y, Wan XF. 2014.
\newblock {Quartet-based methods to reconstruct phylogenetic networks.}
\newblock \emph{BMC systems biology}. 8:21.


\bibitem{gambette2012}
Gambette P, Berry V, Paul C. 2012.
\newblock {Quartets and unrooted phylogenetic networks}.
\newblock \emph{J Bioinform Comput Biol}, 10(4):1250004.


\bibitem{Meng2009}
Meng C, Kubatko LS. 2009.
\newblock {Detecting hybrid speciation in the presence of incomplete lineage
  sorting using gene tree incongruence: a model.}
\newblock \emph{Theoretical population biology}. 75:35--45.



\bibitem{Strimmer2000}
Strimmer K, Moulton V. 2000.
\newblock {Likelihood analysis of phylogenetic networks using directed
  graphical models.}
\newblock \emph{Molecular biology and evolution}. 17:875--81.

\bibitem{Yu2012a}
Yu Y, Degnan JH, Nakhleh L. 2012.
\newblock {The probability of a gene tree topology within a phylogenetic
  network with applications to hybridization detection.}
\newblock \emph{PLoS genetics}. 8:e1002660.

\bibitem{Yu2014}
Yu Y, Dong J, Liu KJ, Nakhleh L. 2014.
\newblock {Maximum Likelihood Inference of Reticulate Evolutionary Histories}.
\newblock \emph{PNAS}. 111:16448--16453.


\bibitem{Nguyen2015}
Nguyen Q, Roos T. 2015.
\newblock {Likelihood-based inference of phylogenetic networks from sequence
  data by PhyloDAG}.
\newblock in \emph{Proc. 2nd International Conference on Algorithms for Computational Biology}, LNBI 9199, Springer, pp126--140.


\bibitem{Cui2013}
Cui R, Schumer M, Kruesi K, Walter R, Andolfatto P, Rosenthal GG. 2013.
\newblock {Phylogenomics reveals extensive reticulate evolution in \textit{Xiphophorus}
  fishes.}
\newblock \emph{Evolution; international journal of organic evolution}.
  67:2166--79.


\bibitem{francisSteel2015}
Francis, AR, Steel, M. 2015.
\newblock {Which Phylogenetic Networks are Merely Trees with Additional Arcs?}
\newblock \emph{Systematic Biology}. 64(5):768--777.


\bibitem{Liu2010}
Liu L, Yu L, Edwards SV. 2010.
\newblock {A maximum pseudo-likelihood approach for estimating species trees
  under the coalescent model.}
\newblock \emph{BMC evolutionary biology}. 10:302.


\bibitem{Baum2007}
Baum DA. 2007.
\newblock {Concordance trees, concordance factors, and the exploration of
  reticulate genealogy}.
\newblock \emph{Taxon}. 56:417--426.


\bibitem{ane2010}
An{\'e} C. 2010.
\newblock {Reconstructing concordance trees and testing the
coalescent model from genome-wide data sets}.
\newblock Chapter 3, p.35-52 in
"Estimating species trees: Practical and theoretical aspects",
L. Knowles and L. Kubatko, eds. Wiley-Blackwell.



\bibitem{allmanDegnanRhodes2011}
Allman ES, Degnan, JH, Rhodes JA. 2011.
\newblock {Identifying the rooted species tree from the
distribution of unrooted gene trees under the coalescent},
\newblock  \emph{J. Math. Biol.}. 62(6):833--862.


\bibitem{macaulay2}
Grayson DR, Stillman ME.
\newblock {Macaulay2, a software system for research in algebraic geometry},
\newblock available at \verb+http://www.math.uiuc.edu/Macaulay2/+.

\bibitem{allman2008}
Allman ES, An\'e C, Rhodes JA. 2008.
\newblock {Identifiability of a Markovian model of molecular evolution with Gamma-distributed rates}.
\newblock \emph{Adv. in Appl. Probab.} 40(1):229--249.


\bibitem{allman2009}
Allman ES, Rhodes JA. 2009.
\newblock {The Identifiability of Covarion Models in Phylogenetics},
\newblock \emph{IEEE/ACM Trans Comput Biol Bioinform},
6(1):76--88.


\bibitem{Pickrell2012}
Pickrell JJK, Pritchard JJK. 2012.
\newblock {Inference of population splits and mixtures from genome-wide allele
  frequency data.}
\newblock \emph{PLoS genetics}. 8:e1002967.


\bibitem{julia}
Bezanson J, Edelman A, Karpinski S, Shah VB. 2014.
\newblock {Julia: A fresh approach to numerical computing}
\newblock \verb+http://arxiv.org/abs/1411.1607+


\bibitem{Mirarab2014}
Mirarab S, Reaz R, Bayzid MS, Zimmermann T, Swenson MS, Warnow T. 2014.
\newblock {ASTRAL: genome-scale coalescent-based species tree estimation}.
\newblock \emph{Bioinformatics}. 30:i541--i548.


\bibitem{Avni2015-qmc}
Avni E, Cohen R and Snir S. 2015.
\newblock {Weighted quartets phylogenetics},
\newblock \emph{Systematic Biology}, 64(2):233--242.


\bibitem{Snir2012-qmc}
Snir S, Rao S. 2012.
\newblock {Quartet MaxCut: A fast algorithm for amalgamating quartet trees},
\newblock \emph{Molecular Phylogenetics and Evolution}, 62(1):1--8.


\bibitem{Huber2015}
Huber KT, Linz S, Moulton V, Wu T. 2015.
\newblock {Spaces of phylogenetic networks from generalized nearest-neighbor
  interchange operations}.
\newblock \emph{Journal of Mathematical Biology}. In press.


\bibitem{Larget2010}
Larget B, Kotha SK, Dewey CN, An\'e C. 2010.
\newblock {BUCKy: Gene tree / species tree reconciliation with Bayesian concordance analysis},
\newblock \emph{Bioinformatics}, 26(22):2910--2911.


\bibitem{Ane2007}
An\'{e} C, Larget B, Baum DA, Smith SD, Rokas A. 2007.
\newblock {Bayesian estimation of concordance among gene trees.}
\newblock \emph{Molecular biology and evolution}. 24:412--26.


\bibitem{mrbayes3}
Ronquist F, Huelsenbeck JP. 2003.
\newblock {MrBayes 3: Bayesian phylogenetic inference under mixed models},
\newblock \emph{Bioinformatics} 19:15721574.


\bibitem{raxml8}
Stamatakis A. 2014.
\newblock {RAxML version 8: a tool for phylogenetic analysis and post-analysis
 of large phylogenies}.
\newblock \emph{Bioinformatics} 30:13121313.


\bibitem{Hudson2002}
Hudson RR. 2002.
\newblock {Generating samples under a Wright-Fisher neutral model of genetic
  variation.}
\newblock \emph{Bioinformatics (Oxford, England)}. 18:337--338.


\bibitem{seqgen}
Rambaut A, Grassly NC. 1997.
\newblock {Seq-Gen: An application for the Monte Carlo simulation of DNA sequence evolution along phylogenetic trees}.
\newblock \emph{Comput. Appl. Biosci.} 13:235-238.


\bibitem{Stenz2015}
Stenz N, Larget B, Baum DA, An\'e C. 2015.
\newblock {Exploring tree-like and non-tree-like patterns using genome
  sequences: and example using the inbreeding plant species Arabidopsis
  thaliana (L.) Heynh.}
\newblock \emph{Systematic Biology}. 64(5):809--823.


\bibitem{Nakhleh2010}
Nakhleh L. 2010.
\newblock {A metric on the space of reduced phylogenetic networks}.
\newblock \emph{IEEE/ACM Transactions on Computational Biology and
  Bioinformatics}. 7:218--222.


\bibitem{Solis-Lemus2015}
Sol\'{i}s-Lemus, C., M. Yang, and C. An\'{e}. 2015.
\newblock {Inconsistency of species-tree methods under gene flow}.
\newblock \emph{Systematic Biology, accepted pending revisions}.



\bibitem{durand-etal-2011}
Durand EY, Patterson N, Reich D, Slatkin M. 2011.
\newblock {Testing for ancient admixture between closely
related populations},
\newblock \emph{Molecular Biology and Evolution}, 28:2239--2252.


\bibitem{clarkMesser2015}
Clark AG, Messer PW. 2015.
\newblock {Conundrum of jumbled mosquito genomes},
\newblock \emph{Science} 347(6217):27--28.


\bibitem{fontaine2015}
Fontaine MC et al. 2015.
\newblock {Extensive introgression in a malaria vector species complex revealed by phylogenomics},
\newblock \emph{Science} 347(6217):1258524


\bibitem{jonsson2015}
J{\'o}nsson MS et al. 2014.
\newblock {Speciation with gene flow in equids despite extensive chromosomal plasticity},
\newblock \emph{PNAS} 111(52):18655--18660


\bibitem{camaraLevineRabadan}
Camara PG, Levine AJ, Rabadan R. 2015.
\newblock {Inference of ancestral recombination graphs through
topological data analysis}.
\newblock \verb+http://arxiv.org/abs/1505.05815v1+


\bibitem{Pardi2015}
Pardi F, Scornavacca C. 2015.
\newblock {Reconstructible Phylogenetic Networks: Do Not Distinguish the
  Indistinguishable}.
\newblock \emph{PLOS Computational Biology}. 11:e1004135.


\bibitem{YuNakhleh2015}
Yu Y, Nakhleh L. 2015.
\newblock {A distance-based method for inferring phylogenetic networks in the presence of incomplete lineage sorting}.
\newblock Bioinformatics Research and Applications, \emph{Lecture Notes in Computer Science} 9096, Springer International Publishing Switzerland. pp.378--389


\bibitem{chifmanKubatko2014}
Chifman J, Kubatko L. 2014.
\newblock {Quartet inference from {SNP} data under the coalescent model},
\newblock \emph{Bioinformatics}, 30(23):3317--3324.

\bibitem{Yu2013}
Yu Y, Barnett RM, Nakhleh L. 2013.
\newblock {Parsimonious inference of hybridization in the presence of
  incomplete lineage sorting.}
\newblock \emph{Systematic biology}. 62:738--51.


\bibitem{kubatko2009}
Kubatko, LS. 2009.
\newblock {Identifying hybridization events in the presence of coalescence via model selection},
\newblock \emph{Systematic Biology} 58(5):478--488


\bibitem{baudry2012}
Baudry J, Maugis C, Michel B. 2012.
\newblock {Slope heuristics: overview and implementation}.
\newblock \emph{Statistics and Computing}, 22(2):455--470


\bibitem{birgeMassart2006}
Birg{\'e} L, Massart P. 2006.
\newblock {Minimal penalties for Gaussian model selection}.
\newblock \emph{Probab. Theory Relat. Fields}, 138:33--73.

\bibitem{yu2015}
Yu Y., Nakhleh, L.
\newblock {A maximum pseudo-likelihood approach for phylogenetic
networks}.
\newblock \emph{BMC Genomics 2015}, 16(Suppl 10): S10.

\bibitem{Bezanson2012}
Bezanson J, Karpinski S, Shah VB, Edelman A. 2012.
\newblock {Julia: A Fast Dynamic Language for Technical Computing}.
\newblock \verb+http://arxiv.org/abs/1209.5145+. 

\bibitem{Cox2007}
Cox D, Little J, O'Shea D. 2007.
\newblock {Ideals, varieties, and algorithms}.
\newblock Springer, third edition.

\bibitem{Feller1950}
Feller W. 1950.
\newblock {Introduction to Probability Theory vol. I}.
\newblock New York, NY: Wiley, third edition.

\bibitem{Huelsenbeck2001}
Huelsenbeck JP, Ronquist F. 2001.
\newblock {MrBayes: Bayesian inference of phylogeny}.
\newblock \emph{Bioinformatics}. 17:754--755.

\bibitem{massart2007}
Massart P. 2007.
\newblock {Concentration inequalities and model selection},
\newblock \emph{\'Ecole d'Et\'e de Probabilit\'es de Saint-Flour}
1896. Springer-Verlag Berlin Heidelberg.



\end{thebibliography}
\end{document}